\documentclass[aps,prb,reprint,superscriptaddress,nobibnotes]{revtex4-1}
\usepackage[version=3]{mhchem}

\usepackage{bm,bbm}   
\usepackage[mathscr]{euscript} 

\usepackage{natmove}

\usepackage{graphicx} 
\usepackage[T1]{fontenc}
\usepackage{balance}
\usepackage{booktabs}
\AtBeginDocument{
    \heavyrulewidth=0.08em
    \lightrulewidth=0.05em
    \cmidrulewidth=0.03em
    \belowrulesep=0.65ex
    \belowbottomsep=0pt
    \aboverulesep=0.4ex
    \abovetopsep=0pt
    \cmidrulesep=\doublerulesep
    \cmidrulekern=0.5em
    \defaultaddspace=0.5em
}

\usepackage{xcolor}
\definecolor{cream}{RGB}{222,217,201} 
\makeatletter 
\newlength{\figrulesep} 
\newcommand{\topfigrule}{\vspace*{-1pt}%
\noindent{\color{cream}\rule[-\figrulesep]{\columnwidth}{1.5pt}} }

\newcommand{\botfigrule}{\vspace*{-2pt}%
\noindent{\color{cream}\rule[\figrulesep]{\columnwidth}{1.5pt}} }

\newcommand{\dblfigrule}{\vspace*{-1pt}%
\noindent{\color{cream}\rule[-\figrulesep]{\textwidth}{1.5pt}} }

\makeatother
\usepackage{hyperref} 

\begin{document}

\title{Analytical nuclear gradients for the range-separated many-body dispersion model of noncovalent interactions}

\author{Martin A. \surname{Blood-Forsythe}}
\thanks{These authors contributed equally to this work.}
\affiliation{\mbox{Department of Chemistry and Chemical Biology, Harvard University, Cambridge, MA, USA.}}

\author{Thomas \surname{Markovich}}
\thanks{These authors contributed equally to this work.}
\affiliation{\mbox{Department of Chemistry and Chemical Biology, Harvard University, Cambridge, MA, USA.}}

\author{Robert A. \surname{DiStasio Jr.}}
\affiliation{\mbox{Department of Chemistry, Princeton University, Princeton, NJ, USA.}}
\affiliation{\mbox{Department of Chemistry and Chemical Biology, Cornell University, Ithaca, NY, USA.}}

\author{Roberto \surname{Car}}
\affiliation{\mbox{Department of Chemistry, Princeton University, Princeton, NJ, USA.}}
\affiliation{\mbox{Princeton Institute for the Science and Technology of Materials, Princeton University, Princeton, NJ, USA.}}
\affiliation{\mbox{Department of Physics, Princeton University, Princeton, NJ, USA.}}
\affiliation{\mbox{Program in Applied and Computational Mathematics, Princeton University, Princeton, NJ, USA.}}

\author{Al\'{a}n \surname{Aspuru-Guzik}}
\email[E-mail: ]{aspuru@chemistry.harvard.edu}

\affiliation{\mbox{Department of Chemistry and Chemical Biology, Harvard University, Cambridge, MA, USA.}}

\begin{abstract}
\noindent Accurate treatment of the long-range electron correlation energy, including van der Waals (vdW) or dispersion interactions, is essential for describing the structure, dynamics, and function of a wide variety of systems. Among the most accurate models for including dispersion into density functional theory (DFT) is the range-separated many-body dispersion (MBD) method [A. Ambrossetti  \textit{et al.}, J. Chem. Phys. \textbf{140}, 18A508 (2014)], in which the correlation energy is modeled at short-range by a semi-local density functional and at long-range by a model system of coupled quantum harmonic oscillators. In this work, we develop analytical gradients of the MBD energy with respect to nuclear coordinates, including all implicit coordinate dependencies arising from the  partitioning of the charge density into Hirshfeld effective volumes. To demonstrate the efficiency and accuracy of these MBD gradients for geometry optimizations of systems with intermolecular and intramolecular interactions, we optimized conformers of the benzene dimer and isolated small peptides with aromatic side-chains. We find excellent agreement with the wavefunction theory reference geometries of these systems (at a fraction of the computational cost) and find that MBD consistently outperforms the popular TS and D3(BJ) dispersion corrections. To demonstrate the performance of the MBD model on a larger system with supramolecular interactions, we optimized the \ce{C60}@\ce{C60H28} buckyball catcher host--guest complex. Finally, we find that neglecting the implicit nuclear coordinate dependence arising from the charge density partitioning, as has been done in prior numerical treatments, leads to an unacceptable error in the MBD forces, with relative errors of $\sim20\%$ (on average) that can extend well beyond 100\%.
\end{abstract}

\maketitle

\section{Introduction \label{sec:introduction}}

A theoretically sound description of noncovalent interactions, such as hydrogen bonding and van der Waals (vdW) or dispersion forces, is often crucial for an accurate and reliable prediction of the structure, stability, and function of many molecular and condensed-phase systems.~\cite{stone_theory_2013,wales_energy_2003,parsegian_van_2006,israelachvili_intermolecular_2011} 
Dispersion interactions are inherently quantum mechanical in nature since they originate from collective non-local electron correlations.  Consequently, they pose a significant challenge for electronic structure theory and often require sophisticated wavefunction-based quantum chemistry methodologies for a quantitatively (and in some cases qualitatively) correct treatment. Over the past decade, this challenge has been addressed by a number of approaches seeking to approximately account for dispersion interactions within the hierarchy of exchange-correlation functional approximations in Kohn-Sham density functional theory (DFT),~\cite{rapcewicz_fluctuation_1991,dobson_constraint_1996,andersson_van_1996,
	elstner_hydrogen_2001,wu_towards_2001,
	wu_empirical_2002,
	dion_2004,lilienfeld_2004,grimme_accurate_2004,
	misquitta_2005, becke_jcp_xdm_2005, becke_jcp_dft_2005, johnson_jcp_hf_2005,
	grimme_2006,becke_jcp_2006,zhao_MO6_2006,
	becke_jcp_2007,becke_jcp_xdm_rev_2007,jurecka_2007,
	silvestrelli_2008,sun_jcp_2008,chai_wB97XD_2008,dilabio_cpl_2008,mackie_2008,
	tkatchenko_2009,roman-perez_2009,vydrov_2009,johnson_jpcc_2009,sato_density_2009,sato_local_2010,
	grimme_d3_2010,tkatchenko_2010,cooper_2010,riley_2010,kannemann_2010,vydrov_2010,lee_2010,
	grimme_review_2011,steinmann_2011,marom_2011,grimme_bj_2011,
	tkatchenko_2012,distasio_pnas_2012,torres_jpcl_2012,
	tkatchenko_jcp_2013,sabatini_prb_2013,silvestrelli_jcp_2013,
	distasio_2014,ambrosetti_2014} 
which is arguably the most successful electronic structure method in widespread use today throughout chemistry, physics, and materials science.~\cite{becke_perspective_2014}

Based on a summation over generalized interatomic London ($C_6/R^6$) dispersion contributions, the class of pairwise-additive dispersion methods provide a simple and computationally efficient avenue for approximately incorporating these ubiquitous long-range interactions within the framework of DFT. (See Ref.~\citenum{klimes_2012} for a recent and comprehensive review of dispersion methods in DFT.) Although these pairwise-additive methods are capable of reliably describing the dispersion interactions in many molecular systems, it is now well known that both quantitative and qualitative failures can occur, as demonstrated recently in the binding energetics of host-guest complexes,~\cite{ambrosetti_jpcl_2014} conformational energetics in polypeptide $\alpha$-helices,~\cite{tkatchenko_2011} cohesive properties in molecular crystals,~\cite{oterodelaroza_jcp_2012,reilly_2013,reilly_jpcl_2013} relative stabilities of (bio)-molecular crystal polymorphs,~\cite{marom_2013,reilly_2014,kronik_2014} and interlayer interaction strengths in layered materials,~\cite{shtogun_many-body_2010,bjorkman_prl_2012} to name a few.  

In each of these cases, the true many-body nature of dispersion interactions becomes important, whether it is due to beyond-pairwise contributions to the dispersion energy, such as the well-known three-body Axilrod-Teller-Muto (ATM) term,~\cite{axilrod_jcp_1943,*muto_jpmsj_1943} electrodynamic response screening effects,~\cite{tkatchenko_2012,dobson_review_2012,reilly_chemsci_2015} 
or the non-additivity of the dynamic polarizability.~\cite{wagner_non-additivity_2014}  One of the most successful models for incorporating these many-body effects into DFT is the many-body dispersion (MBD) model of Tkatchenko \textit{et al.}~\cite{tkatchenko_2012,distasio_pnas_2012,distasio_2014,ambrosetti_2014} which approximates the long-range correlation energy \textit{via} the zero-point energy of a model system of quantum harmonic oscillators (QHOs) coupled to one another in the dipole approximation. The correlation energy derived from diagonalizing the corresponding Hamiltonian of these QHOs is provably equivalent to the random-phase approximation (RPA) correlation energy (through the adiabatic-connection fluctuation-dissipation theorem).~\cite{tkatchenko_jcp_2013,ambrosetti_2014} The MBD model has consistently provided improved qualitative and quantitative agreement with both experimental results and wavefunction-based benchmarks.~\cite{tkatchenko_2012,distasio_pnas_2012}  Notably, MBD correctly predicts the experimentally known relative stabilities of the molecular crystal polymorphs of glycine~\cite{marom_2013} and aspirin,~\cite{reilly_2014} which pairwise methods fail to do. Refs.~\citenum{distasio_2014} and~\citenum{reilly_chemsci_2015} offer recent perspectives on the role of non-additive dispersion effects in molecular materials and the key successes of the MBD model.

In this work, we seek to extend the applicability of the MBD model by deriving and implementing the analytical  gradients of the range-separated many-body dispersion (MBD@rsSCS) energy with respect to nuclear coordinates, thereby enabling  efficient geometry optimizations and molecular dynamics simulations at the DFT+MBD level of theory. This paper is principally divided into a theoretical derivation of the analytical forces in the MBD model (Sec.~\ref{sec:theory}), and a discussion of the first applications of these analytical MBD forces to the optimization of isolated molecular systems (Sec.~\ref{sec:results}). In Secs.~\ref{sec:notation}-\ref{sec:review}, we start by presenting a self-contained summary of the MBD framework to clarify notation and highlight the different dependencies of the MBD energy on the nuclear coordinates. We then derive analytical nuclear gradients of the MBD@rsSCS correlation energy (Sec.~\ref{sec:analytic_derivatives}). In Sec.~\ref{sec:Computational_Details} and Sec.\ref{sec:esi:additional_computational_details} of the accompanying Electronic Supplementary Information (ESI)\footnote{Electronic Supplementary Information (ESI) available: The ESI contains additional derivation details, a comprehensive symbol glossary, and Cartesian coordinates for all structures considered in this work.}, we give computational details. Subsequently, we demonstrate the importance of MBD forces for several representative systems encompassing inter-, intra-, and supra-molecular interactions (Secs.~\ref{sec:benzene}-\ref{sec:catcher}).  We finally examine the role of the implicit nuclear coordinate dependence that arises from the partitioning of the electron density into effective atomic volumes (Sec.~\ref{sec:hirshfeld_gradient}) and conclude with some final remarks on potential avenues for future work.

\section{Theory}\label{sec:theory}

\subsection{Notation employed in this work.~}\label{sec:notation} 

As the theory comprising the MBD model has evolved over the past few years, several notational changes have been required to accommodate the development of a more complete formalism that accounts for the various contributions to the long-range correlation energy in molecular systems and condensed-phase materials. In this section, we provide a current and self-contained review of the MBD@rsSCS model followed by a detailed derivation of the corresponding analytical nuclear gradients (forces). Our discussion most closely follows the notation employed in Refs.~\citenum{ambrosetti_2014,distasio_2014}. To assist in the interpretation of these equations, we have also furnished a glossary of symbols utilized in this work as part of the ESI.\cite{Note1} For a more thorough discussion of the MBD model (including its approximations and physical interpretations), we refer the reader to the original works~\cite{tkatchenko_2012,ambrosetti_2014} as well as a recent review~\cite{distasio_2014} on many-body dispersion interactions in molecules and condensed matter.

Throughout this manuscript, all equations are given in Hartree atomic units ($\hbar=m_e=e=1$) with tensor (vector and matrix) quantities denoted by bold typeface. In this regard, one particularly important bold/normal typeface distinction that will arise below is the difference between the $3\times3$ dipole polarizability tensor,
\begin{equation}
\bm{\alpha} = \begin{pmatrix}
\alpha^{xx} & \alpha^{xy} & \alpha^{xz} \\
\alpha^{yx} & \alpha^{yy} & \alpha^{yz} \\
\alpha^{zx} & \alpha^{zy} & \alpha^{zz}
\end{pmatrix} ,
\end{equation}
and the ``isotropized'' dipole polarizability, a scalar quantity obtained \textit{via}
\begin{equation}
\label{isotropized_alpha}
\alpha = \tfrac{1}{3} {\rm Tr} \big[ \bm{\alpha} \big] .
\end{equation}

The Cartesian components of tensor quantities are indicated by superscript Latin indices $ij$, \textit{i.e.}, ${\rm T}^{\,ij}$ is the $(i,j)^{\, \rm th}$ component of the tensor $\mathbf{T}$. Likewise, Cartesian unit vectors are indicated by $\{\hat{\mathbf{e}}_i,\hat{\mathbf{e}}_j\}$. Atom (or QHO) indices are denoted by subscript Latin indices $abc$. The index $p$ will be used as a dummy index for summation. The imaginary unit is indicated with blackboard bold typeface, $\mathbbm{i}$, to distinguish it from the Cartesian component index $i$. Quantities that arise from the solution of the range-separated self-consistent screening (rsSCS) system of equations introduced by Ambrosetti \textit{et al.}~\cite{ambrosetti_2014} will be denoted by an overline, \textit{i.e.}, $X\rightarrow\overline{X}$. For brevity we will refer to the MBD@rsSCS model (which has also been denoted as MBD* elsewhere) as simply MBD throughout the manuscript. 

The MBD model requires keeping track of several different quantities that are naturally denoted with variants of the letter ``R'', so we highlight these quantities here for the benefit of the reader. Spatial position, such as the argument of the electron density, $\rho(\mathbf{r})$, is indicated by $\mathbf{r}$. The nuclear position of an atom $a$ (or QHO mapped to that atom) is indicated by $\bm{\mathscr{R}}_a$.  The internuclear vector is denoted \mbox{$\mathbf{R}_{ab}= \bm{\mathscr{R}}_a -\bm{\mathscr{R}}_b$,} such that the internuclear distance is given by $R_{ab} = \| \mathbf{R}_{ab} \|$. It follows that the $i^{\, \rm th}$ Cartesian component of this internuclear vector is ${R}_{ab}^i$. Finally, the effective vdW radius of an atom $a$ is indicated by $\mathcal{R}_{~a}^{\,\rm vdW}$.

The dependence of the long-range MBD correlation energy, $E_{\rm MBD}$, on the underlying nuclear positions, $\{\bm{\mathscr{R}}\}=\bm{\mathscr{R}}_a,\bm{\mathscr{R}}_b,\bm{\mathscr{R}}_c,\ldots$, will arise both \textit{explicitly} through the presence of internuclear distance terms, $R_{ab}$, and \textit{implicitly} through the presence of effective atomic volume terms, $V_a=V_a[\{\bm{\mathscr{R}}\}]$, obtained \textit{via} the Hirshfeld partitioning~\cite{hirshfeld_1977} of $\rho(\mathbf{r})$ (see Sec.~\ref{sec:alpha}). As such, these distinct types of dependence on the nuclear positions will be clearly delineated throughout the review of the MBD model and the derivation of the corresponding MBD nuclear forces below. For notational convenience, we will often use $\bm{\partial}_c$ rather than $\bm{\nabla}_{\bm{\mathscr{R}}_c}$ to indicate a derivative with respect to the nuclear position of atom $c$.

\subsection{Review of the many-body dispersion (MBD) model.~}\label{sec:review}
The MBD formalism is based on a one-to-one mapping of the $N$ atoms comprising a molecular system of interest to a collection of $N$ QHOs centered at the nuclear coordinates, each of which is characterized by a \textit{bare} isotropic frequency-dependent dipole polarizability, $\alpha_a(\mathbbm{i}\omega)$.  Derived from the electron density, \textit{i.e.}, $\alpha_a=\alpha_a[\rho(\mathbf{r})]$, these polarizabilities describe the unique local chemical environment surrounding a given atom by accounting for hybridization (coordination number), Pauli repulsion, and other non-trivial exchange-correlation effects (see Sec.~\ref{sec:alpha}). To account for anisotropy in the local chemical environment as well as collective polarization/depolarization effects, the solution of a range-separated Dyson-like self-consistent screening (rsSCS) equation is used to generate \textit{screened} isotropic frequency-dependent dipole polarizabilities for each QHO,  $\overline{\alpha}_a$ (see Sec.~\ref{sec:rsscs}). The MBD model Hamiltonian is then constructed based on these \textit{screened} frequency-dependent dipole polarizabilities. Diagonalization of this Hamiltonian couples this collection of QHOs within the dipole approximation, yielding a set of interacting QHO eigenmodes with corresponding eigenfrequencies $\{\lambda\}$.  The difference between the zero-point energy of these \textit{interacting} QHO eigenmodes and that of the input \textit{non-interacting} modes ($\{\overline{\omega}\}$), is then used to compute the long-range correlation energy at the MBD level of theory (see Sec.~\ref{sec:mbd}), \textit{i.e.},
\begin{equation}
E_{\rm MBD} = \frac{1}{2} \sum_{p=1}^{3N} \sqrt{\lambda_p} - \frac{3}{2} \sum_{a=1}^{N} \overline{\omega}_a .	\label{E_MBD}
\end{equation}

\subsubsection{The MBD starting point: bare dipole polarizabilities.~} \label{sec:alpha}
Mapping the $N$ atoms comprising a molecular system of interest onto a collection of $N$ QHOs is accomplished via a Hirshfeld partitioning\protect\footnote[6]{Although there are numerous schemes for partitioning the electron density, the Hirshfeld prescription (Ref.~\citenum{hirshfeld_1977}) has been shown to result in atomic partitions that most closely resemble the densities of the corresponding free (isolated) atoms (by minimizing the Kullback-Leibler entropy deficiency of information theory) cf. Ref.~\citenum{nalewajski_2000}.} of $\rho(\mathbf{r})$, the ground state electron density. Partitioning $\rho(\mathbf{r})$ into $N$ spherical effective atoms enables assignment of the \textit{bare} frequency-dependent dipole polarizabilities $\bm{\alpha}_a(\mathbbm{i}\omega)$ used to characterize a given QHO. Within the MBD formalism, this assignment is given by the following $0/2$-order Pad\'e approximant applied to the scalar dipole polarizabilities:~\cite{tang_pr_1968}
\begin{equation}\label{eq:alpha_pade}
\alpha_a(\mathbbm{i}\omega) = \frac{\alpha_a(0)}{1-\left(\mathbbm{i}\omega/\omega_a\right)^2} ,
\end{equation}
in which $\alpha_a(0) $ is the \textit{static} dipole polarizability and $\omega_a$ is the characteristic excitation (resonant) frequency for atom $a$.  
The dependence of the \textit{bare} frequency-dependent dipole polarizability in Eq.~\eqref{eq:alpha_pade} on $\rho(\mathbf{r})$ is introduced by considering the direct proportionality between polarizability and atomic volume,~\cite{brinck_1993} an approach that has been very successful in the Tkatchenko-Scheffler (TS) dispersion correction,~\cite{tkatchenko_2009} \textit{i.e.},
\begin{align} \label{TS_polarizability} 
\alpha_a[\rho(\mathbf{r})](0) =& \left ( \frac{V_a[\rho(\mathbf{r})]}{V_a^{\rm free}} \right ) \alpha_a^{\rm free}(0)  \\
=& \left( \frac{\int {\rm d\mathbf{r}} \,  w_a(\mathbf{r}) \rho(\mathbf{r}) r^3}{\int {\rm d\mathbf{r}} \, \rho_a^{\rm free}(\mathbf{r}) r^3 } \right) \alpha_a^{\rm free}(0) , 
\end{align}
in which $V_a^{\rm free}$ and $\alpha_a^{\rm free}$ are the volume and \textit{static} dipole polarizability of the \textit{free} (isolated) atom in vacuo, respectively, obtained from either experiment or high-level quantum mechanical calculations. \textit{Explicit} dependence on $\rho(\mathbf{r})$ resides in the \textit{effective} ``atom-in-a-molecule'' volume, $V_a[\rho(\mathbf{r})]$, obtained \textit{via} Hirshfeld partitioning~\cite{hirshfeld_1977} of $\rho(\mathbf{r})$ into atomic components, in which the weight functions,
\begin{equation}
w_a(\mathbf{r}) = \rho_a^{\rm free}(\mathbf{r}) /\sum_{b} \rho_b^{\rm free}(\mathbf{r}) ,
\end{equation}
are constructed from the set of spherical \textit{free} atom densities, $\{\rho_b^{\rm free}(\mathbf{r})\}$.  At present, we compute the Hirshfeld partitioning and subsequently the MBD energy and forces as an \textit{a posteriori} update to the solution of the non-linear Kohn-Sham equations, \textit{i.e.} without performing self-consistent updates to $\rho(\mathbf{r})$. Future work will address the impacts of computing the Hirshfeld partitioning iteratively~\cite{bultinck_2007} and using the MBD potential to update the Kohn-Sham density self-consistently.  In this regard, recent work on the self-consistent application of the TS method indicates that self-consistency can have a surprisingly large impact on the charge densities, and corresponding work functions, of metallic surfaces,~\cite{ferri_electronic_2015} so we anticipate that self-consistent MBD will be particularly interesting for the study of surfaces and polarizable low-dimensional systems.

For later convenience, we rewrite Eqs.~\eqref{eq:alpha_pade} and~\eqref{TS_polarizability} to collect all quantities that \textit{do not} implicitly depend on the nuclear coordinates through $V_a[\rho(\mathbf{r})]$ into the quantity $\Upsilon_a(\mathbbm{i}\omega)$:
\begin{eqnarray}
\alpha_a[\rho(\mathbf{r})](\mathbbm{i}\omega) &=&  \left[\frac{1}{1-\left(\mathbbm{i}\omega/\omega^{\rm free}_a\right)^2} \frac{\alpha_a^{\rm free}(0) }{V_a^{\rm free}} \right]  V_a [\rho(\mathbf{r})]  \label{TS_polarizability_lambda} \\
&\equiv& \Upsilon_a(\mathbbm{i}\omega)\; V_a[\rho(\mathbf{r})] .   \label{Lambda}
\end{eqnarray}

\subsubsection{Range-separated self-consistent screening (rsSCS).~}\label{sec:rsscs}
Let $\mathbf{A}$ be a $3N\times3N$ block diagonal matrix formed from the frequency-dependent polarizabilities in Eq.~\eqref{TS_polarizability_lambda}:~\footnote[7]{The dipole polarizability tensor $\bm{\alpha}_a$ for a given atom or QHO is formed by populating the diagonal elements ($\alpha_{xx},\alpha_{yy},\alpha_{zz}$) with the \textit{isotropic} dipole polarizability in Eq.~\eqref{TS_polarizability_lambda}.}
\begin{equation}\label{A_TS}
\mathbf{A}(\mathbbm{i}\omega) = \bigoplus_{b=1}^N \bm{\alpha}_b(\mathbbm{i}\omega) = {\rm diag}[\bm{\alpha}_1,\;\bm{\alpha}_2,\;\ldots,\;\bm{\alpha}_N ] .
\end{equation}
This quantity will be referred to as the \textit{bare} system dipole polarizability tensor.
For a given frequency, range-separated self-consistent screening (rsSCS) of $\mathbf{A}(\mathbbm{i}\omega)$ is then accomplished by solving the following matrix equation~\cite{thole_1981,distasio_2014} (see the ESI\cite{Note1} for the detailed derivation of Eq.~\eqref{rsSCS}):
\begin{eqnarray}
\overline{\mathbf{A}} &=& \mathbf{A} - \mathbf{A} \, \mathbf{T}_{\rm SR} \, \overline{\mathbf{A}}   \label{rsSCS_1}\\
\Rightarrow \overline{\mathbf{A}} &=& \Big[\mathbf{A}^{-1} + \mathbf{T}_{\rm SR}\Big]^{-1} ,\label{rsSCS}
\end{eqnarray}
where $\mathbf{T}_{\rm SR}$ is the short-range dipole--dipole interaction tensor, defined below in Sec.~\ref{sec:dipole_interaction_tensor} Eq.~\eqref{T_SR}. The matrix $\overline{\mathbf{A}}$ is the (dense) \textit{screened} non-local polarizability matrix, sometimes called the relay matrix.~\protect\footnote[8]{At this point, it is very important to note a difference in the notation relative to \mbox{Refs.~\citenum{distasio_2014} \&~\citenum{tkatchenko_jcp_2013}:} our matrix $\overline{\mathbf{A}}$ is equivalent to their $\overline{\mathbf{B}}$ or $B$, which was keeping with Thole's original notation for the relay matrix (Ref.~\citenum{thole_1981}).}
 
Partial internal contraction over atomic sub-blocks of $\overline{\mathbf{A}}$ yields the screened and anisotropic \textit{atomic} polarizability tensors (the corresponding \textit{molecular} polarizability is obtained by total internal contraction), \textit{i.e.},
\begin{equation}\label{screened_polarizabilities}
\overline{\bm{\alpha}}_{a}(\mathbbm{i}\omega) = \sum_{b=1}^N \overline{\mathbf{A}}_{ab}(\mathbbm{i}\omega).
\end{equation}
The static ``isotropized'' screened polarizability scalars, $\overline{\alpha}_a(0)$, that appear in the MBD Hamiltonian in Eq.~\eqref{H_MBD} and Sec.~\ref{sec:mbd} below are then calculated from $\overline{\bm{\alpha}}_{a}(0)$ \textit{via}
\begin{equation}\label{screened_static_polarizabilities}
\overline{\alpha}_a(0) = \frac{1}{3} {\rm Tr}\Big[ \overline{\bm{\alpha}}_a(0)  \Big]
\end{equation}
as described above in Eq.~\eqref{isotropized_alpha}. Note that Eqs.~(\ref{rsSCS}-\ref{screened_polarizabilities}) can be solved at any imaginary frequency, $\mathbbm{i}\omega$, so we do not require the Pad\'{e} approximant given in Eq.~\eqref{eq:alpha_pade} to bootstrap from $\overline{\alpha}_a(0)$ to $\overline{\alpha}_a(\mathbbm{i}\omega)$.  However, the relationship between $\overline{\omega}_a$ and $\overline{C}_{6,aa}$, given in Eq.~\eqref{screened_omega}, is one that is derived from the Pad\'{e} approximant for the bare polarizability $\alpha(\mathbbm{i}\omega)$.

In the non-retarded regime, the Casimir-Polder integral relates the effective $C_{6,ab}$ dispersion coefficient to the dipole polarizabilities of QHOs $a$ and $b$ \textit{via} the following integral over imaginary frequencies:\cite{casimir_1948}
\begin{equation}\label{casimir-polder}
C_{6,ab} =\frac{3}{\pi} \int_0^\infty {\rm d \omega }\; \alpha_a(\mathbbm{i}\omega) \alpha_b(\mathbbm{i}\omega) .
\end{equation}
By solving Eqs.~(\ref{rsSCS}-\ref{screened_polarizabilities}) on a grid of imaginary frequencies $\{\mathbbm{i}y_p\}$, a set of screened effective $C_6$ coefficients, $\{\overline{C}_6\}$, can be determined by a Gauss-Legendre quadrature estimate of the integral in Eq.~\eqref{casimir-polder}.
The screened QHO characteristic excitation frequency, $\overline{\omega}_a$, is then calculated as
\begin{equation}
\label{screened_omega}
\overline{\omega}_a = \frac{4}{3} \frac{\overline{C}_{6,aa}}{[\overline{\alpha}_a(0)]^2}  = \frac{4}{\pi}  \sum_{p} g_p \left[ \frac{ \overline{\alpha}_a(\mathbbm{i}y_p) }{ \overline{\alpha}_a(0)  } \right]^2 ,
\end{equation}
where $g_p$ and $y_p$ are the quadrature weights and abscissae, respectively. Scaling of the usual Gauss-Legendre abscissae from $[-1,1]$ to the semi-infinite interval $[0,\infty)$ is discussed in the accompanying ESI.\cite{Note1}

\subsubsection{The MBD model Hamiltonian.~} \label{sec:mbd}
The central concept in the MBD model is the Hamiltonian for a set of coupled QHOs that each fluctuate within an isotropic harmonic potential $U(\mathbf{x}_a) = \frac{1}{2} m_a \omega_a^2  \mathbf{x}_a ^2 $, and acquire instantaneous dipole moments, $\mathbf{d}_a = q_a \mathbf{x}_a$, that are proportional to the displacement, $\mathbf{x}_a$,  from the  equilibrium position and  charge, $q_a$, on each oscillator.
This Hamiltonian defines the so-called coupled fluctuating dipole model (CFDM),~\cite{donchev_2006} and is given by:
\begin{equation}\label{H_CFDM}
\bm{\mathcal{H}}_{\rm CFDM} = - \sum_{a=1}^N  \frac{1}{2} \frac{  \bm{\nabla}^2_{\mathbf{x}_a } }{ m_a}  +  \sum_{a=1}^N   \frac{1}{2} m_a \omega_a^2   \mathbf{x} _a^2
 + \sum_{a>b}^N  \mathbf{d}_a^{\dag} \mathbf{T}_{ab}  \mathbf{d}_b ,
\end{equation}
where $\mathbf{T}_{ab}$ is the dipole--dipole interaction tensor that couples dipoles $a$ and $b$.

In the range-separated MBD model,~\cite{ambrosetti_2014} $\mathbf{T}$ is replaced by a long-range screened interaction tensor, $\overline{\mathbf{T}}_{\;\rm LR}$ (as defined in Sec.~\ref{sec:dipole_interaction_tensor} and Eq.~\eqref{T_LR} below), and the fluctuating point dipoles are replaced with the Gaussian charge densities of  QHOs, with effective masses \mbox{$m_a = \left(\overline{\alpha}_a(0)\; \overline{\omega}_a^2\right)^{-1}$} obtained from their respective static polarizabilities and excitation frequencies. The corresponding range-separated MBD model Hamiltonian is therefore:~\cite{ambrosetti_2014}
\begin{align}\label{H_MBD}
\bm{\mathcal{H}}_{\rm MBD} =& - \sum_{a=1}^N   \frac{1}{2}\bm{\nabla}^2_{\bm{\mu}_a }  + \sum_{a=1}^N \frac{1}{2}    \overline{\omega}_a^2 \bm{\mu}_a^2
\\ &+ \sum_{a>b}^N \overline{\omega}_a \overline{\omega}_b \sqrt{  \overline{\alpha}_a(0) \overline{\alpha}_b(0) }\; \bm{\mu}_a^\dag \overline{\mathbf{T}}^{\;\rm  LR}_{ab}  \bm{\mu}_b, \nonumber
\end{align}
in which $\bm{\mu}_a=\sqrt{m_a}\,\bm{\xi}_a$ is the mass-weighted dipole moment\footnote{Since each QHO is assigned a unit charge ($e=1$), the dipole moment $\bm{\mu}$ is thereby equivalent to the displacement vector $\bm{\xi}$.} of QHO $a$ that has been displaced by $\bm{\xi}_a$ from its equilibrium position. 
The first two terms in Eq.~\eqref{H_MBD} represent the kinetic and potential energy of the individual QHOs, respectively, and the third term is the two-body coupling due to the long-range dipole--dipole interaction tensor, $\overline{\mathbf{T}}^{\;\rm  LR}_{ab}$, defined below in Eq.~\eqref{T_LR}.

By considering the single-particle potential energy and dipole--dipole interaction terms in Eq.~\eqref{H_MBD}, we can construct the \mbox{$3N\times 3N$} MBD interaction matrix, which is comprised of $3\times 3$ subblocks describing the coupling of each pair of QHOs $a$ and $b$:
\begin{equation}\label{C_MBD}
\mathbf{C}^{\rm MBD}_{ab} = \delta_{ab}  \overline{\omega}_a^2 +(1-\delta_{ab})   \overline{\omega}_a  \overline{\omega}_b \sqrt{  \overline{\alpha}_a(0) \overline{\alpha}_b(0)}\; \overline{\mathbf{T}}^{\;\rm LR}_{ab},
\end{equation}
where $\delta_{ab}$ is the Kronecker delta between atomic indices.

The eigenvalues $\{\lambda_p\}$ obtained by diagonalizing $\mathbf{C}^{\rm MBD}$ correspond to the \textit{interacting} (or ``dressed'') QHO modes, while $\overline{\omega}_a$ correspond to the modes of the \textit{non-interacting} reference system of screened oscillators. The MBD correlation energy is then evaluated \textit{via} Eq.~\eqref{E_MBD} as the zero-point energetic difference between the interacting and non-interacting modes. 

For periodic systems, all instances of the dipole--dipole interaction tensor would be replaced by
\begin{equation}
\mathbf{T}_{ab}  \to  \mathbf{T}_{ab}+\sum_{b'} \mathbf{T}_{ab'}
\end{equation}
where the sum over $b'$ indicates a lattice sum over the periodic images of atom $b$. Since this is an additive modification of $\mathbf{T}$, it will not qualitatively modify the expressions for the analytical nuclear derivatives of the MBD energy. Hence, the derivation of the nuclear forces presented herein (and the accompanying chemical applications) will focus on non-periodic (or isolated) systems. We note in passing that the current implementation of the MBD energy and nuclear forces in \textsc{Quantum ESPRESSO} (QE)~\cite{giannozzi_2009} is able to treat both periodic and non-periodic systems. In this regard, a forthcoming paper~\cite{markovich_unpublished} will describe the details of the implementation and discuss the subtleties required to make the computation of well-converged MBD nuclear forces efficient for periodic systems.

\subsubsection{The range-separated dipole--dipole interaction.~}\label{sec:dipole_interaction_tensor}

Prior to range-separation, the $3\times 3$ sub-block $\mathbf{T}_{ab}$ of the dipole--dipole interaction tensor $\mathbf{T}$, which describes the coupling between QHOs $a$ and $b$, is defined as:
\begin{equation}\label{dipole_interaction_tensor}
\mathbf{T}_{ab} = \bm{\nabla}_{\bm{\mathscr{R}}_a} \otimes \bm{\nabla}_{\bm{\mathscr{R}}_b} v_{ab} ,
\end{equation}
where $v_{ab}$ is the frequency-dependent Coulomb interaction between two spherical Gaussian charge distributions.~\cite{helgaker_2000} This frequency-dependent interaction arises due to the fact that the ground state of a QHO has a Gaussian charge density:
\begin{equation}\label{Coulomb_interaction_Gaussians}
v_{ab}(R_{ab},\mathbbm{i}\omega) = \frac{ {\rm erf}\left[\zeta_{ab}(\mathbbm{i}\omega)\right]}{R_{ab}} ,
\end{equation}
where
$R_{ab}=\|\bm{\mathscr{R}}_a-\bm{\mathscr{R}}_b\|$,
\begin{equation}\label{zeta}
\zeta_{ab}(\mathbbm{i}\omega)\equiv R_{ab}/\Sigma_{ab}(\mathbbm{i}\omega)
\end{equation}
and
\begin{equation}\label{Sigma}
\Sigma_{ab}(\mathbbm{i}\omega) = \sqrt{\sigma_a(\mathbbm{i}\omega)^2+\sigma_b(\mathbbm{i}\omega)^2}
\end{equation}
is the effective correlation length of the interaction potential defined by the widths of the QHO Gaussians (see Eq.~\eqref{qho_width}, below). As such, the dependence of $\mathbf{T}$ on both the frequency and (implicitly) on the nuclear coordinates originates from $\Sigma_{ab}(\mathbbm{i}\omega)$ (see also Eqs.~\eqref{TS_polarizability_lambda}-\eqref{Lambda}).

In terms of the bare dipole polarizability, the width of the QHO ground-state Gaussian charge density is given by:
\begin{eqnarray}
\sigma_a(\mathbbm{i}\omega) &=& \left[ \tfrac{1}{3}\sqrt{\tfrac{2}{\pi}}  \alpha_a(\mathbbm{i}\omega)\right]^{1/3} \label{qho_width} \\
 &=& \left[ \tfrac{ 1}{3}\sqrt{\tfrac{2}{\pi}}\Upsilon_a(\mathbbm{i}\omega) \right]^{1/3}  \left[ V_a\right]^{1/3},   \label{sigma_in_terms_of_Veff}
\end{eqnarray}
where $\alpha_a(\mathbbm{i}\omega)= \tfrac{1}{3}{\rm Tr}\left[\bm{\alpha}_a\right] $ is the ``isotropized''  \textit{bare} dipole polarizability
and Eq.~\eqref{Lambda} was used to make the effective volume dependence more explicit.

The Cartesian components of the dipole--dipole interaction tensor in Eq.~\eqref{dipole_interaction_tensor} (with all QHO indices and  frequency-dependence of $\zeta$ suppressed) are given by:
\begin{align} \label{TGG_cartesian}
\mathbf{ T}^{\,ij}(\mathbbm{i}\omega) =&\left[{\rm erf}[\zeta]
-\frac{2\zeta}{\sqrt{\pi}}\exp\left[-\zeta^2\right]\right]  \mathbf{ T}_{\rm dip}^{\,ij}
\\& +\frac{4}{\sqrt{\pi}}\frac{R^i R^j }{ R^5} \zeta^3\exp\left[-\zeta^2\right] , \nonumber
\end{align}
where $R^i =\mathbf{R}_{ab}\cdot \hat{\mathbf{e}}_i$ is the $i^{th}$ Cartesian component of $\mathbf{R}_{ab}$,
and $\mathbf{T}_{\rm dip}$ is the frequency-\textit{independent} interaction between two point dipoles:
\begin{equation}
{\rm T}_{\;\rm dip}^{\,ij} = \frac{-3 R^i R^j +R^2\delta_{\,ij}}{ R^5} ,		\label{T_dip}
\end{equation}
with $\delta_{\,ij}$ indicating the Kronecker delta between Cartesian indices.

The range-separation of the dipole--dipole interaction tensor is accomplished by using a Fermi-type damping function,~\cite{tkatchenko_2009,silvestrelli_2008,grimme_2006} 
\begin{equation}
f( Z_{ab} ) =  \Big[1+ \exp \left[- Z_{ab}  \right]  \Big]^{-1} 		\label{f_damp},
\end{equation}
which depends on $Z_{ab}$, the ratio between $R_{ab}$, the internuclear distance,  and  $S_{ab}$, the scaled sum of the effective vdW radii of atoms $a$ and $b$, $\mathcal{R}_{~a}^{\rm vdW}$ and $\mathcal{R}_{~b}^{\rm vdW}$: 
\begin{eqnarray}
Z_{ab} &\equiv & 6\left[\frac{ R_{ab}  }{ S_{ab} }  -1\right] 		 \label{Z_ab} \\
S_{ab}& \equiv& \beta \left[ \mathcal{R}_{~a}^{\rm vdW} + \mathcal{R}_{~b}^{\rm vdW} \right] . 	\label{S_ab}
\end{eqnarray}
Here, the range-separation parameter $\beta$ is fit once for a given exchange-correlation functional by minimizing the energy deviations with respect to highly accurate reference data.~\cite{ambrosetti_2014}
The short- and long-range components of the dipole--dipole interaction tensor in Eq.~\eqref{TGG_cartesian} are then separated according to:
\begin{equation}
\mathbf{T}_{\rm SR} =\left[1-f(Z)\right] \mathbf{T} 
\end{equation}
and
\begin{equation}
\mathbf{T}_{\rm LR} = f(Z) \mathbf{T}  \label{eq:exact_range_separation_of_T_LR} .
\end{equation}  
However, at long-range, the frequency-dependence in $\mathbf{T}$ dies off quickly, so when evaluating the MBD Hamiltonian we replace Eq.~\eqref{eq:exact_range_separation_of_T_LR} with the approximation
\begin{equation}\label{eq:T_to_Tdip_approx}
\mathbf{T}_{\rm LR} \simeq  f(Z) \mathbf{T} _{\rm dip}
\end{equation}
which is equivalent to taking ${\rm erf}\left[ \zeta \right] \simeq 1$ and $\exp[-\zeta^2]\simeq 0$ in Eq.~\eqref{TGG_cartesian} and~\eqref{eq:exact_range_separation_of_T_LR}.  This has the added benefit of improved computational efficiency since special functions such as the error function and exponential are relatively costly to compute.
As shown in Fig.~\ref{fig:esi:damping_lengthscales} in the ESI,\cite{Note1} these approximations are exact to within machine precision for $\zeta>6$, and thus in practice by the time $f(Z)$ has obtained a substantial value, the frequency dependence in $\mathbf{T}$ has vanished, thereby justifying Eq.~\eqref{eq:T_to_Tdip_approx}.

The rsSCS procedure described in Sec.~\ref{sec:rsscs} adds a further subtlety in that it modifies the effective vdW radii in the definition of the $S_{ab}$ and $Z_{ab}$ quantities above (see Refs.~\citenum{distasio_2014,tkatchenko_2012} for a more detailed discussion of these definitions). For the short-range interaction tensor (\textit{i.e.}, the tensor used in the rsSCS procedure) the damping function utilizes effective vdW radii calculated at the Tkatchenko-Scheffler (TS) level:~\cite{tkatchenko_2009}
\begin{equation}\label{RvdW_TS}
\mathcal{R}_{~a}^{\rm vdW,\,TS} \equiv \left( \frac{V_a}{V_a^{\rm free}} \right)^{1/3} \mathcal{R}_{~a}^{\rm vdW,\, free}
\end{equation}
where $\mathcal{R}_{~a}^{\rm vdW,\, free}$ is the free-atom vdW radius defined in Ref.~\citenum{tkatchenko_2009} using an electron density contour, not the Bondi~\cite{bondi_1964} radius that corresponds to the ``atom-in-a-molecule'' analog of this quantity.
To indicate that the TS-level effective vdW radii are being used, the argument of the damping function for the short-range interaction tensor, used in Eqs.~(\ref{rsSCS_1}-\ref{rsSCS}), will be denoted with $Z^{\rm TS}$ (cf. Eqs. (\ref{RvdW_TS}, \ref{Z_ab}-\ref{S_ab})):
\begin{equation}\label{T_SR}
 \mathbf{T}_{\rm SR} = \Big[1-  f\left(Z^{\rm TS}\right)\Big] \mathbf{T} .
\end{equation}
For the long-range dipole--dipole interaction tensor used in the MBD Hamiltonian in Eq.~\eqref{H_MBD}, the damping function utilizes the self-consistently screened effective vdW radii~\cite{tkatchenko_2012}:
\begin{equation}\label{RvdW_SCS}
 \overline{\mathcal{R}}_{~a}^{\rm vdW} \equiv \left( \frac{\overline{\alpha}_a(0)}{\alpha_a^{\rm free}(0)} \right)^{1/3} \mathcal{R}_{~a}^{\rm vdW,\, free} ,
 \end{equation}
wherein the ratio $\overline{\alpha}(0)/\alpha^{\rm free}(0)$ takes the place of $V/V^{\rm free}$ thereby still exploiting the proportionality between polarizability and volume.~\cite{brinck_1993,distasio_2014}
To indicate that the screened effective vdW radii are being used, the argument of the damping function for the long-range interaction tensor will be denoted with $\overline{Z}$  (cf. Eqs. (\ref{RvdW_SCS}, \ref{Z_ab}-\ref{S_ab})):
\begin{equation}\label{T_LR}
\overline{\mathbf{T}}_{\rm LR} = f\left(\, \overline{Z}\,\right)\mathbf{T}_{\rm dip} .
\end{equation}
This dependence on $\overline{Z}$ is why we use an overline on $\mathbf{T}_{\rm LR}$ above, and in Eqs.~(\ref{H_MBD},\ref{C_MBD}).

\subsection{Derivation of the MBD nuclear forces.~} \label{sec:analytic_derivatives}

With the above definitions in hand, we are now ready to proceed with the derivation of the analytical derivatives of the MBD correlation energy with respect to the nuclear (or nuclear) position $\bm{\mathscr{R}}_c$ of an arbitrary atom $c$. These MBD forces are added to the DFT-based forces. As mentioned above in Sec.~\ref{sec:notation}, two distinct types of nuclear coordinate dependence will arise: \textit{explicit} dependence through $\mathbf{R}_{ab}=\bm{\mathscr{R}}_a-\bm{\mathscr{R}}_b$ and \textit{implicit} dependence through $V[\{\bm{\mathscr{R}}\}]$ (as moving a neighboring atom $c$ will slightly alter the effective volume assigned to atom $a$). Future work will address the effects of the MBD contribution to the exchange-correlation potential when applied self-consistently, which will ultimately impact $\rho(\mathbf{r})$. Our current work neglects these effects, and computes MBD as an \textit{a posteriori} correction to DFT, \textit{i.e.}, \textit{non}-self-consistently.

Having carefully separated out the implicit dependence on $V[\{\bm{\mathscr{R}}\}]$ in the relevant quantities above, the derivation proceeds largely by brute force application of the chain and product rules. The derivative of the MBD correlation energy given in Eq.~\eqref{E_MBD} is governed by:
\begin{eqnarray}
\bm{\partial}_c E_{\rm MBD} &=&  \frac{1}{2} \sum_{p=1}^{3N}  \bm{\partial}_c \sqrt{\lambda_p} - \frac{3}{2} \sum_{a=1}^N \bm{\partial}_c \overline{\omega}_a ,
\end{eqnarray}
hence requiring derivatives of the screened excitation frequencies, $\overline{\omega}_a$, as well as the eigenvalues, $\lambda_p$, of the $\mathbf{C}^{\rm MBD}$ matrix. Since $\mathbf{C}^{\rm MBD}$ is real and symmetric, it has $3N$ orthogonal eigenvectors. We therefore do not concern ourselves here with repeated eigenvalues (see the ESI\cite{Note1} for a more detailed discussion) and take derivatives of $\lambda_p$ as:\cite{nelson_1976}
\begin{eqnarray}
\bm{\partial}_c \sqrt{\lambda_p}   &=&  \frac{ \bm{\partial}_c \lambda_p }{2\sqrt{\lambda_p}}  \\
\bm{\partial}_c \lambda_p &=& \left[ \bm{\mathcal{X}}^{\top} \bm{\partial}_c \mathbf{C}^{\rm MBD}  \bm{\mathcal{X}}  \right]_{pp} \\
\Rightarrow  \sum_{p=1}^N   \bm{\partial}_c \sqrt{ \lambda_p}   &=& \frac{1}{2}   {\rm Tr} \left[ \bm{\Lambda}^{-1/2}  \bm{\mathcal{X}}^{\top} \bm{\partial}_c \mathbf{C}^{\rm MBD}  \bm{\mathcal{X}}  \right].
\label{derivative_of_lambda_distinct_eigenvalues}
\end{eqnarray}
where $\bm{\mathcal{X}}$ is the matrix of eigenvectors of $\mathbf{C}^{\rm MBD}$ and $\bm{\Lambda}={\rm diag}[\lambda_p]$ is the diagonal matrix of eigenvalues.
To evaluate this last line we require the derivative of the $ab$ block of $\mathbf{C}^{\rm MBD}$ (cf. Eq.~\eqref{C_MBD}), 
\begin{widetext}
\begin{eqnarray}
\bm{\partial}_c \mathbf{C}^{\rm MBD}_{ab} &=&  2 \delta_{ab} \overline{\omega}_a \bm{\partial}_c \overline{\omega}_a 
 + (1- \delta_{ab})\, [\overline{\omega}_a \bm{\partial}_c \overline{\omega}_b + \overline{\omega}_b \bm{\partial}_c \overline{\omega}_a] \sqrt{\overline{\alpha}_a(0) \overline{\alpha}_b(0) }\; \overline{\mathbf{T}}^{\;\rm LR}_{ab}  \\
&& + (1-\delta_{ab}) \, \overline{\omega}_a \overline{\omega}_b \frac{\left[ \overline{\alpha}_a(0)\, \bm{\partial}_c \overline{\alpha}_b(0) + \overline{\alpha}_b(0)\, \bm{\partial}_c \overline{\alpha}_a(0) \right]}{2\sqrt{\overline{\alpha}_a(0) \overline{\alpha}_b(0) }} \; \overline{\mathbf{T}}^{\;\rm LR}_{ab}   
 + (1- \delta_{ab})\, \overline{\omega}_a \overline{\omega}_b \sqrt{\overline{\alpha}_a(0) \overline{\alpha}_b(0) }\; \bm{\partial}_c \overline{\mathbf{T}}^{\;\rm LR}_{ab} . \nonumber
\end{eqnarray}
\end{widetext}
To proceed any further we now need the derivatives of $\overline{\omega}$, $\overline{\alpha}$, and $\overline{\mathbf{T}}_{\rm LR}$.
From Eq.~\eqref{screened_omega}, we find that the derivative of the screened excitation frequency, $\overline{\omega}$, requires us to evaluate derivatives of $\overline{\alpha}(\mathbbm{i}\omega)$ (with $\overline{\alpha}(0)$ as a specific case) as follows:
\begin{align}
\bm{\partial}_c \overline{\omega}_a = \frac{8}{\pi} \sum_{p=1}^n g_p  \Bigg[&     \frac{ \overline{\alpha}_a(\mathbbm{i}y_p) \bm{\partial}_c \overline{\alpha}_a(\mathbbm{i}y_p)}{ [\overline{\alpha}_a(0)  ]^{2}}   \\& - \frac{\left[\overline{\alpha}_a(\mathbbm{i}y_p) \right]^2 \bm{\partial}_c \overline{\alpha}_a(0)}{ [\overline{\alpha}_a(0) ]^3} \Bigg].  \nonumber\label{derivative_of_screened_omega}
\end{align}
The derivative of the screened polarizability, $\overline{\alpha}$, Eq.~\eqref{screened_static_polarizabilities}, is calculated from the ``isotropized'' partial contraction of $\overline{\mathbf{A}}$ (with the frequency dependence suppressed):
\begin{equation}
\bm{\partial}_c \overline{\alpha}_a = \frac{1}{3} {\rm Tr}\left[ \sum_{b=1}^N \big[ \bm{\partial}_c \overline{\mathbf{A}}\, \big]_{ab} \right] .  \label{derivative_of_screened_alpha}
\end{equation}
Using Eq.~\eqref{rsSCS} and~\eqref{T_SR} and expanding the derivative of the inverse of a non-singular matrix, we have
\begin{equation}
\bm{\partial}_c \overline{\mathbf{A}} = -\overline{\mathbf{A}}  \left[ - \mathbf{A}^{-1} \left[ \bm{\partial}_c \mathbf{A} \right]\mathbf{A}^{-1} + \bm{\partial}_c \mathbf{T}_{\rm SR}\right] \overline{\mathbf{A}}. \label{long_A_inverse}
\end{equation}
Using Eqs.~\eqref{Lambda} and~\eqref{A_TS}, we compute $\bm{\partial}_c \mathbf{A}$ as:
\begin{equation}
\bm{\partial}_c \mathbf{A} = \bigoplus_{a=1}^N   {\rm diag}\left[  \Upsilon_a \; \bm{\partial}_c V_a  \right] . \label{derivative_of_alpha_TS}
\end{equation}
In Eq.~\eqref{derivative_of_alpha_TS} we have terminated the chain-rule with $\bm{\partial}_c V_a$, which has remaining \textit{implicit} dependence on the nuclear coordinates. We regard $\bm{\partial}_c V_a$ as one of our three fundamental derivatives since the Hirshfeld partitioning is typically computed separately from the rest of the MBD algorithm. Discussion of how to compute $\bm{\partial}_c V_a$ may be found in the ESI.\cite{Note1}
  
In considering the derivatives of the dipole--dipole interaction tensors, we will encounter both implicit and explicit nuclear position dependence through $\zeta_{ab}$, Eq.~\eqref{zeta}.  The derivatives of \mbox{$\mathbf{T}_{\rm SR}$, Eq.~\eqref{T_SR},} and $\overline{\mathbf{T}}_{\rm LR}$, Eq.~\eqref{T_LR}, are fairly complicated, so it will help to consider first the damping function, $f$, in isolation. Here, 
\begin{eqnarray}
\bm{\partial}_c f(R_{ab}) &=& \frac{\exp \left[-Z_{ab}\right]}{\left[1+ \exp \left[-Z_{ab}\right] \right]^2} \; \bm{\partial}_c Z_{ab} , \label{f_derivative}\\
\bm{\partial}_c Z_{ab} &=&  6 \left[ \frac{\bm{\partial}_c R_{ab} }{S_{ab} } -\frac{R_{ab} \bm{\partial}_c S_{ab}} {S^2_{ab}}   \right],   \label{derivative_of_Z} \\
\bm{\partial}_c S_{ab} &=& \beta \left[\bm{\partial}_c \mathcal{R}_{~a}^{\rm vdW}+\bm{\partial}_c \mathcal{R}_{~b}^{\rm vdW}    \right] ,
\end{eqnarray}
where $\bm{\partial}_c R_{ab}$ is calculated as
\begin{equation}
\bm{\partial}_c R_{ab} = \bm{\nabla}_{\bm{\mathscr{R}}_c} \| \mathbf{R}_{ab}\|  =     \left(  \delta_{ac} - \delta_{bc} \right) \frac{\mathbf{R}_{ab}}{ \| \mathbf{R}_{ab} \| } , \label{dRab}
\end{equation}
and the effective vdW radii have only implicit nuclear coordinate dependence.
For the gradient of $\mathbf{T}_{\rm SR}$, Eq.~\eqref{T_SR}, we require the derivative of the TS-level effective vdW radii, Eq.~\eqref{RvdW_TS}:
\begin{equation} \label{vdw_TS_derivative}
\bm{\partial}_c \mathcal{R}_{~a}^{\rm vdW,\,TS} = \frac{ \mathcal{R}_{~a}^{\rm vdW,\,free} }{ \left[V_a^{\rm free}\right]^{1/3} } \frac{ \bm{\partial}_c  V_a }{ 3\left[    V_a  \right]^{2/3}} ,
\end{equation}
while for the gradient of $\overline{\mathbf{T}}_{\rm LR}$, Eq.~\eqref{T_LR}, we require the derivative of the screened effective vdW radii, Eq.~\eqref{RvdW_SCS}:
\begin{equation}\label{vdw_screened_derivative}
\bm{\partial}_c \overline{\mathcal{R}}_{~a}^{\rm vdW} = \frac{ \mathcal{R}_{~a}^{\rm vdW,\,free} }{ \left[\alpha_a^{\rm free}(0)\right]^{1/3} } \frac{ \bm{\partial}_c  \overline{\alpha}_a(0) }{ 3\left[    \overline{\alpha}_a(0) \right]^{2/3}} ,
\end{equation}
which was evaluated using Eqs.~\eqref{derivative_of_screened_alpha}-\eqref{derivative_of_alpha_TS}.

In the following we suppress the $a,b,c$ QHO indices where possible so that the Cartesian indices $i,j$ are highlighted.
First we consider the derivative of $\mathbf{T}_{\rm dip}$, Eq.~\eqref{T_dip}, which is given by:
\begin{equation}
\bm{\partial} {\rm T}^{\,ij}_{\rm dip} = - 3 \left[ \frac{\delta_{\,ij}}{R^4} \bm{\partial} R  + \frac{R^j \bm{\partial} R^i +  R^i  \bm{\partial} R^j} { R^5} - \frac{5 R^i R^j}{ R^6} \bm{\partial} R   \right],
\label{Tdip_derivative}
\end{equation}
where $\bm{\partial} R^i$ is evaluated as:
\begin{equation}
\bm{\partial}_c R_{ab}^{\,i}  =\bm{\nabla}_{\bm{\mathscr{R}}_c}\left( (\bm{\mathscr{R}}_a-\bm{\mathscr{R}}_b) \cdot \hat{\mathbf{e}}_i \right) = (\delta_{ac}-\delta_{bc}) \hat{\mathbf{e}}_i . \label{dRvec_i}
\end{equation}
Since the long-range dipole--dipole interaction tensor is approximated with the frequency-independent $\mathbf{T}_{\rm dip}$ (thereby \mbox{eliminating $\zeta$),} Eqs.~\eqref{f_derivative}-\eqref{vdw_screened_derivative} and~\eqref{Tdip_derivative} provide us with all of the quantities needed to evaluate $\bm{\partial}_c\overline{\mathbf{T}}_{\rm LR}$ as:
\begin{equation}
\bm{\partial}_c \overline{{\rm T}}^{\,ij}_{ab,\;\rm LR} = {\rm T}^{\,ij}_{ab,\;\rm dip} \; \bm{\partial}_c f\left(\overline{Z}_{ab}\right) + f\left(\overline{Z}_{ab}\right) \bm{\partial}_c {\rm T}^{\,ij}_{ab,\;\rm dip} . 
\end{equation}
The derivative of $\mathbf{T}_{\rm SR}$ is more complex since $\mathbf{T}$ depends on $\zeta$:
 \begin{equation}
\bm{\partial}_c {\rm T}^{\,ij}_{ab,\;\rm SR} = - {\rm T}_{ab}^{\,ij} \; \bm{\partial}_c f\left(Z_{ab}^{\rm TS}\right) + \left[1- f\left(Z_{ab}^{\rm TS}\right) \right] \bm{\partial}_c {\rm T}_{ab}^{\,ij},
\end{equation}
in which the derivative of ${\rm T}^{\,ij}$ is given below (see the ESI\cite{Note1} for a detailed derivation):
\begin{align}\label{T_GG}
\bm{\partial} {\rm T}^{\,ij} =& -3 \left[{\rm erf}\left[\zeta\right]   -\frac{h(\zeta)}{2\zeta}   \right]  \bm{\partial} {\rm T}_{\rm dip}^{\,ij}  \\
&  + \zeta\, h(\zeta)  \left[  - \frac{1}{3} \bm{\partial} {\rm T}_{\rm dip}^{\,ij} - \frac{\delta_{\,ij}}{R^4} \bm{\partial} R \right]    \nonumber \\
& +  \left[ {\rm T}_{\rm dip}^{\,ij} + \frac{R^{\,i} R^{\,j}}{ R^5} \Big[ 3  - 2\zeta^2  \Big]\right]  h(\zeta)\,  \bm{\partial} \zeta,   \nonumber
\end{align}
wherein we have defined the following function for compactness,
\begin{equation}\label{h_zeta}
h(\zeta_{ab}) \equiv \frac{4 \zeta_{ab}^2}{\sqrt{\pi}}  \exp[-\zeta_{ab}^2].
\end{equation}
The derivative of $\zeta_{ab}$ is given by (with QHO indices restored to express $\bm{\partial}_c \Sigma_{ab}$ from Eq.~\eqref{Sigma}):
\begin{equation}\label{zeta_derivative}
\bm{\partial}_c \zeta_{ab} = \frac{\zeta_{ab}}{R_{ab}} \bm{\partial}_c R_{ab} - \frac{\zeta_{ab}^3 \left[  \sigma_a\bm{\partial}_c \sigma_a + \sigma_b \bm{\partial}_c \sigma_b \right]}{R_{ab}^2} ,
\end{equation}
where $\bm{\partial}_c \sigma_a$ is computed from Eq.~\eqref{sigma_in_terms_of_Veff} as
\begin{equation}\label{derivative_of_cigma}
\bm{\partial}_c \sigma_a = \left[ \frac{1}{3}\sqrt{\frac{2}{\pi}} \Upsilon_a \right]^{1/3}  \frac{ \bm{\partial}_c V_a }{3 \left[V_a\right]^{2/3}}  .
\end{equation}

We have now reduced the analytical nuclear derivative of the MBD correlation energy to quantities that depend on three fundamental derivatives: $\bm{\partial}_c R_{ab}$, $\bm{\partial}_c R_{ab}^i$ and $\bm{\partial}_c V_a$. The expressions for $\bm{\partial}_c R_{ab}$ and $\bm{\partial}_c R_{ab}^i$ have been given above in Eqs.~\eqref{dRab} and \eqref{dRvec_i}, and are straightforward to implement. The computation of $\bm{\partial}_c V_a$ is outlined briefly in the ESI.\cite{Note1} 

\section{Computational Details \label{sec:Computational_Details}}
We have implemented the MBD energy and analytical nuclear gradients (forces) in a development version of \textsc{Quantum ESPRESSO}~v5.1 (\textsc{QE}).~\cite{giannozzi_2009} A forthcoming publication will discuss the details of this implementation, including the parallelization and algorithmic strategies required to make the method efficient for treating large-scale condensed-phase systems.~\cite{markovich_unpublished}

All calculations were performed with the Perdew, Burke, and Ernzerhof (PBE) exchange-correlation functional,~\cite{PBE_ref1,*PBE_ref2} and Hamann-Schlueter-Chiang-Vanderbilt (HSCV) norm-conserving pseudopotentials.~\cite{hamann_prl_1979,*bachelet_prb_1982,*vanderbilt_1985}  As a point of completeness, it should be noted that in \textsc{QE} the Hirshfeld partitioning has only been implemented for norm-conserving pseudopotentials, and thus the MBD method cannot presently be used with ultrasoft pseudopotentials or projector-augmented wave methods.  To ensure a fair comparison with our implementation of the MBD model, all TS calculations were performed as \textit{a posteriori} corrections to the solution of the non-linear Kohn-Sham equations, \textit{i.e.} we turned off the self-consistent density updates from TS.  Additional computational details, including detailed convergence tolerances and basis sets are given in Sec.~\ref{sec:esi:additional_computational_details} of the ESI.\cite{Note1}
For comparison with the D3(BJ) dispersion correction of Grimme \textit{et al}.~\cite{grimme_d3_2010,grimme_bj_2011} (hereafter abbreviated as D3) we also optimized structures using \textsc{Orca}~v3.03.~\cite{neese_orca_2012}  We used the atom-pairwise version of D3(BJ) since only numerical gradients were available for the three-body term. 

\section{Results and Discussion}\label{sec:results}

To verify our implementation of the MBD energy in \textsc{QE}, we compared against the implementation of the MBD@rsSCS model in the \textsc{FHI-aims} code~\cite{blum_2009,MBD_rsSCS_FHI-aims} and find agreement to within $\rm 10^{-11}~E_h$.  We next verified our implementation of the analytical gradients by computing numerical derivatives \textit{via} the central difference formula and find agreement within the level of expected error given the finite spacing between the grid points describing $\rho(\mathbf{r})$ and error propagation of finite differences of the Hirshfeld effective volume derivatives. 

To demonstrate the efficiency and accuracy of the analytical MBD nuclear gradient, we performed geometry optimizations on representative systems for intermolecular interactions (benzene dimer), intramolecular interactions (polypeptide secondary structure), and supramolecular interactions (buckyball catcher host-guest complex).  We subsequently examined the importance of the implicit nuclear coordinate dependence that arises from the Hirshfeld effective volume gradient $\bm{\partial} V$ in the computation of the MBD forces.

\subsection{Intermolecular interactions: stationary points on the benzene dimer potential energy surface.~}\label{sec:benzene}

\begin{figure*}[!htbp]
\centering
\fbox{\includegraphics[width=0.935\textwidth]{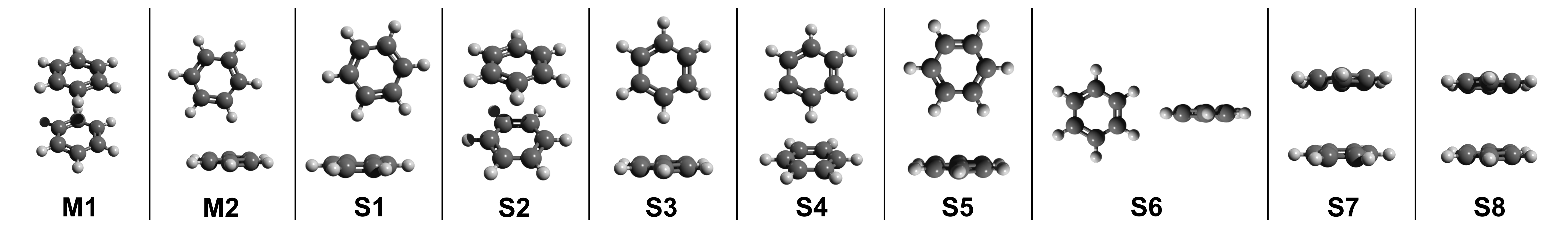}}\\\vspace{2pt}  
\fbox{\includegraphics[width=0.95\columnwidth]{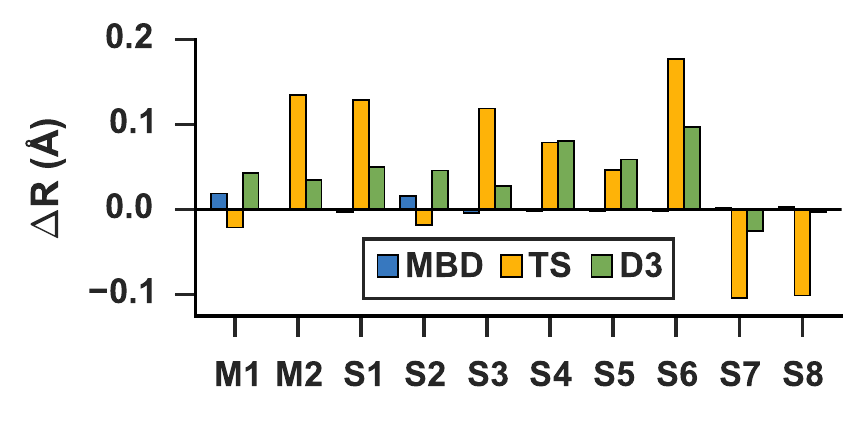}}\vspace{2pt} 
\fbox{\includegraphics[width=0.95\columnwidth]{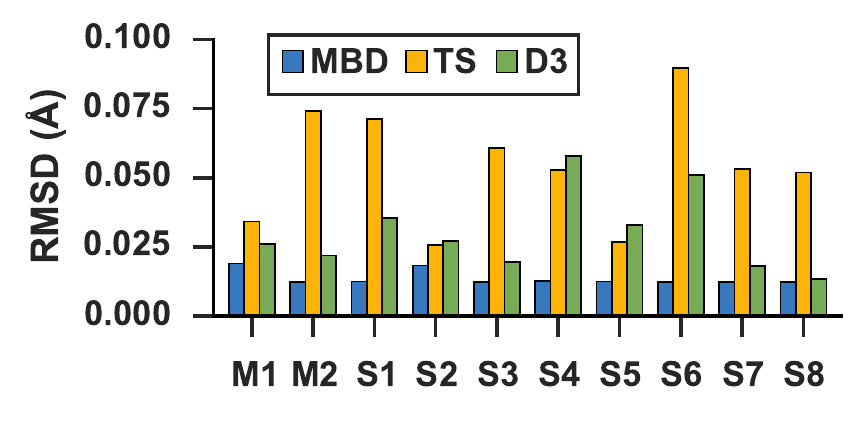}}
\caption{
\textbf{Top:} Graphical depictions of the 10 configurations that correspond to stationary points on the benzene dimer PES, following the nomenclature of Podeszwa \textit{et al.}~\cite{podeszwa_2006} (M$n=$ minima; S$n=$ saddle points).
\textbf{Left:}    Change in inter-monomer distance, $R$, relative to the PBE/CCSD(T) reference for geometries optimized with PBE+vdW methods: MBD (shown in blue), TS (shown in yellow) and D3 (shown in green). PBE+MBD consistently predicts the correct inter-monomer distance. For the stacked configurations (M1, S4, S7, and S8) PBE+TS shortens the inter-monomer distance, while for T-shaped configurations (M2, S1, S2, S3, S5, and S6) the inter-monomer distance is elongated. For all configurations except the stacked S7 and S8 structures PBE+D3 predicts too long an inter-monomer distance.
\textbf{Right:} Root-mean-square-deviations (RMSD) in \AA\ between the PBE+vdW and PBE/CCSD(T)~\cite{podeszwa_2006} optimized geometries of these 10 benzene dimer configurations. The RMSD between the PBE+MBD and reference PBE/CCSD(T) geometries (shown in blue) are uniformly small and consistent across all minima and saddle points on the benzene dimer PES. For several M$n$ and S$n$ configurations, the PBE+D3 optimized geometries (shown in green) agree quite well with the PBE/CCSD(T) reference, while the PBE+TS optimized geometries (shown in yellow) have more significant deviations.
\label{fig:benzene_rmsd} }
\end{figure*}

As the prototypical example of the $\pi-\pi$ interaction, there have been a large number of theoretical studies on the benzene dimer using very high-level wavefunction theory methods.~\cite{hobza_benzene_dimer_1990,
	arunan_benzene_dimer_1993,
	hobza_benzene_dimer_1993,
	hobza_benzene_dimer_1994,
	jaffe_benzene_dimer_1996,
	hobza_benzene_dimer_1996,
	gauss_stanton_2000,
	tsuzuki_benzene_dimer_2002,
	sinnokrot_benzene_dimer_2002,
	sinnokrot_benzene_dimer_2004,
	tauer_additivity_2005,
	podeszwa_2006,
	grant_benzene_dimer_2006,
	distasio_benzene_dimer_2007,
	janowski_benzene_dimer_2007,
	lee_benzene_dimer_2007,
	fernandez_coupled_2007,
	bludsky_2008,
	pavone_benzene_dimer_2008,
	pitonak_benzene_dimer_2008,
	sherrill_benzene_dimer_2009,
	grafenstein_benzene_dimer_2009} 
Since the intermolecular attraction between the benzene dimer arises primarily from a balance between dispersion interactions and quadrupole-quadrupole interactions (depending on the intermolecular binding motif), the interaction energy is quite small $(\sim2-3~\rm kcal/mol)$ and the potential energy surface (PES) is very flat. Consequently, resolving the stationary points of this PES is quite challenging for both theory and experiment. The prediction of the interaction energy in the benzene dimer represents a stringent test of the ability of a given electronic structure theory method to capture and accurately describe non-bonded \textit{intermolecular} interactions. 
	 Historically, three conformers of the dimer have received the most attention, namely the ``sandwich,'' ``parallel-displaced,'' and ``T-shaped'' structures. Using the high-level benchmark interaction energy calculations as a guide, several studies have used a variety of more approximate methods to examine the PES more broadly.~\cite{pitonak_benzene_dimer_2008,grafenstein_benzene_dimer_2009,podeszwa_2006,distasio_benzene_dimer_2007}
By scanning the PES of the benzene dimer with DFT-based symmetry adapted perturbation theory (DFT-SAPT), Podeszwa \textit{et al.}~\cite{podeszwa_2006} identified 10 stationary points, \textit{i.e.}, either minima (M) or saddle points (S) of the interaction energy (see Fig.~\ref{fig:benzene_rmsd}). Most wavefunction studies of the benzene dimer PES have used a fixed monomer geometry, assuming that the weak interactions will produce very little relaxation of the rigid monomer.~\cite{sinnokrot_benzene_dimer_2002}
 Using the highly accurate \textit{fixed} benzene monomer geometry of Gauss and Stanton,~\cite{gauss_stanton_2000} Bludsk\'{y} \textit{et al.}~\cite{bludsky_2008} performed counterpoise-corrected geometry optimizations of these 10 configurations at the PBE/CCSD(T) level of theory, with an aug-cc-pVDZ basis set. The resulting geometries are among the largest molecular dimers to be optimized with a CCSD(T) correction to date and represent the most accurate available structures for the dimer of this classic aromatic system. 

As a first application of the MBD analytical nuclear gradients derived and implemented in this work, we performed geometry optimizations on these 10 benzene dimer configurations at the PBE+MBD, PBE+TS, and PBE+D3 levels of theory. All of the geometry optimizations performed herein minimized the force components on all atomic degrees of freedom according to the thresholds and convergence criteria specified in Sec.~\ref{sec:esi:additional_computational_details} of the ESI\cite{Note1} (\textit{i.e.}, frozen benzene monomers were not employed in these geometry optimizations). The root-mean-square-deviations (RMSD in \AA) between the PBE+MBD, PBE+TS, and PBE+D3 optimized geometries with respect to the reference PBE/CCSD(T) results are depicted in Fig.~\ref{fig:benzene_rmsd}.

From this figure, it is clear that the PBE+MBD method, with a mean RMSD value of 0.01 \AA\ (and a vanishingly small standard deviation of $3\times10^{-4}$ \AA) with respect to the reference PBE/CCSD(T) results, was able to provide uniformly accurate predictions for the geometries of all of the benzene dimer configurations considered. These findings are encouraging and consistent with the fact that the PBE+MBD method yields significantly improved binding energies for the benzene dimer as well as a more accurate quantitative description of the fractional anisotropy in the static dipole polarizability of the benzene monomer.~\cite{distasio_2014} This is also consistent with the finding of von Lilienfeld and Tkatchenko that the three-body ATM term contributes $\sim25\%$ of the binding energy of the benzene dimer in the parallel displaced configuration.~\cite{anatole_2_and_3_2010}

With a mean RMSD value of $0.03\pm0.01$ \AA\ and $0.05\pm0.02$ \AA\ respectively, the PBE+D3 and PBE+TS methods both yielded a less quantitative measure of the benzene dimer geometries with respect to the reference PBE/CCSD(T) data. Of the 7 benzene dimer configurations for which the PBE+TS RMSD values were greater than 0.05 \AA\ (namely M2, S1, S3, S4, S6, S7, and S8), it is difficult to identify a shared intermolecular binding motif among them. Interestingly, PBE+D3 seems to fare better on sandwiched geometries and it is only the T-shaped S4 and S6 which have RMSDs above 0.05 \AA. 

However, analysis of the inter-monomer distance (see Fig.~\ref{fig:benzene_rmsd}) reveals that PBE+TS tends to shorten the inter-monomer distance $R$ for sandwich geometries (M1, S4, S7, and S8) by an average of \mbox{0.03 \AA} relative to the PBE/CCSD(T) results, while it elongates the inter-monomer distance by an average of 0.09 \AA\ for T-shaped structures. The dispersive interaction between the stacked structures (S7 and S8) is stronger than that of the parallel displaced structures (M1 and S4), so PBE+TS shortens the inter-monomer distance more significantly for S7 and S8. Likewise, these are the only two structures for which PBE+D3 shortens the inter-monomer distance.  For all other geometries PBE+D3 elongates the inter-monomer distance by an average of \mbox{0.06 \AA.}  For both sandwich and T-shaped structures, PBE+MBD performs much more consistently, elongating the inter-monomer distance by a scant $5\times10^{-3}$ \AA\ and $1\times 10^{-3}$ \AA\ for sandwich and T-shaped configurations, respectively.

We note that RMSD values in the range of \mbox{0.03--0.08 \AA,} and errors on the inter-monomer distances of \mbox{0.05-0.15 \AA,} in the geometries of small molecular dimers (as found here with the PBE+TS and PBE+D3 methods) are not unacceptably large in magnitude; however, these differences will become even more pronounced as the sizes and polarizabilities of the monomers continue to increase. In this regime, the MBD method---by accounting for both anisotropy and non-additivity in the polarizabilities as well as beyond-pairwise many-body contributions to the long-range correlation energy---is expected to yield accurate and consistent equilibrium geometries for such systems. As such, the combination of DFT+MBD has the potential to emerge as a computationally efficient and accurate electronic structure theory methodology for performing scans of high-dimensional PESs for molecular systems whose overall stability is primarily dictated by long-range intermolecular interactions.

\subsection{Intramolecular interactions: secondary structure of polypeptides.~}\label{sec:peptide}

As a second application, we considered the \textit{intramolecular} interactions that are responsible for the secondary structure in small polypeptide conformations. In particular, we studied 76 conformers of 5 isolated polypeptide sequences (GFA, FGG, GGF, WG, and WGG), which are comprised of the following four amino acids: glycine (G), alanine (A), phenylalanine (F), and tryptophan (W). This set of peptide building blocks includes the simplest amino acids, glycine and alanine (with hydrogen and methyl side chains, respectively), as well as the larger aromatic amino acids, phenylalanine and tryptophan (with benzyl and indole side chains, respectively). Although each of these polypeptides are relatively small (with 34-41 atoms each), a significant amount of conformational flexibility is present due to the non-trivial intramolecular binding motifs found in these systems, such as non-bonded side chain--backbone interactions and intramolecular hydrogen bonding. In fact, it is the presence of these interactions that leads to the formation of $\alpha$-helices and $\beta$-pleated sheets---the main signatures of secondary structure in large polypeptides and proteins.

\begin{figure}[!htbp]
\includegraphics[width=0.9\columnwidth]{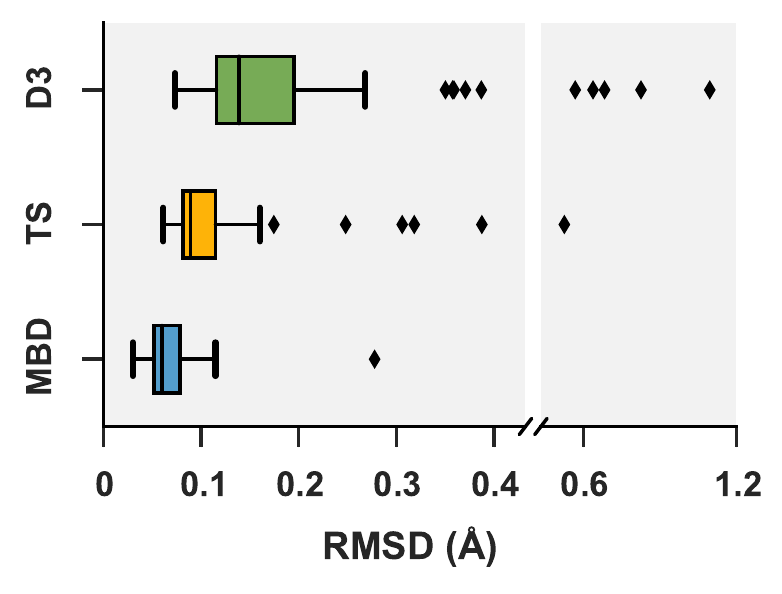} 
\caption{\label{fig:peptide_rmsd}
Box-and-whisker plots showing the distribution of root-mean-square-deviations (RMSDs) in \AA\ between 76 conformers of 5 isolated small peptides optimized with PBE+MBD (blue), PBE+TS (yellow) and PBE+D3 (green) compared against the MP2 reference geometries of Ref.~\citenum{valdes_benchmark_2008}. Whiskers extend to data within 1.5 times the interquartile range.~\cite{tukey_exploratory_1977} Note the need for a broken axis to show the largest RMSDs of PBE+D3. PBE+MBD consistently outperforms both PBE+TS and PBE+D3 in terms of yielding optimized geometries closer to the MP2 reference. Median (maximum) values are: \mbox{0.06 (0.28) \AA} for PBE+MBD, \mbox{0.09 (0.52) \AA} for PBE+TS, and  \mbox{0.14 (1.10) \AA} for PBE+D3.
}
\end{figure}

Following a benchmark study by Valdes \textit{et al.},~\cite{valdes_benchmark_2008} in which the geometries of these 76 conformers were optimized using second-order M{\o}ller-Plesset perturbation theory~\cite{moller_plesset_1934}  (MP2) within the resolution-of-the-identity approximation~\cite{baerends_1973,dunlap_1979, weigend_1998}
\mbox{(RI-MP2)} and the fairly high-quality cc-pVTZ atomic orbital basis set,~\cite{dunning_cc-basis_1989} we performed geometry optimizations on this set of conformers with several vdW-inclusive DFT approaches, namely, PBE+D3, PBE+TS, and PBE+MBD. All of the geometry optimizations performed in this section minimized the force components on \textit{all} atomic degrees of freedom according to the thresholds and convergence criteria specified in the ESI\cite{Note1} Sec.~\ref{sec:esi:additional_computational_details}. Treating the MP2 geometries as our reference, Fig.~\ref{fig:peptide_rmsd} displays box-and-whisker plots of the distributions of root-mean-square deviations (in \AA) obtained from geometry optimizations employing the aforementioned vdW-inclusive DFT methodologies.

Here we find that the PBE+MBD method again yields equilibrium geometries that are consistently in significantly closer agreement with the reference MP2 data than both the PBE+TS and PBE+D3 methodologies. For instance, the RMSDs between the PBE+MBD and MP2 conformers are smaller than 0.12 \AA\ for all but one GGF conformer (\textbf{34}: GGF04), with an overall mean RMSD value of $0.07\pm0.03$ \AA. In contrast to the intermolecular case of the benzene dimer, the PBE+TS method performs significantly better than PBE+D3 on the same benchmark set of polypeptides, with overall mean RMSD values of $0.11\pm0.07$ \AA\ and $0.20\pm0.17$ \AA, respectively. In this regard, the whiskers in Fig.~\ref{fig:peptide_rmsd} extend to RMSD values that are within 1.5 times the interquartile range (\textit{i.e.}, following the original, although arbitrary, convention for determining outliers  suggested by Tukey~\cite{tukey_exploratory_1977}), which highlights the fact that there are several conformers for which both PBE+TS and PBE+D3 predict equilibrium geometries that are significantly different than MP2. 

Although MP2 is the most economical wavefunction-based electronic structure method that can describe dispersion interactions, MP2 tends to grossly overestimate $C_6$ dispersion coefficients and hence the binding energies of dispersion-bound complexes such as the benzene dimer.~\cite{tkatchenko_jcp_mp2_2009} Since PBE+MBD should bind less strongly than MP2, we expect the side-chain to backbone distance to elongate slightly for bent conformers. Conformers where the side chain is extended away from the backbone are expected to show less deviation between MP2 and PBE+MBD as the side-chain to backbone dispersion interaction will be less significant in determining the geometry of the conformer. 

\begin{figure*}[!htbp]
\centering
\fbox{\includegraphics[width=0.45\columnwidth]{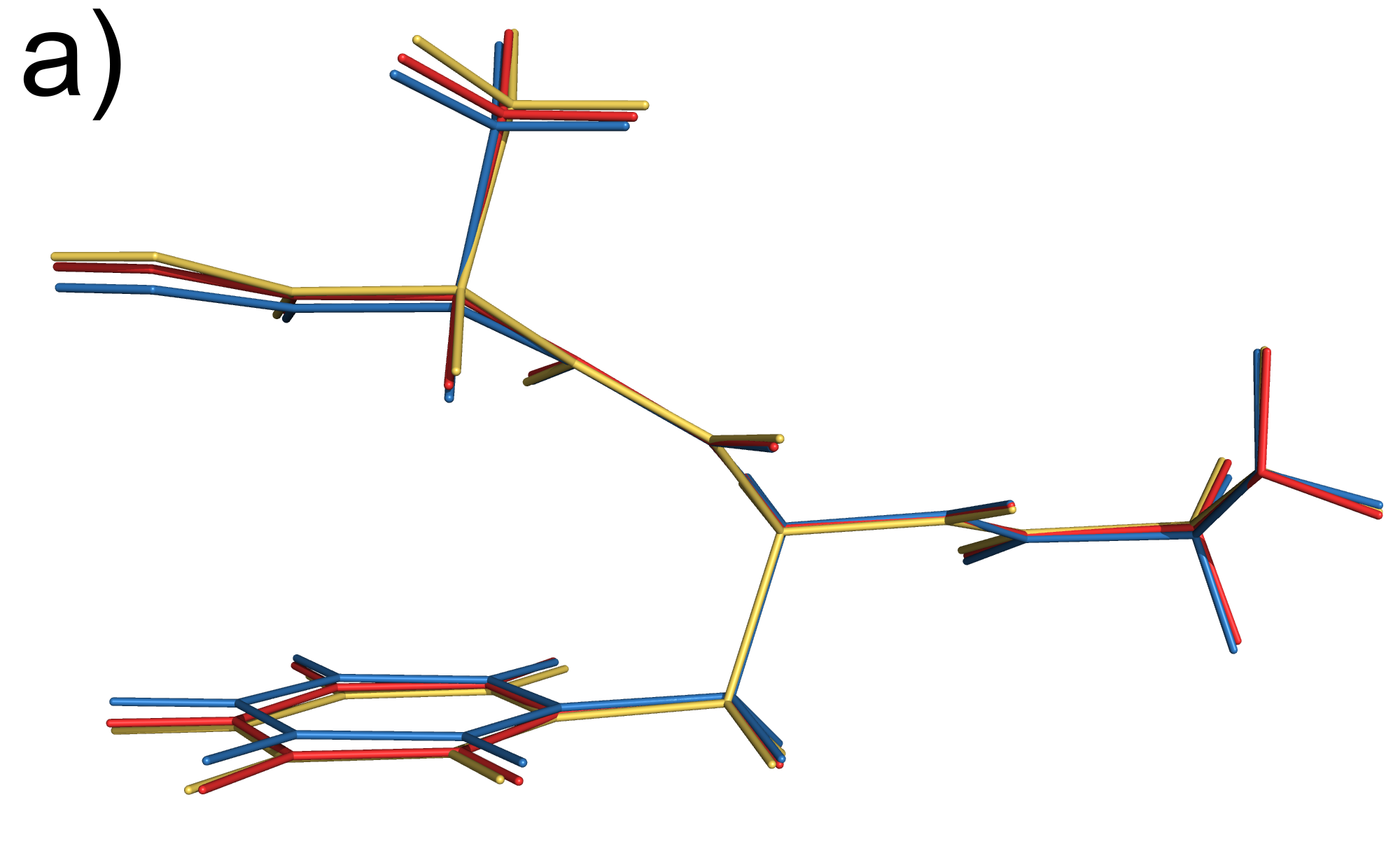}}
\fbox{\includegraphics[width=0.45\columnwidth]{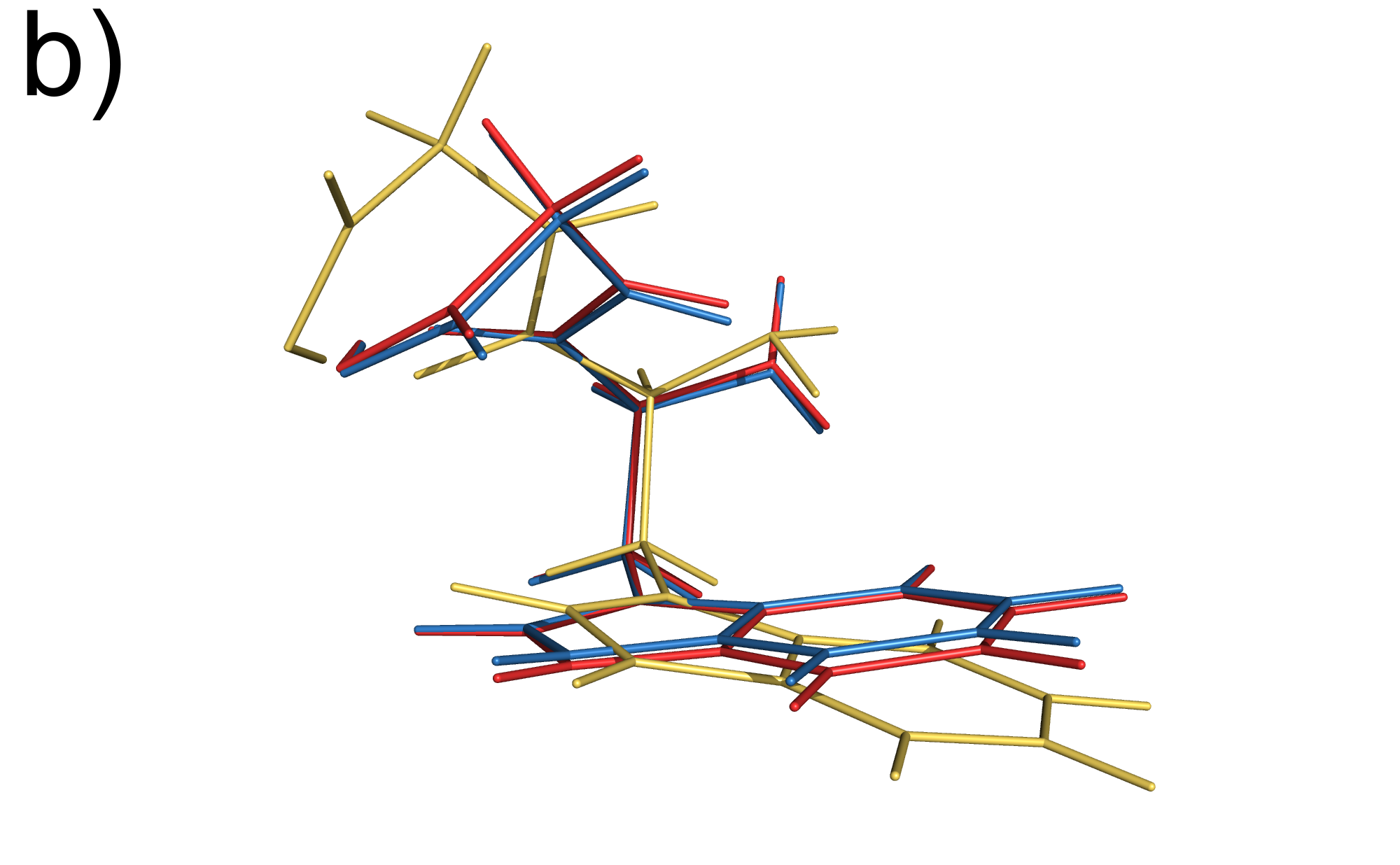}}  
\fbox{\includegraphics[width=0.45\columnwidth]{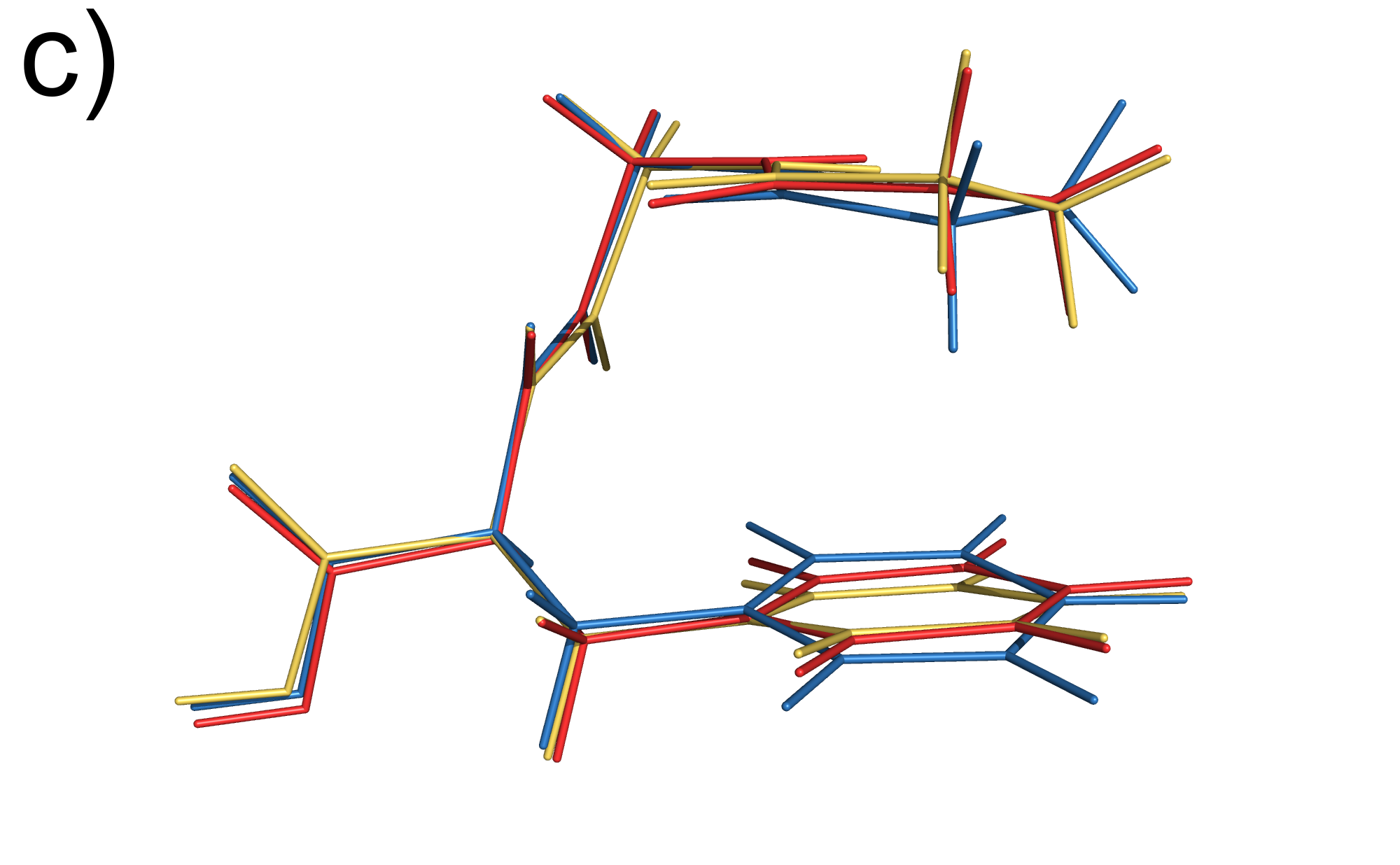}}
\fbox{\includegraphics[width=0.45\columnwidth]{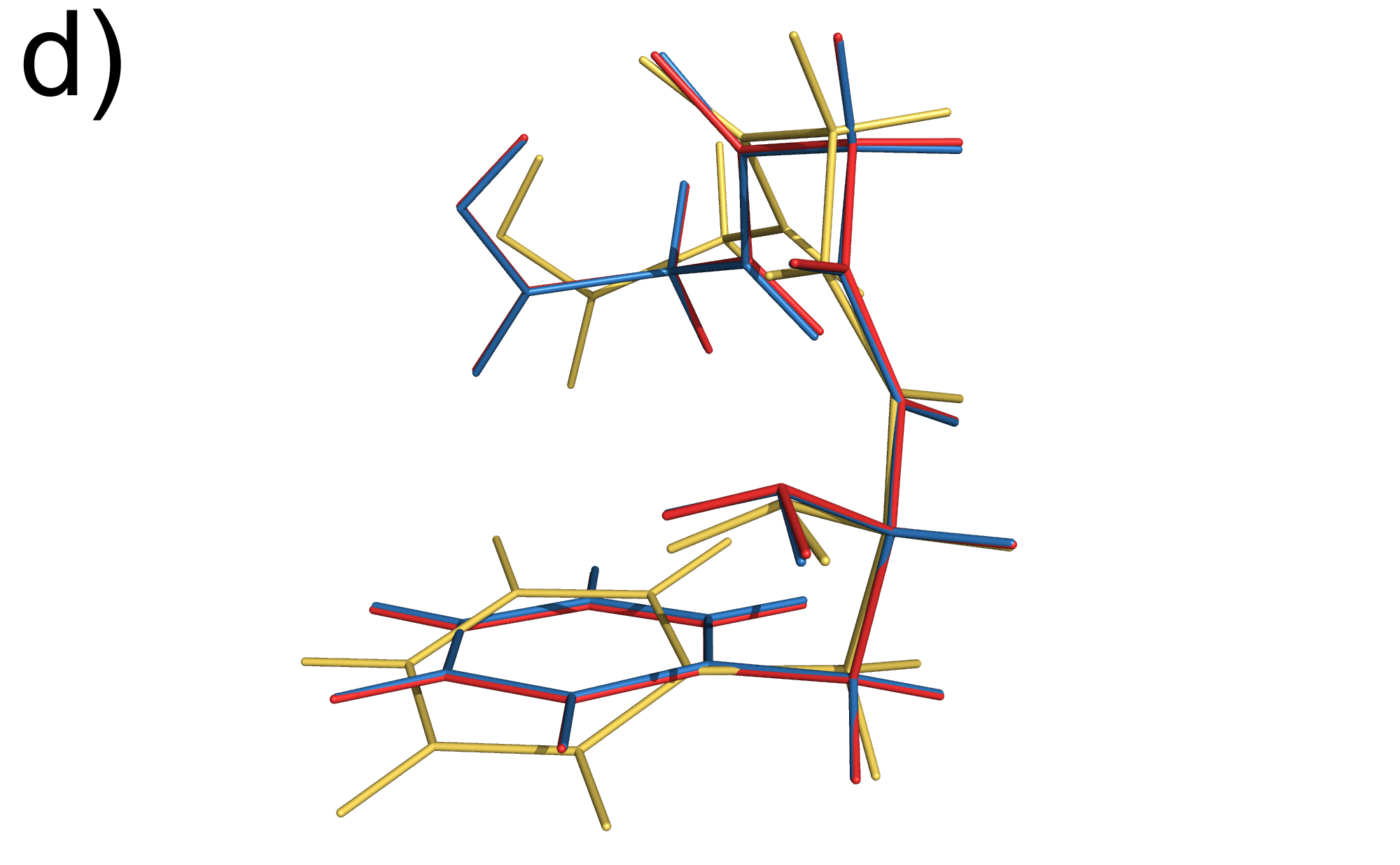}}
\caption{
Overlays of the structures obtained from geometry optimization with MP2 (blue), PBE+MBD (red), and PBE+D3 (yellow).  In both a) GFA03 and b) WG03, the MBD correction opens the cleft between the backbone and aromatic side-chain as MP2 tends to over-bind dispersion interactions. c) In GGF04, PBE+MBD rotates the phenylalanine and alanine groups together. d) In FGG215, since the side-chain is farther away from the backbone, PBE+MBD matches the MP2 geometry almost exactly. 
\label{fig:overlay}}
\end{figure*}

Aside from the noticeable outliers, the structural deviations in most of the conformers correspond to small rotations or deflection of terminal groups and side chains due to dispersion-based interactions, in contrast to the backbone which is constrained by non-rotatable bonds. In Fig.~\ref{fig:overlay} we present representative overlays of this rearrangement, showing the MP2 (blue), PBE+MBD (red), and PBE+D3 (yellow) geometries. 
In a) structure $\textbf{17}$ (GFA03) is a conformer for which both PBE+MBD and PBE+D3 give small/moderate RMSDs with MP2. Both PBE+MBD and PBE+D3 open the cleft between the alanine and phenylalanine, also causing the amine on the backbone to slightly rotate. The relative positioning of these structures is expected, given the tendency of MP2 to over-bind dispersion interactions and the tendency of PBE+D3 to under-bind. In b) structure $\textbf{48}$ (WG03), again shows PBE+MBD agreeing well with MP2, but slightly opening the backbone-side chain distance. However, PBE+D3 is disastrous for this structure, yielding an RMSD of 1.10 \AA\, due to large rotations in both the backbone and indole side-chain.

Structures where the side-chain lies farther off to the side of the backbone, such as $\textbf{4}$ (FGG215) shown in panel d), show the smallest RMSDs between the PBE+MBD and reference MP2 geometries with the PBE+MBD geometry lying almost exactly on top of the MP2 geometry. However, FGG215 is again a structure where D3 does poorly with respect to the MP2 geometry, this time rotating the benzyl side-chain away from the terminal glycine, yielding an RMSD of 0.64 \AA.

The structure for which the PBE+MBD method has the largest RMSD, at 0.28 \AA, is $\textbf{34}$ (GGF04), shown in panel c). As opposed  to opening a cleft like in GFA03, PBE+MBD rotates the phenylalanine and alanine groups together. This rotation occurs because the terminal hydrogen on the glycine is attracted to the $\pi$-system on the phenylalanine. The rigid nature of the glycine combined with the rotatable bond in the phenylalanine, forces the phenylalanine to slightly rotate in response. The motion of the middle glycine solely attempts to minimize molecular strain from these other two interactions. Both PBE+TS and PBE+D3 methods show a similar rotation for this structure, though PBE+D3 rotates the structure even farther than PBE+MBD. This concerted rotation is associated with a very flat potential energy surface, as indicated by the fact that a second optimization run with the same tolerances resulted in a slightly greater rotation.

Following Valdes \textit{et al.}, we classified the structures by the existence of an intramolecular hydrogen-bond between the \ce{-OH} of the terminal carboxyl group and the \ce{C=O} group of the preceding residue. The mean RMSD is strongly influenced by the high outliers, so the median RMSD is a more representative measure for comparing these two groups of conformers.  The median RMSD for \ce{CO2H}$_{\rm free}$ (\ce{CO2H}$_{\rm bonded}$) structures is: \mbox{0.06 (0.07) \AA} for PBE+MBD, \mbox{0.09 (0.09) \AA} for PBE+TS, and \mbox{0.14 (0.14) \AA} for PBE+D3.
Overall, we find that the presence of this intramolecular hydrogen bond does not strongly correlate with which structures deviate more from the MP2 geometries. This finding was somewhat unexpected since Valdes \textit{et al.} asserted that dispersion interactions are more important in determining the structure of the \ce{CO2H}$_{\rm free}$ family of conformers due to tendency of the peptide backbone to lie over the aromatic side chain.

Overall, we find excellent agreement between the MP2 and PBE+MBD geometries.  Where PBE+MBD deviates, we find agreement with physical and chemical intuition when we take into account the well known over-binding for dispersion interactions present in MP2. The agreement between PBE+MBD and MP2 geometries is in marked contrast to the inconsistent performance of PBE+D3 and PBE+TS, which both yielded numerous outliers.  Although computational cost is not directly comparable between a Gaussian-type-orbital code and a planewave code, we are greatly encouraged by the accuracy of our PBE+MBD geometry optimizations since such calculations with a generalized gradient approximation functional like PBE are substantially cheaper than with RI-MP2.

\subsection{Supramolecular interactions: the buckyball catcher host--guest complex.~}\label{sec:catcher}
Noncovalent interactions are particularly important in supramolecular chemistry, where non-bonded interactions, including dispersion, stabilize molecular assemblies. The large size of supramolecular host-guest complexes typically places them outside the reach of high-level quantum chemical methodologies and necessitates the use of DFT for geometry optimizations and energy computations. However, the large polarizable surfaces that interact in these systems requires a many-body treatment of dispersion to achieve a chemically accurate description of supramolecular binding energies.~\cite{tkatchenko_catcher_2012,ambrosetti_jpcl_2014}   The \ce{C60} ``buckyball catcher'' host--guest complex (also referred to as \ce{C60}@\ce{C60H28}) in particular has received considerable attention as a benchmark supramolecular system in the hope that it is prototypical of dispersion-driven supramolecular systems, and it has been studied extensively both experimentally~\cite{sygula_catcher_2007,muck-lichtenfeld_catcher_2010, le_thermodynamics_catcher_2014, zabula_catcher_2014} 
and theoretically.~\cite{zhao_catcher_2008,
    muck-lichtenfeld_catcher_2010,
    waller_catcher_2012,
    podeszwa_catcher_2012,
    grimme_supramolecular_2012,
    tkatchenko_catcher_2012,
    denis_catcher_2013,
    risthaus_supramolecular_2013,
    ambrosetti_jpcl_2014}  
The \ce{C60} buckyball catcher (denoted as \textbf{4a} by Grimme) is one of the most well studied members of the S12L test set of noncovalently bound supramolecular complexes.~\cite{grimme_supramolecular_2012}  

Much of the past computational work has focused on modeling the interaction energy of the \ce{C60} buckyball catcher and comparing these results to the experimental data on thermodynamic association constants that have been extracted from titration experiments.~\cite{sygula_catcher_2007,muck-lichtenfeld_catcher_2010,le_thermodynamics_catcher_2014} This  complex is a challenging system for most dispersion correction methods since the three-body term contributes approximately 10\% of the interaction energy.~\cite{risthaus_supramolecular_2013,ambrosetti_jpcl_2014} Motivated by this large contribution of beyond-pairwise dispersion, we optimized the \ce{C60}@\ce{C60H28} complex with PBE+MBD, PBE+TS and PBE+D3 to see how significantly many-body effects impact the geometry. Containing 148 atoms, this system also represents a structure that would be too large to optimize with numerical MBD gradients or high-level wavefunction based methodologies. All theoretical calculations reported herein are for an isolated, \textit{i.e.} gas-phase, host-guest complex at the classical equilibrium geometry at zero temperature, while the experimental values listed in Table~\ref{table:host_complex_distances} correspond to X-ray determined crystal structures measured at finite temperature. Since the base of the buckyball catcher host is quite flexible,~\cite{muck-lichtenfeld_catcher_2010} we expect the packing environment in the solid state to potentially impact the reported conformation.

\begin{table}
\caption{Selected distances of DFT gas-phase optimized geometries of the \ce{C60}@\ce{C60H28} host--guest complex and conformer \textbf{a} of the host alone compared to X-ray crystal structures of \ce{C60}@\ce{C60H28}$\cdot$2PhMe~\cite{sygula_catcher_2007} and the unsolvated buckyball catcher.~\cite{zabula_catcher_2014} The TPSS functional does not identify conformer \textbf{a}, so these entries are left blank\label{table:host_complex_distances}}
\centering
\begin{tabular}{l  ccc  cc}
\toprule
 & \multicolumn{3}{c}{ Complex}  & \multicolumn{2}{c}{Host \textbf{a}} \\ \cmidrule(r){2-4} \cmidrule(r){5-6}
Method 	& $R_c$  (\AA) & $R_p$ (\AA) &   $R_t$  (\AA)   & $R_p$ (\AA) & $R_t$  (\AA) \\
\midrule
PBE+MBD 	& 8.312		& 12.992	& 6.303	 &	13.263	&  6.394		\\
PBE+TS		& 8.361		& 12.974	& 6.337	 &	12.969	&  6.080		\\
PBE+D3  		& 8.454		& 12.987	& 6.286	 &	11.640	&  6.215		\\
TPSS+D3 	& 8.392		& 12.748	& 6.288	 &	 --  		& 	-- 		\\
TPSS+D3$^a$ & 8.361		& 12.822	& 6.303	 &	 --		&	--	        	\\ 
B97-D$^b$ 	& 8.335   		& 12.798  	& 6.299	 &	11.152	& 6.216		\\ 
M06-2L$^c$	& 8.136		& 12.703  	& 6.382	 &	11.844	& 6.322		\\ 
\midrule
X-ray$^{d,e}$		& 8.484(3)	& 12.811(4)& 	6.418(5)	& 	9.055(2)   &  6.44(3) 	\\
\bottomrule
\multicolumn{6}{c}{
$^a$ Ref.~\citenum{grimme_supramolecular_2012},     
$^b$ Ref.~\citenum{muck-lichtenfeld_catcher_2010},   
$^c$ Ref.~\citenum{zhao_catcher_2008},                     
$^d$ Ref.~\citenum{sygula_catcher_2007}, 
$^e$ Ref.~\citenum{zabula_catcher_2014}  
}
\end{tabular}
\end{table}

\begin{figure}[!htbp]
\centering
\includegraphics[width=0.85\columnwidth]{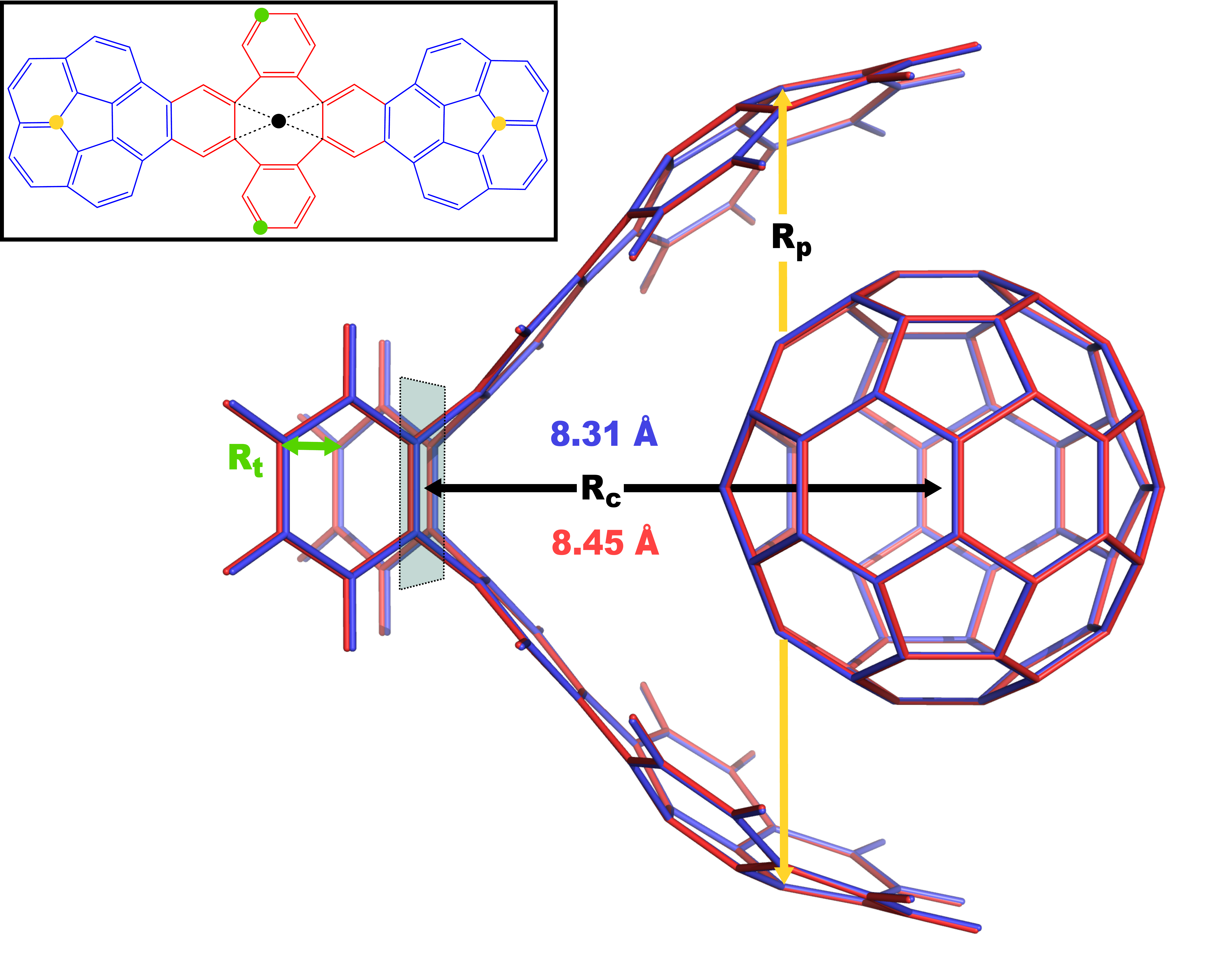}
\caption{Overlay between the geometry of the \ce{C60}@\ce{C60H28} host--guest complex optimized with PBE+D3 (red) and PBE+MBD (blue). The distance, $R_c$, between the \ce{C60} centroid and the plane bisecting the tetrabenzocyclooctatetraene (TBCOT) tether (transparent green) is reduced from $8.45~\text{\AA}$ with PBE+D3 to $8.31~\text{\AA}$ with PBE+MBD.  The green arrow shows that the $R_t$ distance is measured between terminal carbon atoms on the TBCOT tether. The yellow arrow shows that the $R_p$ distance is measured between the most separated carbon atoms of the central five-membered rings of both corannulene subunits.  \textbf{Inset:} The 2D molecular structure of the \ce{C60H28}  buckyball catcher host, with corannulene subunits shown in blue and the TBCOT tether shown in red. Atoms used to define the $R_t$ and $R_p$ distances are marked in green and yellow respectively.  The black dot shows the centroid of the four atoms on the TBCOT tether used to define the $R_c$ distance.\label{fig:supramolecular:overlay}}
\end{figure}

The buckyball catcher host is made of a tetrabenzocyclooctatetraene (TBCOT) tether and two corannulene pincers (cf. Fig.~\ref{fig:esi:catcherstructure} in the ESI\cite{Note1} and Fig.~\ref{fig:supramolecular:overlay} herein).   The conformation of the catcher is determined by a competition between the attractive dispersion interactions between the corannulene pincers and the strain induced by deformation of the TBCOT tether.~\cite{muck-lichtenfeld_catcher_2010} The two lowest energy ``open'' conformers of the catcher have the corannulene bowls in a convex--convex ``catching'' motif or in a convex--concave ``waterwheel'' motif; following the notation of Refs.~\citenum{sygula_catcher_2007, muck-lichtenfeld_catcher_2010, waller_catcher_2012}, we term the ``catching'' motif \textbf{a} and the ``waterwheel'' motif \textbf{b}.      

To compare the size of the cleft between the corannulene pincers when the buckyball catcher is optimized with various DFT+vdW methods, we report the distance between the most separated carbon atoms of the central five-membered rings of both corannulene subunits as a measure of the size of the cleft; we denote this distance as $R_p$ (cf. Fig.~\ref{fig:supramolecular:overlay}).  Closing of the cleft tends to be accompanied by outward deflection of the TBCOT tether, so we also measure the distance between terminal carbons on the tether; we denote this distance as $R_t$ (cf. Fig.~\ref{fig:supramolecular:overlay}). Likewise, we measure the distance between the centroid of the \ce{C60} and the plane that bisects the TBCOT tether at the base of the buckyball catcher (cf. Fig.~\ref{fig:supramolecular:overlay}); we denote this distance as $R_c$.  Interestingly, several of the functionals that have been used to study the buckyball catcher do not identify all four conformers. Notably, TPSS-D3 is prone to drive conformer \textbf{a} to a closed variant that has $R_p = 5.53~\text{\AA}$. With regard to the balance between dispersion and strain, conformer \textbf{a} results when the \ce{C60} is removed from the pincers and the host is allowed to relax. We will focus our discussion on the relaxed conformer \textbf{a} and the optimized complex, but we also provide optimized structures of conformer \textbf{b} in the ESI.\cite{Note1}

Upon optimization with PBE+MBD we find that the corannulene pincers deflect outward, as seen by the increased $R_p$ distance relative to the starting TPSS+D3/def2TZVP geometry from the S12L dataset.~\cite{grimme_supramolecular_2012}  The $R_p$ distance predicted by PBE+MBD is larger than other results from vdW-inclusive functionals (see Table~\ref{table:host_complex_distances}), which may be consistent with previous reports of three-body and higher order terms substantially decreasing the binding energy of the \ce{C60}@\ce{C60H28} host--guest complex.~\cite{risthaus_supramolecular_2013,ambrosetti_jpcl_2014}  However, this deflection is accompanied by a reduction of the buckyball--catcher distance $R_c$, which would suggest a tighter binding.  Just as with the reduced cleft distances in the peptides and the inter-monomer distance in the benzene dimer, we find that the host--guest distance predicted by  \mbox{PBE+MBD ($R_c=8.31~\text{\AA}$)} is smaller than that predicted by  \mbox{PBE+D3 ($R_c=8.45~\text{\AA}$)} and \mbox{PBE+TS ($R_c=8.36~\text{\AA}$).} For comparison, we also optimized the complex with TPSS+D3/def2TZVP and found a buckyball--catcher distance of  \mbox{$R_c=8.39~\text{\AA}$,} which is slightly larger than the  \mbox{$R_c=8.36~\text{\AA}$}  in the previously reported TPSS+D3/def2TZVP geometry in the S12L dataset.~\cite{grimme_supramolecular_2012}  These results are reported in Table~\ref{table:host_complex_distances} together with a comparison to previous vdW-inclusive DFT results and the corresponding distances from the X-ray determined crystal structures.  

The X-ray structure for the complex is taken from \ce{C60}@\ce{C60H28} co-crystallized with two disordered toluene molecules, \textit{i.e.} \ce{C60}@\ce{C60H28}$\cdot$2PhMe.~\cite{sygula_catcher_2007}  In the solid state, the fullerenes form columns along the \textit{a}-axis, while the buckyball catcher aligns back-to-back in the \textit{bc}-plane. These back-to-back interactions have fewer atoms that are in van der Waals contact, but could still push the corannulene units together slightly. Zabula \textit{et al.} recently obtained an X-ray crystal structure of the unsolvated buckyball catcher which adopts an inter-locked structure similar to conformer \textbf{a}.~\cite{zabula_catcher_2014}  This inter-locked structure provides an attractive vdW interaction between corannulene units, which causes the cleft to close  \mbox{($R_p=9.055(2)~\text{\AA}$),} with a corresponding outward deflection of the TBCOT tether  \mbox{($R_t=6.44(3)~\text{\AA}$).}

Perhaps the most unusual trend in Table~\ref{table:host_complex_distances} is the substantial opening of the cleft between the corannulene subunits, and the accompanying outward deflection of the TBCOT tether, when the isolated host is optimized with the PBE+MBD method. Comparing the $R_p$ and $R_t$ distances, we find an ordering of PBE+MBD $>$ PBE+TS $>$ PBE+D3. M\"{u}ck-Lichtenfeld \textit{et al.} previously found that the TBCOT tether is quite flexible, resulting in a shallow bending potential (see Fig. 2 of Ref.~\citenum{muck-lichtenfeld_catcher_2010}) as the $R_p$ distance is varied; using the B97-D functional and 6-31G$^\star$ basis set, the energy of conformer \textbf{b} varies by only $\rm\sim1.3~kcal/mol$ as $R_p$ is scanned from 10-14 \AA.~\cite{muck-lichtenfeld_catcher_2010}  Comparing the energy of the buckyball catcher in the strained conformer that it adopts when hosting the buckyball, to its energy when fully relaxed, we see that at the PBE+D3/def2TZVP level this strain energy is $\rm 1.02~kcal/mol$. This is consistent with the shallow bending potential found by  M\"{u}ck-Lichtenfeld \textit{et al.}  Given how flat this PES is, it is less surprising that the three vdW corrections considered give such different relaxed $R_p$ distances for the isolated host.
 
The structure of the \ce{C60} buckyball does not vary significantly between different vdW-inclusive functionals.  The PBE+MBD optimized structure of \ce{C60} has C-C bond lengths of  $1.45192(5)~\text{\AA}$ for bonds within five-membered rings (fusing pentagons and hexagons), and $1.39804(3)~\text{\AA}$ for bonds fusing hexagonal rings; which compares favorably to the well known gas-phase electron diffraction results of $1.458(6)~\text{\AA}$ and $1.401(10)~\text{\AA}$.~\cite{hedberg_c60_1991} This result is consistent with the short-range behavior of the range-separated PBE+MBD method, which essentially reduces to the bare PBE functional and does a good job of predicting C-C bond lengths.

On the whole we find that the PBE+MBD method yields structures that are comparable to other vdW-inclusive functionals but deviates more significantly from the X-ray determined crystal structure than the PBE+D3 results.  Since we do not have an experimentally determined gas-phase structure or a wavefunction theory reference for the \ce{C60}@\ce{C60H28} host--guest complex, the deviation of the gas-phase PBE+MBD optimization from the experimental crystal structure should not be taken as a benchmark comparison. Future work will address the optimization of this full crystal structure.  

In light of the lack of high-level wavefunction-based geometries to compare against, we conclude with a few comments about the computational efficiency of our method.  Starting from the TPSS/def2TZVP structures from the S12L dataset, we were able to optimize the 148-atom complex with the PBE+MBD method in 68 BFGS steps in about 415 cpu hours, while the PBE+D3 optimization in \textsc{Orca} took 34 BFGS steps in about 450 cpu hours.\footnote[9]{The PBE+MBD optimization was run in about 2.75 hours on 170 Intel Xeon E5-2680 processors while the PBE+D3 optimization was run in about 14 hours on 32 AMD Opteron 6376 Abu Dhabi processors.}  Given that \textsc{Orca} uses redundant internal coordinates for geometry optimizations and the D3 correction is almost instantaneous to calculate, it is worth noting that the Cartesian coordinates optimization in \textsc{QE} with the much more costly MBD correction is roughly competitive.

\subsection{The importance of \texorpdfstring{$\bm{\partial} V$}{dV}.~}\label{sec:hirshfeld_gradient}
Our derivation of the nuclear MBD forces placed considerable emphasis on the importance of including the implicit coordinate dependence arising from the gradients of the Hirshfeld effective atomic volumes. To test how large of a contribution that the $\bm{\partial} V$ terms make to the MBD forces, we re-optimized the benzene dimers, this time setting $\bm{\partial} V =0$ explicitly.  As shown in Fig.~\ref{fig:esi:benzene_intermonomer_distance} in the ESI,\cite{Note1} neglect of the Hirshfeld volume gradients does not have a large  impact for this system, in which the dispersion forces are intermolecular; the mean RMSD becomes $(16\pm5)\times10^{-4}$ \AA.  This result is expected for this system because the Hirshfeld effective atomic volumes only change when nearest neighbor atoms are moved. Not only is the benzene monomer fairly rigid, but the range separation employed in MBD means that the long-range tensor $\mathbf{T}_{\rm LR}$, and correspondingly the MBD correction, is largely turned off within the benzene monomer (see Fig.~\ref{fig:esi:damping_lengthscales} in the ESI\cite{Note1}). 

\begin{figure*}[!htbp]
\centering
\fbox{\includegraphics[width=0.9\columnwidth]{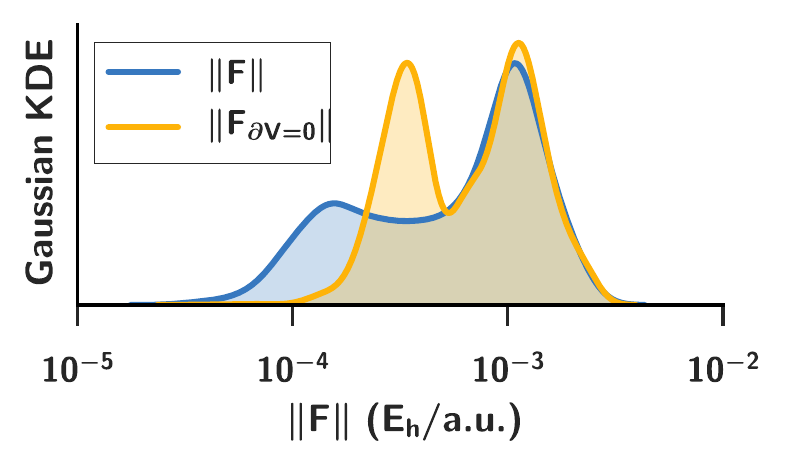}}\vspace{2pt}
\fbox{\includegraphics[width=0.9\columnwidth]{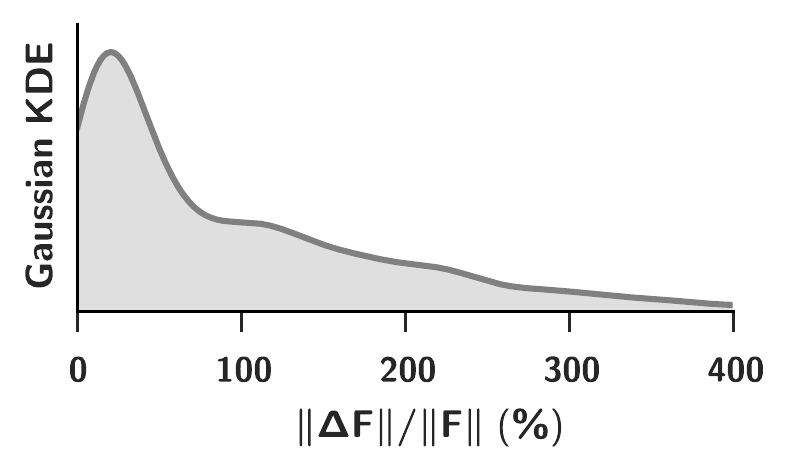}}
\caption{\textbf{Left:} Gaussian kernel density estimate of the distributions of the norm $\|\cdot\|$ of MBD forces $\mathbf{F}_{\rm MBD}$ acting on each atom at the optimized geometries of 76 tripeptide structures. In blue, the MBD forces were computed with full Hirshfeld \mbox{gradients ($\|\mathbf{F}\|$);} in yellow, the forces were computed with the Hirshfeld gradients $\bm{\partial} V$ set to zero ($\|\mathbf{F}_{\bm{\partial} V=0}\|$). 
\textbf{Right:} Gaussian kernel density estimate of the distribution of relative percentage error $\|\Delta \mathbf{F}\|/\|\mathbf{F}\|$ where $\Delta \mathbf{F} \equiv \mathbf{F}-\mathbf{F}_{\bm{\partial} V=0}$ is the error incurred by setting the Hirshfeld gradients to zero. The distribution is peaked at approximately 20\% but extends to values much greater than 100\%.
\label{fig_kde_Fnorm} }
\end{figure*}

We expect a larger impact from Hirshfeld volume gradients for systems that are flexible and large enough for the damping function to have ``turned on'' the MBD correction. The case of polypeptide intramolecular dispersion interactions matches both of these criteria. We computed the MBD forces on the final optimized geometries of all 76 peptide structures and analyzed the atom by atom difference in the forces computed with and without the Hirshfeld volume gradients. As shown in Figure~\ref{fig_kde_Fnorm}, neglect of the Hirshfeld gradient causes a significant shift in the distribution of the MBD forces in the peptides, with a tendency to increase the forces from the lower peak from $\sim 2\times 10^{-4} ~\rm E_h/a.u.$ to $\sim 4\times 10^{-4}~\rm E_h/a.u.$. Comparing the Cartesian components of the MBD forces across all atoms in all 76 structures we find that the deviations between MBD forces with and without the Hirshfeld volume gradients $(\mathbf{F}-\mathbf{F}_{\bm{\partial}V=0})$ are approximately normally distributed with zero mean and a standard deviation of $2\times 10^{-4} ~\rm E_h/a.u.$ (see Fig.~\ref{fig:esi:kde} in the ESI\cite{Note1}). This leads to the norm of the force difference $(\Delta \| \mathbf{F} - \mathbf{F}_{\partial V} \|)$ having a mean of $ (3.2\pm 1.7) \times 10^{-4}~\rm E_h/a.u.$, and a mean of the difference of norms of $ \|\mathbf{F}\|-\|\mathbf{F}_{\rm \partial V=0}\| = (-5\pm17)\times10^{-5}~\rm E_h/a.u.$   Overall, neglect of the Hirshfeld gradients increases forces and causes a long-tailed distribution of relative error, that is peaked at $\sim 20\%$, but extends up to $400\%$. This large distribution of relative errors has the potential to significantly impact the deterministic nature of \textit{ab initio} molecular dynamics (AIMD) simulations run at the MBD level of theory that do not properly account for the analytical gradients of the Hirshfeld effective volumes. Given that this error would accumulate at every time step, combined with the fact that the MBD correction was found to be quite important in the geometry optimizations of the systems considered herein, we find the neglect of the Hirshfeld effective volume gradients to be an unacceptable approximation in AIMD. This finding is particularly true for large flexible molecular systems with significant intramolecular dispersion interactions since this error can cooperatively increase along any extended direction, \textit{i.e.}, along an alkane chain or polypeptide backbone.

\section{Conclusions and Future Research}
By developing analytical energy gradients of the range-separated MBD energy with respect to nuclear coordinates, we have enabled the first applications of MBD to full nuclear relaxations. By treating the gradients of the MBD energy correction analytically, rather than numerically, we have reduced the number of self-consistent calculations that must be performed from $2\times(3N-6)$ to $1$, enabling treatment of much larger systems. Our derivation and implementation includes all implicit coordinate dependencies arising from the Hirshfeld charge density partitioning. In the isolated molecule optimizations that we considered herein, the implicit coordinate dependencies that arise from the Hirshfeld volume gradients resulted in significant changes to the MBD forces.  The long-tailed distribution of relative error that we observed indicates that any future AIMD simulations employing MBD forces must include full treatment of the Hirshfeld volume gradients, or the accumulation of error will negatively impact the simulation dynamics. Our careful treatment of these volume gradients paves the wave for future work to address how a self-consistent implementation of the MBD model will impact the electronic band structures of layered materials and intermolecular charge transfer couplings in molecular crystals. A fully self-consistent treatment of MBD will likely be required for energy conservation in AIMD simulations.

Consistent with previous findings that a many-body description of dispersion improves the binding energies of even small molecular dimers,~\cite{distasio_2014} we find that MBD forces significantly improve the structures of isolated dispersively bound molecular systems displaying both intermolecular and intramolecular interactions. We find excellent agreement between PBE+MBD optimized structures and reference PBE/CCSD(T) and MP2 geometries. Notably, PBE+MBD consistently outperformed the pairwise PBE+D3(BJ), and effectively pairwise PBE+TS optimizations.

The first applications of MBD forces in this paper were restricted to gas-phase systems because computation of MBD gradients in the condensed phase, where periodic images of the unit cell must be considered, is substantially more challenging from a computational perspective. Converging the MBD energy in the condensed phase is demanding (from both the memory and computational point of view) due to a real-space supercell procedure that is required to support long-wavelength normal modes of $\mathbf{C}^{\rm MBD}$. A forthcoming publication will describe the details of our implementation of MBD forces for periodic systems, including careful treatment of parallelization and convergence criteria.~\cite{markovich_unpublished}

\newpage
Since MBD forces are very efficient to evaluate for gas-phase molecules, we are eager to explore the application of MBD to AIMD simulations.  Many-body effects have previously been shown to be significant in modeling solvation and aggregation in solution~\cite{donchev_2006} and can lead to soft collective fluctuations that impact hydrophobic association,~\cite{godec_soft_2013} and the entropic stabilization of hydrogen-bonded molecular crystals.~\cite{reilly_2014}  We therefore anticipate that our many-body forces will be of interest for solvated simulations, such as estimates of the thermodynamic properties of metabolites~\cite{jinich_quantum_2014} and modeling novel electrolytes.~\cite{er_computational_2015}  

\section{Acknowledgments}
We thank Alexandre Tkatchenko and Alberto Ambrosetti for useful discussions and for providing source code for the \textsc{FHI-aims} implementation of MBD. 
This research used resources of the Odyssey cluster supported by the FAS Division of Science, Research Computing Group at Harvard University, the National Energy Research Scientific Computing Center (NERSC), a DOE Office of Science User Facility supported by the Office of Science of the U.S. Department of Energy under Contract No. DE-AC02-05CH11231, the Texas Advanced Computing Center (TACC) at The University of Texas at Austin, and the Extreme Science and Engineering Discovery Environment (XSEDE),~\cite{xsede_2014} which is supported by National Science Foundation Grant No. \mbox{ACI-1053575.} 
M.A.B.-F. acknowledges support from the DOE Office of Science Graduate Fellowship Program, made possible in part by the American Recovery and Reinvestment Act of 2009, administered by ORISE-ORAU under Contract No. \mbox{DE-AC05-06OR23100.} 
T.M. acknowledges support from the National Science Foundation (NSF) Graduate Research Fellowship Program.  
R.A.D. and R.C. acknowledge support from the Scientific Discovery through Advanced Computing (SciDAC) program through the Department of Energy under Grant No. \mbox{DE-SC0008626.}
A.A.G. acknowledges support from the STC Center for Integrated Quantum Materials, NSF Grant No. DMR-1231319. All opinions expressed in this paper are the authors' and do not necessarily reflect the policies and views of \mbox{DOE, ORAU, ORISE, or NSF.}

\bibliographystyle{aipnum4-1.bst}
\bibliography{./bibliography/bibliography.bib}

\begin{thebibliography}{163}%
\makeatletter
\providecommand \@ifxundefined [1]{%
 \@ifx{#1\undefined}
}%
\providecommand \@ifnum [1]{%
 \ifnum #1\expandafter \@firstoftwo
 \else \expandafter \@secondoftwo
 \fi
}%
\providecommand \@ifx [1]{%
 \ifx #1\expandafter \@firstoftwo
 \else \expandafter \@secondoftwo
 \fi
}%
\providecommand \natexlab [1]{#1}%
\providecommand \enquote  [1]{``#1''}%
\providecommand \bibnamefont  [1]{#1}%
\providecommand \bibfnamefont [1]{#1}%
\providecommand \citenamefont [1]{#1}%
\providecommand \href@noop [0]{\@secondoftwo}%
\providecommand \href [0]{\begingroup \@sanitize@url \@href}%
\providecommand \@href[1]{\@@startlink{#1}\@@href}%
\providecommand \@@href[1]{\endgroup#1\@@endlink}%
\providecommand \@sanitize@url [0]{\catcode `\\12\catcode `\$12\catcode
  `\&12\catcode `\#12\catcode `\^12\catcode `\_12\catcode `\%12\relax}%
\providecommand \@@startlink[1]{}%
\providecommand \@@endlink[0]{}%
\providecommand \url  [0]{\begingroup\@sanitize@url \@url }%
\providecommand \@url [1]{\endgroup\@href {#1}{\urlprefix }}%
\providecommand \urlprefix  [0]{URL }%
\providecommand \Eprint [0]{\href }%
\providecommand \doibase [0]{http://dx.doi.org/}%
\providecommand \selectlanguage [0]{\@gobble}%
\providecommand \bibinfo  [0]{\@secondoftwo}%
\providecommand \bibfield  [0]{\@secondoftwo}%
\providecommand \translation [1]{[#1]}%
\providecommand \BibitemOpen [0]{}%
\providecommand \bibitemStop [0]{}%
\providecommand \bibitemNoStop [0]{.\EOS\space}%
\providecommand \EOS [0]{\spacefactor3000\relax}%
\providecommand \BibitemShut  [1]{\csname bibitem#1\endcsname}%
\let\auto@bib@innerbib\@empty
\bibitem [{\citenamefont {Stone}(2013)}]{stone_theory_2013}%
  \BibitemOpen
  \bibfield  {author} {\bibinfo {author} {\bibfnamefont {A.~J.}\ \bibnamefont
  {Stone}},\ }\href@noop {} {\emph {\bibinfo {title} {The theory of
  intermolecular forces}}},\ \bibinfo {edition} {2nd}\ ed.\ (\bibinfo
  {publisher} {Oxford University Press},\ \bibinfo {address} {Oxford},\
  \bibinfo {year} {2013})\BibitemShut {NoStop}%
\bibitem [{\citenamefont {Wales}(2003)}]{wales_energy_2003}%
  \BibitemOpen
  \bibfield  {author} {\bibinfo {author} {\bibfnamefont {D.~J.}\ \bibnamefont
  {Wales}},\ }\href@noop {} {\emph {\bibinfo {title} {Energy landscapes}}},\
  Cambridge molecular science\ (\bibinfo  {publisher} {Cambridge University
  Press},\ \bibinfo {address} {Cambridge, UK; New York},\ \bibinfo {year}
  {2003})\BibitemShut {NoStop}%
\bibitem [{\citenamefont {Parsegian}(2006)}]{parsegian_van_2006}%
  \BibitemOpen
  \bibfield  {author} {\bibinfo {author} {\bibfnamefont {V.~A.}\ \bibnamefont
  {Parsegian}},\ }\href@noop {} {\emph {\bibinfo {title} {Van der {Waals}
  forces: a handbook for biologists, chemists, engineers, and physicists}}}\
  (\bibinfo  {publisher} {Cambridge University Press},\ \bibinfo {address} {New
  York},\ \bibinfo {year} {2006})\BibitemShut {NoStop}%
\bibitem [{\citenamefont
  {Israelachvili}(2011)}]{israelachvili_intermolecular_2011}%
  \BibitemOpen
  \bibfield  {author} {\bibinfo {author} {\bibfnamefont {J.~N.}\ \bibnamefont
  {Israelachvili}},\ }\href
  {http://public.eblib.com/choice/publicfullrecord.aspx?p=630012} {\emph
  {\bibinfo {title} {Intermolecular and surface forces}}},\ \bibinfo {edition}
  {3rd}\ ed.\ (\bibinfo  {publisher} {Academic Press},\ \bibinfo {address}
  {Burlington, MA},\ \bibinfo {year} {2011})\BibitemShut {NoStop}%
\bibitem [{\citenamefont {Rapcewicz}\ and\ \citenamefont
  {Ashcroft}(1991)}]{rapcewicz_fluctuation_1991}%
  \BibitemOpen
  \bibfield  {author} {\bibinfo {author} {\bibfnamefont {K.}~\bibnamefont
  {Rapcewicz}}\ and\ \bibinfo {author} {\bibfnamefont {N.~W.}\ \bibnamefont
  {Ashcroft}},\ }\href {\doibase 10.1103/PhysRevB.44.4032} {\bibfield
  {journal} {\bibinfo  {journal} {Phys. Rev. B}\ }\textbf {\bibinfo {volume}
  {44}},\ \bibinfo {pages} {4032} (\bibinfo {year} {1991})}\BibitemShut
  {NoStop}%
\bibitem [{\citenamefont {Dobson}\ and\ \citenamefont
  {Dinte}(1996)}]{dobson_constraint_1996}%
  \BibitemOpen
  \bibfield  {author} {\bibinfo {author} {\bibfnamefont {J.~F.}\ \bibnamefont
  {Dobson}}\ and\ \bibinfo {author} {\bibfnamefont {B.~P.}\ \bibnamefont
  {Dinte}},\ }\href {\doibase 10.1103/PhysRevLett.76.1780} {\bibfield
  {journal} {\bibinfo  {journal} {Phys. Rev. Lett.}\ }\textbf {\bibinfo
  {volume} {76}},\ \bibinfo {pages} {1780} (\bibinfo {year}
  {1996})}\BibitemShut {NoStop}%
\bibitem [{\citenamefont {Andersson}, \citenamefont {Langreth},\ and\
  \citenamefont {Lundqvist}(1996)}]{andersson_van_1996}%
  \BibitemOpen
  \bibfield  {author} {\bibinfo {author} {\bibfnamefont {Y.}~\bibnamefont
  {Andersson}}, \bibinfo {author} {\bibfnamefont {D.~C.}\ \bibnamefont
  {Langreth}}, \ and\ \bibinfo {author} {\bibfnamefont {B.~I.}\ \bibnamefont
  {Lundqvist}},\ }\href {\doibase 10.1103/PhysRevLett.76.102} {\bibfield
  {journal} {\bibinfo  {journal} {Phys. Rev. Lett.}\ }\textbf {\bibinfo
  {volume} {76}},\ \bibinfo {pages} {102} (\bibinfo {year} {1996})}\BibitemShut
  {NoStop}%
\bibitem [{\citenamefont {Elstner}\ \emph {et~al.}(2001)\citenamefont
  {Elstner}, \citenamefont {Hobza}, \citenamefont {Frauenheim}, \citenamefont
  {Suhai},\ and\ \citenamefont {Kaxiras}}]{elstner_hydrogen_2001}%
  \BibitemOpen
  \bibfield  {author} {\bibinfo {author} {\bibfnamefont {M.}~\bibnamefont
  {Elstner}}, \bibinfo {author} {\bibfnamefont {P.}~\bibnamefont {Hobza}},
  \bibinfo {author} {\bibfnamefont {T.}~\bibnamefont {Frauenheim}}, \bibinfo
  {author} {\bibfnamefont {S.}~\bibnamefont {Suhai}}, \ and\ \bibinfo {author}
  {\bibfnamefont {E.}~\bibnamefont {Kaxiras}},\ }\href {\doibase
  10.1063/1.1329889} {\bibfield  {journal} {\bibinfo  {journal} {J. Chem.
  Phys.}\ }\textbf {\bibinfo {volume} {114}},\ \bibinfo {pages} {5149}
  (\bibinfo {year} {2001})}\BibitemShut {NoStop}%
\bibitem [{\citenamefont {Wu}\ \emph {et~al.}(2001)\citenamefont {Wu},
  \citenamefont {Vargas}, \citenamefont {Nayak}, \citenamefont {Lotrich},\ and\
  \citenamefont {Scoles}}]{wu_towards_2001}%
  \BibitemOpen
  \bibfield  {author} {\bibinfo {author} {\bibfnamefont {X.}~\bibnamefont
  {Wu}}, \bibinfo {author} {\bibfnamefont {M.~C.}\ \bibnamefont {Vargas}},
  \bibinfo {author} {\bibfnamefont {S.}~\bibnamefont {Nayak}}, \bibinfo
  {author} {\bibfnamefont {V.}~\bibnamefont {Lotrich}}, \ and\ \bibinfo
  {author} {\bibfnamefont {G.}~\bibnamefont {Scoles}},\ }\href {\doibase
  10.1063/1.1412004} {\bibfield  {journal} {\bibinfo  {journal} {J. of Chem.
  Phys.}\ }\textbf {\bibinfo {volume} {115}},\ \bibinfo {pages} {8748}
  (\bibinfo {year} {2001})}\BibitemShut {NoStop}%
\bibitem [{\citenamefont {Wu}\ and\ \citenamefont
  {Yang}(2002)}]{wu_empirical_2002}%
  \BibitemOpen
  \bibfield  {author} {\bibinfo {author} {\bibfnamefont {Q.}~\bibnamefont
  {Wu}}\ and\ \bibinfo {author} {\bibfnamefont {W.}~\bibnamefont {Yang}},\
  }\href {\doibase 10.1063/1.1424928} {\bibfield  {journal} {\bibinfo
  {journal} {J. Chem. Phys.}\ }\textbf {\bibinfo {volume} {116}},\ \bibinfo
  {pages} {515} (\bibinfo {year} {2002})}\BibitemShut {NoStop}%
\bibitem [{\citenamefont {Dion}\ \emph {et~al.}(2004)\citenamefont {Dion},
  \citenamefont {Rydberg}, \citenamefont {{Schr\"{o}der}}, \citenamefont
  {Langreth},\ and\ \citenamefont {Lundqvist}}]{dion_2004}%
  \BibitemOpen
  \bibfield  {author} {\bibinfo {author} {\bibfnamefont {M.}~\bibnamefont
  {Dion}}, \bibinfo {author} {\bibfnamefont {H.}~\bibnamefont {Rydberg}},
  \bibinfo {author} {\bibfnamefont {E.}~\bibnamefont {{Schr\"{o}der}}},
  \bibinfo {author} {\bibfnamefont {D.~C.}\ \bibnamefont {Langreth}}, \ and\
  \bibinfo {author} {\bibfnamefont {B.~I.}\ \bibnamefont {Lundqvist}},\ }\href
  {\doibase 10.1103/physrevlett.92.246401} {\bibfield  {journal} {\bibinfo
  {journal} {Phys. Rev. Lett.}\ }\textbf {\bibinfo {volume} {92}},\ \bibinfo
  {pages} {246401} (\bibinfo {year} {2004})}\BibitemShut {NoStop}%
\bibitem [{\citenamefont {{O. A. von Lilienfeld}}\ \emph
  {et~al.}(2004)\citenamefont {{O. A. von Lilienfeld}}, \citenamefont
  {Tavernelli}, \citenamefont {Rothlisberger},\ and\ \citenamefont
  {Sebastiani}}]{lilienfeld_2004}%
  \BibitemOpen
  \bibfield  {author} {\bibinfo {author} {\bibnamefont {{O. A. von
  Lilienfeld}}}, \bibinfo {author} {\bibfnamefont {I.}~\bibnamefont
  {Tavernelli}}, \bibinfo {author} {\bibfnamefont {U.}~\bibnamefont
  {Rothlisberger}}, \ and\ \bibinfo {author} {\bibfnamefont {D.}~\bibnamefont
  {Sebastiani}},\ }\href {\doibase 10.1103/physrevlett.93.153004} {\bibfield
  {journal} {\bibinfo  {journal} {Phys. Rev. Lett.}\ }\textbf {\bibinfo
  {volume} {93}},\ \bibinfo {pages} {153004} (\bibinfo {year}
  {2004})}\BibitemShut {NoStop}%
\bibitem [{\citenamefont {Grimme}(2004)}]{grimme_accurate_2004}%
  \BibitemOpen
  \bibfield  {author} {\bibinfo {author} {\bibfnamefont {S.}~\bibnamefont
  {Grimme}},\ }\href {\doibase 10.1002/jcc.20078} {\bibfield  {journal}
  {\bibinfo  {journal} {J. Comput. Chem.}\ }\textbf {\bibinfo {volume} {25}},\
  \bibinfo {pages} {1463} (\bibinfo {year} {2004})}\BibitemShut {NoStop}%
\bibitem [{\citenamefont {Misquitta}\ \emph {et~al.}(2005)\citenamefont
  {Misquitta}, \citenamefont {Podeszwa}, \citenamefont {Jeziorski},\ and\
  \citenamefont {Szalewicz}}]{misquitta_2005}%
  \BibitemOpen
  \bibfield  {author} {\bibinfo {author} {\bibfnamefont {A.~J.}\ \bibnamefont
  {Misquitta}}, \bibinfo {author} {\bibfnamefont {R.}~\bibnamefont {Podeszwa}},
  \bibinfo {author} {\bibfnamefont {B.}~\bibnamefont {Jeziorski}}, \ and\
  \bibinfo {author} {\bibfnamefont {K.}~\bibnamefont {Szalewicz}},\ }\href
  {\doibase 10.1063/1.2135288} {\bibfield  {journal} {\bibinfo  {journal} {J.
  Chem. Phys.}\ }\textbf {\bibinfo {volume} {123}},\ \bibinfo {pages} {214103}
  (\bibinfo {year} {2005})}\BibitemShut {NoStop}%
\bibitem [{\citenamefont {Becke}\ and\ \citenamefont
  {Johnson}(2005{\natexlab{a}})}]{becke_jcp_xdm_2005}%
  \BibitemOpen
  \bibfield  {author} {\bibinfo {author} {\bibfnamefont {A.~D.}\ \bibnamefont
  {Becke}}\ and\ \bibinfo {author} {\bibfnamefont {E.~R.}\ \bibnamefont
  {Johnson}},\ }\href {\doibase 10.1063/1.1884601} {\bibfield  {journal}
  {\bibinfo  {journal} {J. Chem. Phys.}\ }\textbf {\bibinfo {volume} {122}},\
  \bibinfo {pages} {154104} (\bibinfo {year} {2005}{\natexlab{a}})}\BibitemShut
  {NoStop}%
\bibitem [{\citenamefont {Becke}\ and\ \citenamefont
  {Johnson}(2005{\natexlab{b}})}]{becke_jcp_dft_2005}%
  \BibitemOpen
  \bibfield  {author} {\bibinfo {author} {\bibfnamefont {A.~D.}\ \bibnamefont
  {Becke}}\ and\ \bibinfo {author} {\bibfnamefont {E.~R.}\ \bibnamefont
  {Johnson}},\ }\href {\doibase 10.1063/1.2065267} {\bibfield  {journal}
  {\bibinfo  {journal} {J. Chem. Phys.}\ }\textbf {\bibinfo {volume} {123}},\
  \bibinfo {pages} {154101} (\bibinfo {year} {2005}{\natexlab{b}})}\BibitemShut
  {NoStop}%
\bibitem [{\citenamefont {Johnson}\ and\ \citenamefont
  {Becke}(2005)}]{johnson_jcp_hf_2005}%
  \BibitemOpen
  \bibfield  {author} {\bibinfo {author} {\bibfnamefont {E.~R.}\ \bibnamefont
  {Johnson}}\ and\ \bibinfo {author} {\bibfnamefont {A.~D.}\ \bibnamefont
  {Becke}},\ }\href {\doibase 10.1063/1.1949201} {\bibfield  {journal}
  {\bibinfo  {journal} {J. Chem. Phys.}\ }\textbf {\bibinfo {volume} {123}},\
  \bibinfo {pages} {024101} (\bibinfo {year} {2005})}\BibitemShut {NoStop}%
\bibitem [{\citenamefont {Grimme}(2006)}]{grimme_2006}%
  \BibitemOpen
  \bibfield  {author} {\bibinfo {author} {\bibfnamefont {S.}~\bibnamefont
  {Grimme}},\ }\href {\doibase 10.1002/jcc.20495} {\bibfield  {journal}
  {\bibinfo  {journal} {J. Comp. Chem.}\ }\textbf {\bibinfo {volume} {27}},\
  \bibinfo {pages} {1787} (\bibinfo {year} {2006})}\BibitemShut {NoStop}%
\bibitem [{\citenamefont {Becke}\ and\ \citenamefont
  {Johnson}(2006)}]{becke_jcp_2006}%
  \BibitemOpen
  \bibfield  {author} {\bibinfo {author} {\bibfnamefont {A.~D.}\ \bibnamefont
  {Becke}}\ and\ \bibinfo {author} {\bibfnamefont {E.~R.}\ \bibnamefont
  {Johnson}},\ }\href {\doibase 10.1063/1.2139668} {\bibfield  {journal}
  {\bibinfo  {journal} {J. Chem. Phys.}\ }\textbf {\bibinfo {volume} {124}},\
  \bibinfo {pages} {014104} (\bibinfo {year} {2006})}\BibitemShut {NoStop}%
\bibitem [{\citenamefont {Zhao}\ and\ \citenamefont
  {Truhlar}(2006)}]{zhao_MO6_2006}%
  \BibitemOpen
  \bibfield  {author} {\bibinfo {author} {\bibfnamefont {Y.}~\bibnamefont
  {Zhao}}\ and\ \bibinfo {author} {\bibfnamefont {D.~G.}\ \bibnamefont
  {Truhlar}},\ }\href {\doibase 10.1063/1.2370993} {\bibfield  {journal}
  {\bibinfo  {journal} {J. Chem. Phys.}\ }\textbf {\bibinfo {volume} {125}},\
  \bibinfo {pages} {194101} (\bibinfo {year} {2006})}\BibitemShut {NoStop}%
\bibitem [{\citenamefont {Becke}\ and\ \citenamefont
  {Johnson}(2007{\natexlab{a}})}]{becke_jcp_2007}%
  \BibitemOpen
  \bibfield  {author} {\bibinfo {author} {\bibfnamefont {A.~D.}\ \bibnamefont
  {Becke}}\ and\ \bibinfo {author} {\bibfnamefont {E.~R.}\ \bibnamefont
  {Johnson}},\ }\href {\doibase 10.1063/1.2768530} {\bibfield  {journal}
  {\bibinfo  {journal} {J. Chem. Phys.}\ }\textbf {\bibinfo {volume} {127}},\
  \bibinfo {pages} {124108} (\bibinfo {year} {2007}{\natexlab{a}})}\BibitemShut
  {NoStop}%
\bibitem [{\citenamefont {Becke}\ and\ \citenamefont
  {Johnson}(2007{\natexlab{b}})}]{becke_jcp_xdm_rev_2007}%
  \BibitemOpen
  \bibfield  {author} {\bibinfo {author} {\bibfnamefont {A.~D.}\ \bibnamefont
  {Becke}}\ and\ \bibinfo {author} {\bibfnamefont {E.~R.}\ \bibnamefont
  {Johnson}},\ }\href {\doibase 10.1063/1.2795701} {\bibfield  {journal}
  {\bibinfo  {journal} {J. Chem. Phys.}\ }\textbf {\bibinfo {volume} {127}},\
  \bibinfo {pages} {154108} (\bibinfo {year} {2007}{\natexlab{b}})}\BibitemShut
  {NoStop}%
\bibitem [{\citenamefont {{Jure\v{c}ka}}\ \emph {et~al.}(2007)\citenamefont
  {{Jure\v{c}ka}}, \citenamefont {{\v{C}ern\'{y}}}, \citenamefont {Hobza},\
  and\ \citenamefont {Salahub}}]{jurecka_2007}%
  \BibitemOpen
  \bibfield  {author} {\bibinfo {author} {\bibfnamefont {P.}~\bibnamefont
  {{Jure\v{c}ka}}}, \bibinfo {author} {\bibfnamefont {J.}~\bibnamefont
  {{\v{C}ern\'{y}}}}, \bibinfo {author} {\bibfnamefont {P.}~\bibnamefont
  {Hobza}}, \ and\ \bibinfo {author} {\bibfnamefont {D.~R.}\ \bibnamefont
  {Salahub}},\ }\href {\doibase 10.1002/jcc.20570} {\bibfield  {journal}
  {\bibinfo  {journal} {J. Comput. Chem.}\ }\textbf {\bibinfo {volume} {28}},\
  \bibinfo {pages} {555} (\bibinfo {year} {2007})}\BibitemShut {NoStop}%
\bibitem [{\citenamefont {Silvestrelli}(2008)}]{silvestrelli_2008}%
  \BibitemOpen
  \bibfield  {author} {\bibinfo {author} {\bibfnamefont {P.~L.}\ \bibnamefont
  {Silvestrelli}},\ }\href {\doibase 10.1103/physrevlett.100.053002} {\bibfield
   {journal} {\bibinfo  {journal} {Phys. Rev. Lett.}\ }\textbf {\bibinfo
  {volume} {100}},\ \bibinfo {pages} {053002} (\bibinfo {year}
  {2008})}\BibitemShut {NoStop}%
\bibitem [{\citenamefont {Sun}\ \emph {et~al.}(2008)\citenamefont {Sun},
  \citenamefont {Kim}, \citenamefont {Lee},\ and\ \citenamefont
  {Zhang}}]{sun_jcp_2008}%
  \BibitemOpen
  \bibfield  {author} {\bibinfo {author} {\bibfnamefont {Y.~Y.}\ \bibnamefont
  {Sun}}, \bibinfo {author} {\bibfnamefont {Y.-H.}\ \bibnamefont {Kim}},
  \bibinfo {author} {\bibfnamefont {K.}~\bibnamefont {Lee}}, \ and\ \bibinfo
  {author} {\bibfnamefont {S.~B.}\ \bibnamefont {Zhang}},\ }\href {\doibase
  10.1063/1.2992078} {\bibfield  {journal} {\bibinfo  {journal} {J. Chem.
  Phys.}\ }\textbf {\bibinfo {volume} {129}},\ \bibinfo {pages} {154102}
  (\bibinfo {year} {2008})}\BibitemShut {NoStop}%
\bibitem [{\citenamefont {Chai}\ and\ \citenamefont
  {Head-Gordon}(2008)}]{chai_wB97XD_2008}%
  \BibitemOpen
  \bibfield  {author} {\bibinfo {author} {\bibfnamefont {J.-D.}\ \bibnamefont
  {Chai}}\ and\ \bibinfo {author} {\bibfnamefont {M.}~\bibnamefont
  {Head-Gordon}},\ }\href {\doibase 10.1039/b810189b} {\bibfield  {journal}
  {\bibinfo  {journal} {Phys. Chem. Chem. Phys.}\ }\textbf {\bibinfo {volume}
  {10}},\ \bibinfo {pages} {6615} (\bibinfo {year} {2008})}\BibitemShut
  {NoStop}%
\bibitem [{\citenamefont {DiLabio}(2008)}]{dilabio_cpl_2008}%
  \BibitemOpen
  \bibfield  {author} {\bibinfo {author} {\bibfnamefont {G.~A.}\ \bibnamefont
  {DiLabio}},\ }\href {\doibase 10.1016/j.cplett.2008.02.110} {\bibfield
  {journal} {\bibinfo  {journal} {Chem. Phys. Lett.}\ }\textbf {\bibinfo
  {volume} {455}},\ \bibinfo {pages} {348} (\bibinfo {year}
  {2008})}\BibitemShut {NoStop}%
\bibitem [{\citenamefont {Mackie}\ and\ \citenamefont
  {DiLabio}(2008)}]{mackie_2008}%
  \BibitemOpen
  \bibfield  {author} {\bibinfo {author} {\bibfnamefont {I.~D.}\ \bibnamefont
  {Mackie}}\ and\ \bibinfo {author} {\bibfnamefont {G.~A.}\ \bibnamefont
  {DiLabio}},\ }\href {\doibase 10.1021/jp806162t} {\bibfield  {journal}
  {\bibinfo  {journal} {J. Phys. Chem. A}\ }\textbf {\bibinfo {volume} {112}},\
  \bibinfo {pages} {10968} (\bibinfo {year} {2008})}\BibitemShut {NoStop}%
\bibitem [{\citenamefont {Tkatchenko}\ and\ \citenamefont
  {Scheffler}(2009)}]{tkatchenko_2009}%
  \BibitemOpen
  \bibfield  {author} {\bibinfo {author} {\bibfnamefont {A.}~\bibnamefont
  {Tkatchenko}}\ and\ \bibinfo {author} {\bibfnamefont {M.}~\bibnamefont
  {Scheffler}},\ }\href {\doibase 10.1103/PhysRevLett.102.073005} {\bibfield
  {journal} {\bibinfo  {journal} {Phys. Rev. Lett.}\ }\textbf {\bibinfo
  {volume} {102}},\ \bibinfo {pages} {073005} (\bibinfo {year}
  {2009})}\BibitemShut {NoStop}%
\bibitem [{\citenamefont {{Rom\'{a}n-P\'{e}rez}}\ and\ \citenamefont
  {Soler}(2009)}]{roman-perez_2009}%
  \BibitemOpen
  \bibfield  {author} {\bibinfo {author} {\bibfnamefont {G.}~\bibnamefont
  {{Rom\'{a}n-P\'{e}rez}}}\ and\ \bibinfo {author} {\bibfnamefont {J.~M.}\
  \bibnamefont {Soler}},\ }\href {\doibase 10.1103/PhysRevLett.103.096102}
  {\bibfield  {journal} {\bibinfo  {journal} {Phys. Rev. Lett.}\ }\textbf
  {\bibinfo {volume} {103}},\ \bibinfo {pages} {096102} (\bibinfo {year}
  {2009})}\BibitemShut {NoStop}%
\bibitem [{\citenamefont {Vydrov}\ and\ \citenamefont
  {Van~Voorhis}(2009)}]{vydrov_2009}%
  \BibitemOpen
  \bibfield  {author} {\bibinfo {author} {\bibfnamefont {O.~A.}\ \bibnamefont
  {Vydrov}}\ and\ \bibinfo {author} {\bibfnamefont {T.}~\bibnamefont
  {Van~Voorhis}},\ }\href {\doibase 10.1103/PhysRevLett.103.063004} {\bibfield
  {journal} {\bibinfo  {journal} {Phys. Rev. Lett.}\ }\textbf {\bibinfo
  {volume} {103}},\ \bibinfo {pages} {063004} (\bibinfo {year}
  {2009})}\BibitemShut {NoStop}%
\bibitem [{\citenamefont {Johnson}\ and\ \citenamefont
  {DiLabio}(2009)}]{johnson_jpcc_2009}%
  \BibitemOpen
  \bibfield  {author} {\bibinfo {author} {\bibfnamefont {E.~R.}\ \bibnamefont
  {Johnson}}\ and\ \bibinfo {author} {\bibfnamefont {G.~A.}\ \bibnamefont
  {DiLabio}},\ }\href {\doibase 10.1021/jp8105056} {\bibfield  {journal}
  {\bibinfo  {journal} {J. Phys. Chem. C}\ }\textbf {\bibinfo {volume} {113}},\
  \bibinfo {pages} {5681} (\bibinfo {year} {2009})}\BibitemShut {NoStop}%
\bibitem [{\citenamefont {Sato}\ and\ \citenamefont
  {Nakai}(2009)}]{sato_density_2009}%
  \BibitemOpen
  \bibfield  {author} {\bibinfo {author} {\bibfnamefont {T.}~\bibnamefont
  {Sato}}\ and\ \bibinfo {author} {\bibfnamefont {H.}~\bibnamefont {Nakai}},\
  }\href {\doibase 10.1063/1.3269802} {\bibfield  {journal} {\bibinfo
  {journal} {J. Chem. Phys.}\ }\textbf {\bibinfo {volume} {131}},\ \bibinfo
  {pages} {224104} (\bibinfo {year} {2009})}\BibitemShut {NoStop}%
\bibitem [{\citenamefont {Sato}\ and\ \citenamefont
  {Nakai}(2010)}]{sato_local_2010}%
  \BibitemOpen
  \bibfield  {author} {\bibinfo {author} {\bibfnamefont {T.}~\bibnamefont
  {Sato}}\ and\ \bibinfo {author} {\bibfnamefont {H.}~\bibnamefont {Nakai}},\
  }\href {\doibase 10.1063/1.3503040} {\bibfield  {journal} {\bibinfo
  {journal} {J. Chem. Phys.}\ }\textbf {\bibinfo {volume} {133}},\ \bibinfo
  {pages} {194101} (\bibinfo {year} {2010})}\BibitemShut {NoStop}%
\bibitem [{\citenamefont {Grimme}\ \emph {et~al.}(2010)\citenamefont {Grimme},
  \citenamefont {Antony}, \citenamefont {Ehrlich},\ and\ \citenamefont
  {Krieg}}]{grimme_d3_2010}%
  \BibitemOpen
  \bibfield  {author} {\bibinfo {author} {\bibfnamefont {S.}~\bibnamefont
  {Grimme}}, \bibinfo {author} {\bibfnamefont {J.}~\bibnamefont {Antony}},
  \bibinfo {author} {\bibfnamefont {S.}~\bibnamefont {Ehrlich}}, \ and\
  \bibinfo {author} {\bibfnamefont {H.}~\bibnamefont {Krieg}},\ }\href
  {\doibase 10.1063/1.3382344} {\bibfield  {journal} {\bibinfo  {journal} {J.
  Chem. Phys.}\ }\textbf {\bibinfo {volume} {132}},\ \bibinfo {pages} {154104}
  (\bibinfo {year} {2010})}\BibitemShut {NoStop}%
\bibitem [{\citenamefont {Tkatchenko}\ \emph {et~al.}(2010)\citenamefont
  {Tkatchenko}, \citenamefont {Romaner}, \citenamefont {Hofmann}, \citenamefont
  {Zojer}, \citenamefont {Ambrosch-Draxl},\ and\ \citenamefont
  {Scheffler}}]{tkatchenko_2010}%
  \BibitemOpen
  \bibfield  {author} {\bibinfo {author} {\bibfnamefont {A.}~\bibnamefont
  {Tkatchenko}}, \bibinfo {author} {\bibfnamefont {L.}~\bibnamefont {Romaner}},
  \bibinfo {author} {\bibfnamefont {O.~T.}\ \bibnamefont {Hofmann}}, \bibinfo
  {author} {\bibfnamefont {E.}~\bibnamefont {Zojer}}, \bibinfo {author}
  {\bibfnamefont {C.}~\bibnamefont {Ambrosch-Draxl}}, \ and\ \bibinfo {author}
  {\bibfnamefont {M.}~\bibnamefont {Scheffler}},\ }\href {\doibase
  10.1557/mrs2010.581} {\bibfield  {journal} {\bibinfo  {journal} {MRS Bull.}\
  }\textbf {\bibinfo {volume} {35}},\ \bibinfo {pages} {435} (\bibinfo {year}
  {2010})}\BibitemShut {NoStop}%
\bibitem [{\citenamefont {Cooper}, \citenamefont {Kong},\ and\ \citenamefont
  {Langreth}(2010)}]{cooper_2010}%
  \BibitemOpen
  \bibfield  {author} {\bibinfo {author} {\bibfnamefont {V.~R.}\ \bibnamefont
  {Cooper}}, \bibinfo {author} {\bibfnamefont {L.}~\bibnamefont {Kong}}, \ and\
  \bibinfo {author} {\bibfnamefont {D.~C.}\ \bibnamefont {Langreth}},\ }\href
  {\doibase 10.1016/j.phpro.2010.01.201} {\bibfield  {journal} {\bibinfo
  {journal} {Phys. Procedia}\ }\textbf {\bibinfo {volume} {3}},\ \bibinfo
  {pages} {1417} (\bibinfo {year} {2010})}\BibitemShut {NoStop}%
\bibitem [{\citenamefont {Riley}\ \emph {et~al.}(2010)\citenamefont {Riley},
  \citenamefont {{Pito\v{n}\'{a}k}}, \citenamefont {{Jure\v{c}ka}},\ and\
  \citenamefont {Hobza}}]{riley_2010}%
  \BibitemOpen
  \bibfield  {author} {\bibinfo {author} {\bibfnamefont {K.~E.}\ \bibnamefont
  {Riley}}, \bibinfo {author} {\bibfnamefont {M.}~\bibnamefont
  {{Pito\v{n}\'{a}k}}}, \bibinfo {author} {\bibfnamefont {P.}~\bibnamefont
  {{Jure\v{c}ka}}}, \ and\ \bibinfo {author} {\bibfnamefont {P.}~\bibnamefont
  {Hobza}},\ }\href {\doibase 10.1021/cr1000173} {\bibfield  {journal}
  {\bibinfo  {journal} {Chem. Rev.}\ }\textbf {\bibinfo {volume} {110}},\
  \bibinfo {pages} {5023} (\bibinfo {year} {2010})}\BibitemShut {NoStop}%
\bibitem [{\citenamefont {Kannemann}\ and\ \citenamefont
  {Becke}(2010)}]{kannemann_2010}%
  \BibitemOpen
  \bibfield  {author} {\bibinfo {author} {\bibfnamefont {F.~O.}\ \bibnamefont
  {Kannemann}}\ and\ \bibinfo {author} {\bibfnamefont {A.~D.}\ \bibnamefont
  {Becke}},\ }\href {\doibase 10.1021/ct900699r} {\bibfield  {journal}
  {\bibinfo  {journal} {J. Chem. Theory Comput.}\ }\textbf {\bibinfo {volume}
  {6}},\ \bibinfo {pages} {1081} (\bibinfo {year} {2010})}\BibitemShut
  {NoStop}%
\bibitem [{\citenamefont {Vydrov}\ and\ \citenamefont
  {Voorhis}(2010)}]{vydrov_2010}%
  \BibitemOpen
  \bibfield  {author} {\bibinfo {author} {\bibfnamefont {O.~A.}\ \bibnamefont
  {Vydrov}}\ and\ \bibinfo {author} {\bibfnamefont {T.~V.}\ \bibnamefont
  {Voorhis}},\ }\href {\doibase 10.1063/1.3521275} {\bibfield  {journal}
  {\bibinfo  {journal} {J. Chem. Phys.}\ }\textbf {\bibinfo {volume} {133}},\
  \bibinfo {pages} {244103} (\bibinfo {year} {2010})}\BibitemShut {NoStop}%
\bibitem [{\citenamefont {Lee}\ \emph {et~al.}(2010)\citenamefont {Lee},
  \citenamefont {Murray}, \citenamefont {Kong}, \citenamefont {Lundqvist},\
  and\ \citenamefont {Langreth}}]{lee_2010}%
  \BibitemOpen
  \bibfield  {author} {\bibinfo {author} {\bibfnamefont {K.}~\bibnamefont
  {Lee}}, \bibinfo {author} {\bibfnamefont {{\'{E}amonn}.~D.}\ \bibnamefont
  {Murray}}, \bibinfo {author} {\bibfnamefont {L.}~\bibnamefont {Kong}},
  \bibinfo {author} {\bibfnamefont {B.~I.}\ \bibnamefont {Lundqvist}}, \ and\
  \bibinfo {author} {\bibfnamefont {D.~C.}\ \bibnamefont {Langreth}},\ }\href
  {\doibase 10.1103/PhysRevB.82.081101} {\bibfield  {journal} {\bibinfo
  {journal} {Phys. Rev. B}\ }\textbf {\bibinfo {volume} {82}},\ \bibinfo
  {pages} {081101(R)} (\bibinfo {year} {2010})}\BibitemShut {NoStop}%
\bibitem [{\citenamefont {Grimme}(2011)}]{grimme_review_2011}%
  \BibitemOpen
  \bibfield  {author} {\bibinfo {author} {\bibfnamefont {S.}~\bibnamefont
  {Grimme}},\ }\href {\doibase 10.1002/wcms.30} {\bibfield  {journal} {\bibinfo
   {journal} {{WIREs} Comput. Mol. Sci.}\ }\textbf {\bibinfo {volume} {1}},\
  \bibinfo {pages} {211} (\bibinfo {year} {2011})}\BibitemShut {NoStop}%
\bibitem [{\citenamefont {Steinmann}\ and\ \citenamefont
  {Corminboeuf}(2011)}]{steinmann_2011}%
  \BibitemOpen
  \bibfield  {author} {\bibinfo {author} {\bibfnamefont {S.~N.}\ \bibnamefont
  {Steinmann}}\ and\ \bibinfo {author} {\bibfnamefont {C.}~\bibnamefont
  {Corminboeuf}},\ }\href {\doibase 10.1021/ct200602x} {\bibfield  {journal}
  {\bibinfo  {journal} {J. Chem. Theory Comput.}\ }\textbf {\bibinfo {volume}
  {7}},\ \bibinfo {pages} {3567} (\bibinfo {year} {2011})}\BibitemShut
  {NoStop}%
\bibitem [{\citenamefont {Marom}\ \emph {et~al.}(2011)\citenamefont {Marom},
  \citenamefont {Tkatchenko}, \citenamefont {Rossi}, \citenamefont {Gobre},
  \citenamefont {Hod}, \citenamefont {Scheffler},\ and\ \citenamefont
  {Kronik}}]{marom_2011}%
  \BibitemOpen
  \bibfield  {author} {\bibinfo {author} {\bibfnamefont {N.}~\bibnamefont
  {Marom}}, \bibinfo {author} {\bibfnamefont {A.}~\bibnamefont {Tkatchenko}},
  \bibinfo {author} {\bibfnamefont {M.}~\bibnamefont {Rossi}}, \bibinfo
  {author} {\bibfnamefont {V.~V.}\ \bibnamefont {Gobre}}, \bibinfo {author}
  {\bibfnamefont {O.}~\bibnamefont {Hod}}, \bibinfo {author} {\bibfnamefont
  {M.}~\bibnamefont {Scheffler}}, \ and\ \bibinfo {author} {\bibfnamefont
  {L.}~\bibnamefont {Kronik}},\ }\href {\doibase 10.1021/ct2005616} {\bibfield
  {journal} {\bibinfo  {journal} {J. Chem. Theory Comput.}\ }\textbf {\bibinfo
  {volume} {7}},\ \bibinfo {pages} {3944} (\bibinfo {year} {2011})}\BibitemShut
  {NoStop}%
\bibitem [{\citenamefont {Grimme}, \citenamefont {Ehrlich},\ and\ \citenamefont
  {Goerigk}(2011)}]{grimme_bj_2011}%
  \BibitemOpen
  \bibfield  {author} {\bibinfo {author} {\bibfnamefont {S.}~\bibnamefont
  {Grimme}}, \bibinfo {author} {\bibfnamefont {S.}~\bibnamefont {Ehrlich}}, \
  and\ \bibinfo {author} {\bibfnamefont {L.}~\bibnamefont {Goerigk}},\ }\href
  {\doibase 10.1002/jcc.21759} {\bibfield  {journal} {\bibinfo  {journal} {J.
  Comp. Chem.}\ }\textbf {\bibinfo {volume} {32}},\ \bibinfo {pages} {1456}
  (\bibinfo {year} {2011})}\BibitemShut {NoStop}%
\bibitem [{\citenamefont {Tkatchenko}\ \emph {et~al.}(2012)\citenamefont
  {Tkatchenko}, \citenamefont {{R. A. DiStasio Jr.}}, \citenamefont {Car},\
  and\ \citenamefont {Scheffler}}]{tkatchenko_2012}%
  \BibitemOpen
  \bibfield  {author} {\bibinfo {author} {\bibfnamefont {A.}~\bibnamefont
  {Tkatchenko}}, \bibinfo {author} {\bibnamefont {{R. A. DiStasio Jr.}}},
  \bibinfo {author} {\bibfnamefont {R.}~\bibnamefont {Car}}, \ and\ \bibinfo
  {author} {\bibfnamefont {M.}~\bibnamefont {Scheffler}},\ }\href {\doibase
  10.1103/physrevlett.108.236402} {\bibfield  {journal} {\bibinfo  {journal}
  {Phys. Rev. Lett.}\ }\textbf {\bibinfo {volume} {108}},\ \bibinfo {pages}
  {236402} (\bibinfo {year} {2012})}\BibitemShut {NoStop}%
\bibitem [{\citenamefont {{R. A. DiStasio Jr.}}, \citenamefont {{O. A. von
  Lilienfeld}},\ and\ \citenamefont {Tkatchenko}(2012)}]{distasio_pnas_2012}%
  \BibitemOpen
  \bibfield  {author} {\bibinfo {author} {\bibnamefont {{R. A. DiStasio Jr.}}},
  \bibinfo {author} {\bibnamefont {{O. A. von Lilienfeld}}}, \ and\ \bibinfo
  {author} {\bibfnamefont {A.}~\bibnamefont {Tkatchenko}},\ }\href {\doibase
  10.1073/pnas.1208121109} {\bibfield  {journal} {\bibinfo  {journal} {Proc.
  Natl. Acad. Sci. USA}\ }\textbf {\bibinfo {volume} {109}},\ \bibinfo {pages}
  {14791} (\bibinfo {year} {2012})}\BibitemShut {NoStop}%
\bibitem [{\citenamefont {Torres}\ and\ \citenamefont
  {DiLabio}(2012)}]{torres_jpcl_2012}%
  \BibitemOpen
  \bibfield  {author} {\bibinfo {author} {\bibfnamefont {E.}~\bibnamefont
  {Torres}}\ and\ \bibinfo {author} {\bibfnamefont {G.~A.}\ \bibnamefont
  {DiLabio}},\ }\href {\doibase 10.1021/jz300554y} {\bibfield  {journal}
  {\bibinfo  {journal} {J. Phys. Chem. Lett.}\ }\textbf {\bibinfo {volume}
  {3}},\ \bibinfo {pages} {1738} (\bibinfo {year} {2012})}\BibitemShut
  {NoStop}%
\bibitem [{\citenamefont {Tkatchenko}, \citenamefont {Ambrosetti},\ and\
  \citenamefont {{R. A. DiStasio Jr.}}(2013)}]{tkatchenko_jcp_2013}%
  \BibitemOpen
  \bibfield  {author} {\bibinfo {author} {\bibfnamefont {A.}~\bibnamefont
  {Tkatchenko}}, \bibinfo {author} {\bibfnamefont {A.}~\bibnamefont
  {Ambrosetti}}, \ and\ \bibinfo {author} {\bibnamefont {{R. A. DiStasio
  Jr.}}},\ }\href {\doibase 10.1063/1.4789814} {\bibfield  {journal} {\bibinfo
  {journal} {J. Chem. Phys.}\ }\textbf {\bibinfo {volume} {138}},\ \bibinfo
  {pages} {074106} (\bibinfo {year} {2013})}\BibitemShut {NoStop}%
\bibitem [{\citenamefont {Sabatini}, \citenamefont {Gorni},\ and\ \citenamefont
  {{S. de Gironcoli}}(2013)}]{sabatini_prb_2013}%
  \BibitemOpen
  \bibfield  {author} {\bibinfo {author} {\bibfnamefont {R.}~\bibnamefont
  {Sabatini}}, \bibinfo {author} {\bibfnamefont {T.}~\bibnamefont {Gorni}}, \
  and\ \bibinfo {author} {\bibnamefont {{S. de Gironcoli}}},\ }\href {\doibase
  10.1103/PhysRevB.87.041108} {\bibfield  {journal} {\bibinfo  {journal} {Phys.
  Rev. B}\ }\textbf {\bibinfo {volume} {87}},\ \bibinfo {pages} {041108(R)}
  (\bibinfo {year} {2013})}\BibitemShut {NoStop}%
\bibitem [{\citenamefont {Silvestrelli}(2013)}]{silvestrelli_jcp_2013}%
  \BibitemOpen
  \bibfield  {author} {\bibinfo {author} {\bibfnamefont {P.~L.}\ \bibnamefont
  {Silvestrelli}},\ }\href {\doibase 10.1063/1.4816964} {\bibfield  {journal}
  {\bibinfo  {journal} {J. Chem. Phys.}\ }\textbf {\bibinfo {volume} {139}},\
  \bibinfo {pages} {054106} (\bibinfo {year} {2013})}\BibitemShut {NoStop}%
\bibitem [{\citenamefont {{R. A. DiStasio Jr.}}, \citenamefont {Gobre},\ and\
  \citenamefont {Tkatchenko}(2014)}]{distasio_2014}%
  \BibitemOpen
  \bibfield  {author} {\bibinfo {author} {\bibnamefont {{R. A. DiStasio Jr.}}},
  \bibinfo {author} {\bibfnamefont {V.~V.}\ \bibnamefont {Gobre}}, \ and\
  \bibinfo {author} {\bibfnamefont {A.}~\bibnamefont {Tkatchenko}},\ }\href
  {\doibase 10.1088/0953-8984/26/21/213202} {\bibfield  {journal} {\bibinfo
  {journal} {J. Phys.: Condens. Matter}\ }\textbf {\bibinfo {volume} {26}},\
  \bibinfo {pages} {213202} (\bibinfo {year} {2014})}\BibitemShut {NoStop}%
\bibitem [{\citenamefont {Ambrosetti}\ \emph
  {et~al.}(2014{\natexlab{a}})\citenamefont {Ambrosetti}, \citenamefont
  {Reilly}, \citenamefont {{R. A. DiStasio Jr.}},\ and\ \citenamefont
  {Tkatchenko}}]{ambrosetti_2014}%
  \BibitemOpen
  \bibfield  {author} {\bibinfo {author} {\bibfnamefont {A.}~\bibnamefont
  {Ambrosetti}}, \bibinfo {author} {\bibfnamefont {A.~M.}\ \bibnamefont
  {Reilly}}, \bibinfo {author} {\bibnamefont {{R. A. DiStasio Jr.}}}, \ and\
  \bibinfo {author} {\bibfnamefont {A.}~\bibnamefont {Tkatchenko}},\ }\href
  {\doibase 10.1063/1.4865104} {\bibfield  {journal} {\bibinfo  {journal} {J.
  Chem. Phys.}\ }\textbf {\bibinfo {volume} {140}},\ \bibinfo {pages} {18A508}
  (\bibinfo {year} {2014}{\natexlab{a}})}\BibitemShut {NoStop}%
\bibitem [{\citenamefont {Becke}(2014)}]{becke_perspective_2014}%
  \BibitemOpen
  \bibfield  {author} {\bibinfo {author} {\bibfnamefont {A.~D.}\ \bibnamefont
  {Becke}},\ }\href {\doibase 10.1063/1.4869598} {\bibfield  {journal}
  {\bibinfo  {journal} {J. Chem. Phys.}\ }\textbf {\bibinfo {volume} {140}},\
  \bibinfo {pages} {18A301} (\bibinfo {year} {2014})}\BibitemShut {NoStop}%
\bibitem [{\citenamefont {Klime\v{s}}\ and\ \citenamefont
  {Michaelides}(2012)}]{klimes_2012}%
  \BibitemOpen
  \bibfield  {author} {\bibinfo {author} {\bibfnamefont {J.}~\bibnamefont
  {Klime\v{s}}}\ and\ \bibinfo {author} {\bibfnamefont {A.}~\bibnamefont
  {Michaelides}},\ }\href {\doibase 10.1063/1.4754130} {\bibfield  {journal}
  {\bibinfo  {journal} {J. Chem. Phys.}\ }\textbf {\bibinfo {volume} {137}},\
  \bibinfo {pages} {120901} (\bibinfo {year} {2012})}\BibitemShut {NoStop}%
\bibitem [{\citenamefont {Ambrosetti}\ \emph
  {et~al.}(2014{\natexlab{b}})\citenamefont {Ambrosetti}, \citenamefont {{D.
  Alf\'{e}}}, \citenamefont {{R. A. DiStasio Jr.}},\ and\ \citenamefont
  {Tkatchenko}}]{ambrosetti_jpcl_2014}%
  \BibitemOpen
  \bibfield  {author} {\bibinfo {author} {\bibfnamefont {A.}~\bibnamefont
  {Ambrosetti}}, \bibinfo {author} {\bibnamefont {{D. Alf\'{e}}}}, \bibinfo
  {author} {\bibnamefont {{R. A. DiStasio Jr.}}}, \ and\ \bibinfo {author}
  {\bibfnamefont {A.}~\bibnamefont {Tkatchenko}},\ }\href {\doibase
  10.1021/jz402663k} {\bibfield  {journal} {\bibinfo  {journal} {J. Phys. Chem.
  Lett.}\ }\textbf {\bibinfo {volume} {5}},\ \bibinfo {pages} {849} (\bibinfo
  {year} {2014}{\natexlab{b}})}\BibitemShut {NoStop}%
\bibitem [{\citenamefont {Tkatchenko}\ \emph {et~al.}(2011)\citenamefont
  {Tkatchenko}, \citenamefont {Rossi}, \citenamefont {Blum}, \citenamefont
  {Ireta},\ and\ \citenamefont {Scheffler}}]{tkatchenko_2011}%
  \BibitemOpen
  \bibfield  {author} {\bibinfo {author} {\bibfnamefont {A.}~\bibnamefont
  {Tkatchenko}}, \bibinfo {author} {\bibfnamefont {M.}~\bibnamefont {Rossi}},
  \bibinfo {author} {\bibfnamefont {V.}~\bibnamefont {Blum}}, \bibinfo {author}
  {\bibfnamefont {J.}~\bibnamefont {Ireta}}, \ and\ \bibinfo {author}
  {\bibfnamefont {M.}~\bibnamefont {Scheffler}},\ }\href {\doibase
  10.1103/physrevlett.106.118102} {\bibfield  {journal} {\bibinfo  {journal}
  {Phys. Rev. Lett.}\ }\textbf {\bibinfo {volume} {106}},\ \bibinfo {pages}
  {118102} (\bibinfo {year} {2011})}\BibitemShut {NoStop}%
\bibitem [{\citenamefont {{A. Otero de la Roza}}\ and\ \citenamefont
  {Johnson}(2012)}]{oterodelaroza_jcp_2012}%
  \BibitemOpen
  \bibfield  {author} {\bibinfo {author} {\bibnamefont {{A. Otero de la
  Roza}}}\ and\ \bibinfo {author} {\bibfnamefont {E.~R.}\ \bibnamefont
  {Johnson}},\ }\href {\doibase 10.1063/1.4738961} {\bibfield  {journal}
  {\bibinfo  {journal} {J. Chem. Phys.}\ }\textbf {\bibinfo {volume} {137}},\
  \bibinfo {pages} {054103} (\bibinfo {year} {2012})}\BibitemShut {NoStop}%
\bibitem [{\citenamefont {Reilly}\ and\ \citenamefont
  {Tkatchenko}(2013{\natexlab{a}})}]{reilly_2013}%
  \BibitemOpen
  \bibfield  {author} {\bibinfo {author} {\bibfnamefont {A.~M.}\ \bibnamefont
  {Reilly}}\ and\ \bibinfo {author} {\bibfnamefont {A.}~\bibnamefont
  {Tkatchenko}},\ }\href {\doibase 10.1063/1.4812819} {\bibfield  {journal}
  {\bibinfo  {journal} {J. Chem. Phys.}\ }\textbf {\bibinfo {volume} {139}},\
  \bibinfo {pages} {024705} (\bibinfo {year} {2013}{\natexlab{a}})}\BibitemShut
  {NoStop}%
\bibitem [{\citenamefont {Reilly}\ and\ \citenamefont
  {Tkatchenko}(2013{\natexlab{b}})}]{reilly_jpcl_2013}%
  \BibitemOpen
  \bibfield  {author} {\bibinfo {author} {\bibfnamefont {A.~M.}\ \bibnamefont
  {Reilly}}\ and\ \bibinfo {author} {\bibfnamefont {A.}~\bibnamefont
  {Tkatchenko}},\ }\href {\doibase 10.1021/jz400226x} {\bibfield  {journal}
  {\bibinfo  {journal} {J. Phys. Chem. Lett.}\ }\textbf {\bibinfo {volume}
  {4}},\ \bibinfo {pages} {1028} (\bibinfo {year}
  {2013}{\natexlab{b}})}\BibitemShut {NoStop}%
\bibitem [{\citenamefont {Marom}\ \emph {et~al.}(2013)\citenamefont {Marom},
  \citenamefont {{R. A. DiStasio Jr.}}, \citenamefont {Atalla}, \citenamefont
  {Levchenko}, \citenamefont {Reilly}, \citenamefont {Chelikowsky},
  \citenamefont {Leiserowitz},\ and\ \citenamefont {Tkatchenko}}]{marom_2013}%
  \BibitemOpen
  \bibfield  {author} {\bibinfo {author} {\bibfnamefont {N.}~\bibnamefont
  {Marom}}, \bibinfo {author} {\bibnamefont {{R. A. DiStasio Jr.}}}, \bibinfo
  {author} {\bibfnamefont {V.}~\bibnamefont {Atalla}}, \bibinfo {author}
  {\bibfnamefont {S.}~\bibnamefont {Levchenko}}, \bibinfo {author}
  {\bibfnamefont {A.~M.}\ \bibnamefont {Reilly}}, \bibinfo {author}
  {\bibfnamefont {J.~R.}\ \bibnamefont {Chelikowsky}}, \bibinfo {author}
  {\bibfnamefont {L.}~\bibnamefont {Leiserowitz}}, \ and\ \bibinfo {author}
  {\bibfnamefont {A.}~\bibnamefont {Tkatchenko}},\ }\href {\doibase
  10.1002/anie.201301938} {\bibfield  {journal} {\bibinfo  {journal} {Angew.
  Chem. Int. Ed.}\ }\textbf {\bibinfo {volume} {52}},\ \bibinfo {pages} {6629}
  (\bibinfo {year} {2013})}\BibitemShut {NoStop}%
\bibitem [{\citenamefont {Reilly}\ and\ \citenamefont
  {Tkatchenko}(2014)}]{reilly_2014}%
  \BibitemOpen
  \bibfield  {author} {\bibinfo {author} {\bibfnamefont {A.~M.}\ \bibnamefont
  {Reilly}}\ and\ \bibinfo {author} {\bibfnamefont {A.}~\bibnamefont
  {Tkatchenko}},\ }\href {\doibase 10.1103/physrevlett.113.055701} {\bibfield
  {journal} {\bibinfo  {journal} {Phys. Rev. Lett.}\ }\textbf {\bibinfo
  {volume} {113}},\ \bibinfo {pages} {055701} (\bibinfo {year}
  {2014})}\BibitemShut {NoStop}%
\bibitem [{\citenamefont {Kronik}\ and\ \citenamefont
  {Tkatchenko}(2014)}]{kronik_2014}%
  \BibitemOpen
  \bibfield  {author} {\bibinfo {author} {\bibfnamefont {L.}~\bibnamefont
  {Kronik}}\ and\ \bibinfo {author} {\bibfnamefont {A.}~\bibnamefont
  {Tkatchenko}},\ }\href {\doibase 10.1021/ar500144s} {\bibfield  {journal}
  {\bibinfo  {journal} {Acc. Chem. Res.}\ }\textbf {\bibinfo {volume} {47}},\
  \bibinfo {pages} {3208} (\bibinfo {year} {2014})}\BibitemShut {NoStop}%
\bibitem [{\citenamefont {Shtogun}\ and\ \citenamefont
  {Woods}(2010)}]{shtogun_many-body_2010}%
  \BibitemOpen
  \bibfield  {author} {\bibinfo {author} {\bibfnamefont {Y.~V.}\ \bibnamefont
  {Shtogun}}\ and\ \bibinfo {author} {\bibfnamefont {L.~M.}\ \bibnamefont
  {Woods}},\ }\href {\doibase 10.1021/jz100309m} {\bibfield  {journal}
  {\bibinfo  {journal} {J. Phys. Chem. Lett.}\ }\textbf {\bibinfo {volume}
  {1}},\ \bibinfo {pages} {1356} (\bibinfo {year} {2010})}\BibitemShut
  {NoStop}%
\bibitem [{\citenamefont {{Bj\"{o}rkman}}\ \emph {et~al.}(2012)\citenamefont
  {{Bj\"{o}rkman}}, \citenamefont {Gulans}, \citenamefont {Krasheninnikov},\
  and\ \citenamefont {Nieminen}}]{bjorkman_prl_2012}%
  \BibitemOpen
  \bibfield  {author} {\bibinfo {author} {\bibfnamefont {T.}~\bibnamefont
  {{Bj\"{o}rkman}}}, \bibinfo {author} {\bibfnamefont {A.}~\bibnamefont
  {Gulans}}, \bibinfo {author} {\bibfnamefont {A.}~\bibnamefont
  {Krasheninnikov}}, \ and\ \bibinfo {author} {\bibfnamefont {R.~M.}\
  \bibnamefont {Nieminen}},\ }\href {\doibase 10.1103/PhysRevLett.108.235502}
  {\bibfield  {journal} {\bibinfo  {journal} {Phys. Rev. Lett.}\ }\textbf
  {\bibinfo {volume} {108}},\ \bibinfo {pages} {235502} (\bibinfo {year}
  {2012})}\BibitemShut {NoStop}%
\bibitem [{\citenamefont {Axilrod}\ and\ \citenamefont
  {Teller}(1943)}]{axilrod_jcp_1943}%
  \BibitemOpen
  \bibfield  {author} {\bibinfo {author} {\bibfnamefont {B.~M.}\ \bibnamefont
  {Axilrod}}\ and\ \bibinfo {author} {\bibfnamefont {E.}~\bibnamefont
  {Teller}},\ }\href {\doibase 10.1063/1.1723844} {\bibfield  {journal}
  {\bibinfo  {journal} {J. Chem. Phys.}\ }\textbf {\bibinfo {volume} {11}},\
  \bibinfo {pages} {299} (\bibinfo {year} {1943})}\BibitemShut {NoStop}%
\bibitem [{\citenamefont {Muto}(1943)}]{muto_jpmsj_1943}%
  \BibitemOpen
  \bibfield  {author} {\bibinfo {author} {\bibfnamefont {Y.}~\bibnamefont
  {Muto}},\ }\href@noop {} {\bibfield  {journal} {\bibinfo  {journal} {J.
  Phys.-Math. Soc. Japan}\ }\textbf {\bibinfo {volume} {17}},\ \bibinfo {pages}
  {629} (\bibinfo {year} {1943})}\BibitemShut {NoStop}%
\bibitem [{\citenamefont {Dobson}\ and\ \citenamefont
  {Gould}(2012)}]{dobson_review_2012}%
  \BibitemOpen
  \bibfield  {author} {\bibinfo {author} {\bibfnamefont {J.~F.}\ \bibnamefont
  {Dobson}}\ and\ \bibinfo {author} {\bibfnamefont {T.}~\bibnamefont {Gould}},\
  }\href {\doibase 10.1088/0953-8984/24/7/073201} {\bibfield  {journal}
  {\bibinfo  {journal} {J. Phys. Condens. Matter}\ }\textbf {\bibinfo {volume}
  {24}},\ \bibinfo {pages} {073201} (\bibinfo {year} {2012})}\BibitemShut
  {NoStop}%
\bibitem [{\citenamefont {Reilly}\ and\ \citenamefont
  {Tkatchenko}(2015)}]{reilly_chemsci_2015}%
  \BibitemOpen
  \bibfield  {author} {\bibinfo {author} {\bibfnamefont {A.~M.}\ \bibnamefont
  {Reilly}}\ and\ \bibinfo {author} {\bibfnamefont {A.}~\bibnamefont
  {Tkatchenko}},\ }\href {\doibase 10.1039/C5SC00410A} {\bibfield  {journal}
  {\bibinfo  {journal} {Chem. Sci.}\ }\textbf {\bibinfo {volume} {6}},\
  \bibinfo {pages} {3289} (\bibinfo {year} {2015})}\BibitemShut {NoStop}%
\bibitem [{\citenamefont {Wagner}\ \emph {et~al.}(2014)\citenamefont {Wagner},
  \citenamefont {Fournier}, \citenamefont {Ruiz}, \citenamefont {Li},
  \citenamefont {{M\"{u}llen}}, \citenamefont {Rohlfing}, \citenamefont
  {Tkatchenko}, \citenamefont {Temirov},\ and\ \citenamefont
  {Tautz}}]{wagner_non-additivity_2014}%
  \BibitemOpen
  \bibfield  {author} {\bibinfo {author} {\bibfnamefont {C.}~\bibnamefont
  {Wagner}}, \bibinfo {author} {\bibfnamefont {N.}~\bibnamefont {Fournier}},
  \bibinfo {author} {\bibfnamefont {V.~G.}\ \bibnamefont {Ruiz}}, \bibinfo
  {author} {\bibfnamefont {C.}~\bibnamefont {Li}}, \bibinfo {author}
  {\bibfnamefont {K.}~\bibnamefont {{M\"{u}llen}}}, \bibinfo {author}
  {\bibfnamefont {M.}~\bibnamefont {Rohlfing}}, \bibinfo {author}
  {\bibfnamefont {A.}~\bibnamefont {Tkatchenko}}, \bibinfo {author}
  {\bibfnamefont {R.}~\bibnamefont {Temirov}}, \ and\ \bibinfo {author}
  {\bibfnamefont {F.~S.}\ \bibnamefont {Tautz}},\ }\href {\doibase
  10.1038/ncomms6568} {\bibfield  {journal} {\bibinfo  {journal} {Nature
  Commun.}\ }\textbf {\bibinfo {volume} {5}},\ \bibinfo {pages} {5568}
  (\bibinfo {year} {2014})}\BibitemShut {NoStop}%
\bibitem [{Note1()}]{Note1}%
  \BibitemOpen
  \bibinfo {note} {Electronic Supplementary Information (ESI) available: The
  ESI contains additional derivation details, a comprehensive symbol glossary,
  and Cartesian coordinates for all structures considered in this
  work.}\BibitemShut {Stop}%
\bibitem [{\citenamefont {Hirshfeld}(1977)}]{hirshfeld_1977}%
  \BibitemOpen
  \bibfield  {author} {\bibinfo {author} {\bibfnamefont {F.~L.}\ \bibnamefont
  {Hirshfeld}},\ }\href {\doibase 10.1007/BF00549096} {\bibfield  {journal}
  {\bibinfo  {journal} {Theor. Chim. Acta}\ }\textbf {\bibinfo {volume} {44}},\
  \bibinfo {pages} {129} (\bibinfo {year} {1977})}\BibitemShut {NoStop}%
\bibitem [{Note6()}]{Note6}%
  \BibitemOpen
  \bibinfo {note} {Although there are numerous schemes for partitioning the
  electron density, the Hirshfeld prescription (Ref.~\protect \citenum
  {hirshfeld_1977}) has been shown to result in atomic partitions that most
  closely resemble the densities of the corresponding free (isolated) atoms (by
  minimizing the Kullback-Leibler entropy deficiency of information theory) cf.
  Ref.~\protect \citenum {nalewajski_2000}.}\BibitemShut {Stop}%
\bibitem [{\citenamefont {Tang}\ and\ \citenamefont
  {Karplus}(1968)}]{tang_pr_1968}%
  \BibitemOpen
  \bibfield  {author} {\bibinfo {author} {\bibfnamefont {K.~T.}\ \bibnamefont
  {Tang}}\ and\ \bibinfo {author} {\bibfnamefont {M.}~\bibnamefont {Karplus}},\
  }\href {\doibase 10.1103/PhysRev.171.70} {\bibfield  {journal} {\bibinfo
  {journal} {Phys. Rev.}\ }\textbf {\bibinfo {volume} {171}},\ \bibinfo {pages}
  {70} (\bibinfo {year} {1968})}\BibitemShut {NoStop}%
\bibitem [{\citenamefont {Brinck}, \citenamefont {Murray},\ and\ \citenamefont
  {Politzer}(1993)}]{brinck_1993}%
  \BibitemOpen
  \bibfield  {author} {\bibinfo {author} {\bibfnamefont {T.}~\bibnamefont
  {Brinck}}, \bibinfo {author} {\bibfnamefont {J.~S.}\ \bibnamefont {Murray}},
  \ and\ \bibinfo {author} {\bibfnamefont {P.}~\bibnamefont {Politzer}},\
  }\href {\doibase 10.1063/1.465038} {\bibfield  {journal} {\bibinfo  {journal}
  {J. Chem. Phys.}\ }\textbf {\bibinfo {volume} {98}},\ \bibinfo {pages} {4305}
  (\bibinfo {year} {1993})}\BibitemShut {NoStop}%
\bibitem [{\citenamefont {Bultinck}\ \emph {et~al.}(2007)\citenamefont
  {Bultinck}, \citenamefont {Van~Alsenoy}, \citenamefont {Ayers},\ and\
  \citenamefont {Carb\'{o}-Dorca}}]{bultinck_2007}%
  \BibitemOpen
  \bibfield  {author} {\bibinfo {author} {\bibfnamefont {P.}~\bibnamefont
  {Bultinck}}, \bibinfo {author} {\bibfnamefont {C.}~\bibnamefont
  {Van~Alsenoy}}, \bibinfo {author} {\bibfnamefont {P.~W.}\ \bibnamefont
  {Ayers}}, \ and\ \bibinfo {author} {\bibfnamefont {R.}~\bibnamefont
  {Carb\'{o}-Dorca}},\ }\href {\doibase 10.1063/1.2715563} {\bibfield
  {journal} {\bibinfo  {journal} {J. Chem. Phys.}\ }\textbf {\bibinfo {volume}
  {126}},\ \bibinfo {pages} {144111} (\bibinfo {year} {2007})}\BibitemShut
  {NoStop}%
\bibitem [{\citenamefont {Ferri}\ \emph {et~al.}(2015)\citenamefont {Ferri},
  \citenamefont {{R. A. DiStasio Jr.}}, \citenamefont {Ambrosetti},
  \citenamefont {Car},\ and\ \citenamefont
  {Tkatchenko}}]{ferri_electronic_2015}%
  \BibitemOpen
  \bibfield  {author} {\bibinfo {author} {\bibfnamefont {N.}~\bibnamefont
  {Ferri}}, \bibinfo {author} {\bibnamefont {{R. A. DiStasio Jr.}}}, \bibinfo
  {author} {\bibfnamefont {A.}~\bibnamefont {Ambrosetti}}, \bibinfo {author}
  {\bibfnamefont {R.}~\bibnamefont {Car}}, \ and\ \bibinfo {author}
  {\bibfnamefont {A.}~\bibnamefont {Tkatchenko}},\ }\href {\doibase
  10.1103/PhysRevLett.114.176802} {\bibfield  {journal} {\bibinfo  {journal}
  {Phys. Rev. Lett.}\ }\textbf {\bibinfo {volume} {114}},\ \bibinfo {pages}
  {176802} (\bibinfo {year} {2015})}\BibitemShut {NoStop}%
\bibitem [{Note7()}]{Note7}%
  \BibitemOpen
  \bibinfo {note} {The dipole polarizability tensor $\protect \bm {\alpha }_a$
  for a given atom or QHO is formed by populating the diagonal elements
  ($\alpha _{xx},\alpha _{yy},\alpha _{zz}$) with the \protect \textit
  {isotropic} dipole polarizability in Eq.~\protect \textup {\hbox
  {\mathsurround \z@ \protect \normalfont (\ignorespaces \ref
  {TS_polarizability_lambda}\unskip \@@italiccorr )}}.}\BibitemShut {Stop}%
\bibitem [{\citenamefont {Thole}(1981)}]{thole_1981}%
  \BibitemOpen
  \bibfield  {author} {\bibinfo {author} {\bibfnamefont {B.}~\bibnamefont
  {Thole}},\ }\href {\doibase 10.1016/0301-0104(81)85176-2} {\bibfield
  {journal} {\bibinfo  {journal} {Chem. Phys.}\ }\textbf {\bibinfo {volume}
  {59}},\ \bibinfo {pages} {341} (\bibinfo {year} {1981})}\BibitemShut
  {NoStop}%
\bibitem [{Note8()}]{Note8}%
  \BibitemOpen
  \bibinfo {note} {At this point, it is very important to note a difference in
  the notation relative to \unhbox \voidb@x \hbox {Refs.~\protect \citenum
  {distasio_2014} \&~\protect \citenum {tkatchenko_jcp_2013}:} our matrix
  $\protect \overline {\protect \mathbf {A}}$ is equivalent to their $\protect
  \overline {\protect \mathbf {B}}$ or $B$, which was keeping with Thole's
  original notation for the relay matrix (Ref.~\protect \citenum
  {thole_1981}).}\BibitemShut {Stop}%
\bibitem [{\citenamefont {Casimir}\ and\ \citenamefont
  {Polder}(1948)}]{casimir_1948}%
  \BibitemOpen
  \bibfield  {author} {\bibinfo {author} {\bibfnamefont {H.~B.~G.}\
  \bibnamefont {Casimir}}\ and\ \bibinfo {author} {\bibfnamefont
  {D.}~\bibnamefont {Polder}},\ }\href {\doibase 10.1103/PhysRev.73.360}
  {\bibfield  {journal} {\bibinfo  {journal} {Phys. Rev.}\ }\textbf {\bibinfo
  {volume} {73}},\ \bibinfo {pages} {360} (\bibinfo {year} {1948})}\BibitemShut
  {NoStop}%
\bibitem [{\citenamefont {Donchev}(2006)}]{donchev_2006}%
  \BibitemOpen
  \bibfield  {author} {\bibinfo {author} {\bibfnamefont {A.~G.}\ \bibnamefont
  {Donchev}},\ }\href {\doibase 10.1063/1.2337283} {\bibfield  {journal}
  {\bibinfo  {journal} {J. Chem. Phys.}\ }\textbf {\bibinfo {volume} {125}},\
  \bibinfo {pages} {074713} (\bibinfo {year} {2006})}\BibitemShut {NoStop}%
\bibitem [{Note2()}]{Note2}%
  \BibitemOpen
  \bibinfo {note} {Since each QHO is assigned a unit charge ($e=1$), the dipole
  moment $\protect \bm {\mu }$ is thereby equivalent to the displacement vector
  $\protect \bm {\xi }$.}\BibitemShut {Stop}%
\bibitem [{\citenamefont {Giannozzi}\ \emph {et~al.}(2009)\citenamefont
  {Giannozzi}, \citenamefont {Baroni}, \citenamefont {Bonini}, \citenamefont
  {Calandra}, \citenamefont {Car}, \citenamefont {Cavazzoni}, \citenamefont
  {Ceresoli}, \citenamefont {Chiarotti}, \citenamefont {Cococcioni},
  \citenamefont {Dabo}, \citenamefont {Dal~Corso}, \citenamefont
  {de~Gironcoli}, \citenamefont {Fabris}, \citenamefont {Fratesi},
  \citenamefont {Gebauer}, \citenamefont {Gerstmann}, \citenamefont
  {Gougoussis}, \citenamefont {Kokalj}, \citenamefont {Lazzeri}, \citenamefont
  {Martin-Samos}, \citenamefont {Marzari}, \citenamefont {Mauri}, \citenamefont
  {Mazzarello}, \citenamefont {Paolini}, \citenamefont {Pasquarello},
  \citenamefont {Paulatto}, \citenamefont {Sbraccia}, \citenamefont {Scandolo},
  \citenamefont {Sclauzero}, \citenamefont {Seitsonen}, \citenamefont
  {Smogunov}, \citenamefont {Umari},\ and\ \citenamefont
  {Wentzcovitch}}]{giannozzi_2009}%
  \BibitemOpen
  \bibfield  {author} {\bibinfo {author} {\bibfnamefont {P.}~\bibnamefont
  {Giannozzi}}, \bibinfo {author} {\bibfnamefont {S.}~\bibnamefont {Baroni}},
  \bibinfo {author} {\bibfnamefont {N.}~\bibnamefont {Bonini}}, \bibinfo
  {author} {\bibfnamefont {M.}~\bibnamefont {Calandra}}, \bibinfo {author}
  {\bibfnamefont {R.}~\bibnamefont {Car}}, \bibinfo {author} {\bibfnamefont
  {C.}~\bibnamefont {Cavazzoni}}, \bibinfo {author} {\bibfnamefont
  {D.}~\bibnamefont {Ceresoli}}, \bibinfo {author} {\bibfnamefont {G.~L.}\
  \bibnamefont {Chiarotti}}, \bibinfo {author} {\bibfnamefont {M.}~\bibnamefont
  {Cococcioni}}, \bibinfo {author} {\bibfnamefont {I.}~\bibnamefont {Dabo}},
  \bibinfo {author} {\bibfnamefont {A.}~\bibnamefont {Dal~Corso}}, \bibinfo
  {author} {\bibfnamefont {S.}~\bibnamefont {de~Gironcoli}}, \bibinfo {author}
  {\bibfnamefont {S.}~\bibnamefont {Fabris}}, \bibinfo {author} {\bibfnamefont
  {G.}~\bibnamefont {Fratesi}}, \bibinfo {author} {\bibfnamefont
  {R.}~\bibnamefont {Gebauer}}, \bibinfo {author} {\bibfnamefont
  {U.}~\bibnamefont {Gerstmann}}, \bibinfo {author} {\bibfnamefont
  {C.}~\bibnamefont {Gougoussis}}, \bibinfo {author} {\bibfnamefont
  {A.}~\bibnamefont {Kokalj}}, \bibinfo {author} {\bibfnamefont
  {M.}~\bibnamefont {Lazzeri}}, \bibinfo {author} {\bibfnamefont
  {L.}~\bibnamefont {Martin-Samos}}, \bibinfo {author} {\bibfnamefont
  {N.}~\bibnamefont {Marzari}}, \bibinfo {author} {\bibfnamefont
  {F.}~\bibnamefont {Mauri}}, \bibinfo {author} {\bibfnamefont
  {R.}~\bibnamefont {Mazzarello}}, \bibinfo {author} {\bibfnamefont
  {S.}~\bibnamefont {Paolini}}, \bibinfo {author} {\bibfnamefont
  {A.}~\bibnamefont {Pasquarello}}, \bibinfo {author} {\bibfnamefont
  {L.}~\bibnamefont {Paulatto}}, \bibinfo {author} {\bibfnamefont
  {C.}~\bibnamefont {Sbraccia}}, \bibinfo {author} {\bibfnamefont
  {S.}~\bibnamefont {Scandolo}}, \bibinfo {author} {\bibfnamefont
  {G.}~\bibnamefont {Sclauzero}}, \bibinfo {author} {\bibfnamefont {A.~P.}\
  \bibnamefont {Seitsonen}}, \bibinfo {author} {\bibfnamefont {A.}~\bibnamefont
  {Smogunov}}, \bibinfo {author} {\bibfnamefont {P.}~\bibnamefont {Umari}}, \
  and\ \bibinfo {author} {\bibfnamefont {R.~M.}\ \bibnamefont {Wentzcovitch}},\
  }\href {\doibase 10.1088/0953-8984/21/39/395502} {\bibfield  {journal}
  {\bibinfo  {journal} {J. Phys.: Condens. Matter}\ }\textbf {\bibinfo {volume}
  {21}},\ \bibinfo {pages} {395502} (\bibinfo {year} {2009})}\BibitemShut
  {NoStop}%
\bibitem [{\citenamefont {Markovich}\ \emph {et~al.}()\citenamefont
  {Markovich}, \citenamefont {Blood-Forsythe}, \citenamefont {{R. A. DiStasio
  Jr.}},\ and\ \citenamefont {Aspuru-Guzik}}]{markovich_unpublished}%
  \BibitemOpen
  \bibfield  {author} {\bibinfo {author} {\bibfnamefont {T.}~\bibnamefont
  {Markovich}}, \bibinfo {author} {\bibfnamefont {M.~A.}\ \bibnamefont
  {Blood-Forsythe}}, \bibinfo {author} {\bibnamefont {{R. A. DiStasio Jr.}}}, \
  and\ \bibinfo {author} {\bibfnamefont {A.}~\bibnamefont {Aspuru-Guzik}},\
  }\href@noop {} {\enquote {\bibinfo {title} {Enabling large-scale simulation
  of many-body dispersion forces in condensed phase systems},}\ }\bibinfo
  {note} {Unpublished}\BibitemShut {NoStop}%
\bibitem [{\citenamefont {Helgaker}, \citenamefont {{J{\o}rgensen}},\ and\
  \citenamefont {Olsen}(2000)}]{helgaker_2000}%
  \BibitemOpen
  \bibfield  {author} {\bibinfo {author} {\bibfnamefont {T.}~\bibnamefont
  {Helgaker}}, \bibinfo {author} {\bibfnamefont {P.}~\bibnamefont
  {{J{\o}rgensen}}}, \ and\ \bibinfo {author} {\bibfnamefont {J.}~\bibnamefont
  {Olsen}},\ }\href@noop {} {\emph {\bibinfo {title} {Molecular
  electronic-structure theory}}}\ (\bibinfo  {publisher} {Wiley},\ \bibinfo
  {address} {Chichester; New York},\ \bibinfo {year} {2000})\BibitemShut
  {NoStop}%
\bibitem [{\citenamefont {Bondi}(1964)}]{bondi_1964}%
  \BibitemOpen
  \bibfield  {author} {\bibinfo {author} {\bibfnamefont {A.}~\bibnamefont
  {Bondi}},\ }\href {\doibase 10.1021/j100785a001} {\bibfield  {journal}
  {\bibinfo  {journal} {J. Phys. Chem.}\ }\textbf {\bibinfo {volume} {68}},\
  \bibinfo {pages} {441} (\bibinfo {year} {1964})}\BibitemShut {NoStop}%
\bibitem [{\citenamefont {Nelson}(1976)}]{nelson_1976}%
  \BibitemOpen
  \bibfield  {author} {\bibinfo {author} {\bibfnamefont {R.~B.}\ \bibnamefont
  {Nelson}},\ }\href {\doibase 10.2514/3.7211} {\bibfield  {journal} {\bibinfo
  {journal} {{AIAA} Journal}\ }\textbf {\bibinfo {volume} {14}},\ \bibinfo
  {pages} {1201} (\bibinfo {year} {1976})}\BibitemShut {NoStop}%
\bibitem [{\citenamefont {Perdew}, \citenamefont {Burke},\ and\ \citenamefont
  {Ernzerhof}(1996)}]{PBE_ref1}%
  \BibitemOpen
  \bibfield  {author} {\bibinfo {author} {\bibfnamefont {J.~P.}\ \bibnamefont
  {Perdew}}, \bibinfo {author} {\bibfnamefont {K.}~\bibnamefont {Burke}}, \
  and\ \bibinfo {author} {\bibfnamefont {M.}~\bibnamefont {Ernzerhof}},\ }\href
  {\doibase 10.1103/PhysRevLett.77.3865} {\bibfield  {journal} {\bibinfo
  {journal} {Phys. Rev. Lett.}\ }\textbf {\bibinfo {volume} {77}},\ \bibinfo
  {pages} {3865} (\bibinfo {year} {1996})}\BibitemShut {NoStop}%
\bibitem [{\citenamefont {Perdew}, \citenamefont {Burke},\ and\ \citenamefont
  {Ernzerhof}(1997)}]{PBE_ref2}%
  \BibitemOpen
  \bibfield  {author} {\bibinfo {author} {\bibfnamefont {J.~P.}\ \bibnamefont
  {Perdew}}, \bibinfo {author} {\bibfnamefont {K.}~\bibnamefont {Burke}}, \
  and\ \bibinfo {author} {\bibfnamefont {M.}~\bibnamefont {Ernzerhof}},\ }\href
  {\doibase 10.1103/PhysRevLett.78.1396} {\bibfield  {journal} {\bibinfo
  {journal} {Phys. Rev. Lett.}\ }\textbf {\bibinfo {volume} {78}},\ \bibinfo
  {pages} {1396} (\bibinfo {year} {1997})}\BibitemShut {NoStop}%
\bibitem [{\citenamefont {Hamann}, \citenamefont {{Schl\"{u}ter}},\ and\
  \citenamefont {Chiang}(1979)}]{hamann_prl_1979}%
  \BibitemOpen
  \bibfield  {author} {\bibinfo {author} {\bibfnamefont {D.~R.}\ \bibnamefont
  {Hamann}}, \bibinfo {author} {\bibfnamefont {M.}~\bibnamefont
  {{Schl\"{u}ter}}}, \ and\ \bibinfo {author} {\bibfnamefont {C.}~\bibnamefont
  {Chiang}},\ }\href {\doibase 10.1103/PhysRevLett.43.1494} {\bibfield
  {journal} {\bibinfo  {journal} {Phys. Rev. Lett.}\ }\textbf {\bibinfo
  {volume} {43}},\ \bibinfo {pages} {1494} (\bibinfo {year}
  {1979})}\BibitemShut {NoStop}%
\bibitem [{\citenamefont {Bachelet}, \citenamefont {Hamann},\ and\
  \citenamefont {{Schl\"{u}ter}}(1982)}]{bachelet_prb_1982}%
  \BibitemOpen
  \bibfield  {author} {\bibinfo {author} {\bibfnamefont {G.~B.}\ \bibnamefont
  {Bachelet}}, \bibinfo {author} {\bibfnamefont {D.~R.}\ \bibnamefont
  {Hamann}}, \ and\ \bibinfo {author} {\bibfnamefont {M.}~\bibnamefont
  {{Schl\"{u}ter}}},\ }\href {\doibase 10.1103/PhysRevB.26.4199} {\bibfield
  {journal} {\bibinfo  {journal} {Phys. Rev. B}\ }\textbf {\bibinfo {volume}
  {26}},\ \bibinfo {pages} {4199} (\bibinfo {year} {1982})}\BibitemShut
  {NoStop}%
\bibitem [{\citenamefont {Vanderbilt}(1985)}]{vanderbilt_1985}%
  \BibitemOpen
  \bibfield  {author} {\bibinfo {author} {\bibfnamefont {D.}~\bibnamefont
  {Vanderbilt}},\ }\href {\doibase 10.1103/PhysRevB.32.8412} {\bibfield
  {journal} {\bibinfo  {journal} {Phys. Rev. B}\ }\textbf {\bibinfo {volume}
  {32}},\ \bibinfo {pages} {8412} (\bibinfo {year} {1985})}\BibitemShut
  {NoStop}%
\bibitem [{\citenamefont {Neese}(2012)}]{neese_orca_2012}%
  \BibitemOpen
  \bibfield  {author} {\bibinfo {author} {\bibfnamefont {F.}~\bibnamefont
  {Neese}},\ }\href {\doibase 10.1002/wcms.81} {\bibfield  {journal} {\bibinfo
  {journal} {WIREs Comput. Mol. Sci.}\ }\textbf {\bibinfo {volume} {2}},\
  \bibinfo {pages} {73} (\bibinfo {year} {2012})}\BibitemShut {NoStop}%
\bibitem [{\citenamefont {Blum}\ \emph {et~al.}(2009)\citenamefont {Blum},
  \citenamefont {Gehrke}, \citenamefont {Hanke}, \citenamefont {Havu},
  \citenamefont {Havu}, \citenamefont {Ren}, \citenamefont {Reuter},\ and\
  \citenamefont {Scheffler}}]{blum_2009}%
  \BibitemOpen
  \bibfield  {author} {\bibinfo {author} {\bibfnamefont {V.}~\bibnamefont
  {Blum}}, \bibinfo {author} {\bibfnamefont {R.}~\bibnamefont {Gehrke}},
  \bibinfo {author} {\bibfnamefont {F.}~\bibnamefont {Hanke}}, \bibinfo
  {author} {\bibfnamefont {P.}~\bibnamefont {Havu}}, \bibinfo {author}
  {\bibfnamefont {V.}~\bibnamefont {Havu}}, \bibinfo {author} {\bibfnamefont
  {X.}~\bibnamefont {Ren}}, \bibinfo {author} {\bibfnamefont {K.}~\bibnamefont
  {Reuter}}, \ and\ \bibinfo {author} {\bibfnamefont {M.}~\bibnamefont
  {Scheffler}},\ }\href {\doibase 10.1016/j.cpc.2009.06.022} {\bibfield
  {journal} {\bibinfo  {journal} {Comput. Phys. Commun.}\ }\textbf {\bibinfo
  {volume} {180}},\ \bibinfo {pages} {2175} (\bibinfo {year}
  {2009})}\BibitemShut {NoStop}%
\bibitem [{\citenamefont {Tkatchenko}(2014)}]{MBD_rsSCS_FHI-aims}%
  \BibitemOpen
  \bibfield  {author} {\bibinfo {author} {\bibfnamefont {A.}~\bibnamefont
  {Tkatchenko}},\ }\href {{http://th.fhi-berlin.mpg.de/~tkatchen/MBD}}
  {\enquote {\bibinfo {title} {{MBD@rsSCS}},}\ } (\bibinfo {year} {Accessed
  Oct. 1, 2014})\BibitemShut {NoStop}%
\bibitem [{\citenamefont {Hobza}, \citenamefont {Selzle},\ and\ \citenamefont
  {Schlag}(1990)}]{hobza_benzene_dimer_1990}%
  \BibitemOpen
  \bibfield  {author} {\bibinfo {author} {\bibfnamefont {P.}~\bibnamefont
  {Hobza}}, \bibinfo {author} {\bibfnamefont {H.~L.}\ \bibnamefont {Selzle}}, \
  and\ \bibinfo {author} {\bibfnamefont {E.~W.}\ \bibnamefont {Schlag}},\
  }\href {\doibase 10.1063/1.459587} {\bibfield  {journal} {\bibinfo  {journal}
  {J. Chem. Phys.}\ }\textbf {\bibinfo {volume} {93}},\ \bibinfo {pages} {5893}
  (\bibinfo {year} {1990})}\BibitemShut {NoStop}%
\bibitem [{\citenamefont {Arunan}\ and\ \citenamefont
  {Gutowsky}(1993)}]{arunan_benzene_dimer_1993}%
  \BibitemOpen
  \bibfield  {author} {\bibinfo {author} {\bibfnamefont {E.}~\bibnamefont
  {Arunan}}\ and\ \bibinfo {author} {\bibfnamefont {H.~S.}\ \bibnamefont
  {Gutowsky}},\ }\href {\doibase 10.1063/1.465035} {\bibfield  {journal}
  {\bibinfo  {journal} {J. Chem. Phys.}\ }\textbf {\bibinfo {volume} {98}},\
  \bibinfo {pages} {4294} (\bibinfo {year} {1993})}\BibitemShut {NoStop}%
\bibitem [{\citenamefont {Hobza}, \citenamefont {Selzle},\ and\ \citenamefont
  {Schlag}(1993)}]{hobza_benzene_dimer_1993}%
  \BibitemOpen
  \bibfield  {author} {\bibinfo {author} {\bibfnamefont {P.}~\bibnamefont
  {Hobza}}, \bibinfo {author} {\bibfnamefont {H.~L.}\ \bibnamefont {Selzle}}, \
  and\ \bibinfo {author} {\bibfnamefont {E.~W.}\ \bibnamefont {Schlag}},\
  }\href {\doibase 10.1021/j100118a002} {\bibfield  {journal} {\bibinfo
  {journal} {J. Phys. Chem.}\ }\textbf {\bibinfo {volume} {97}},\ \bibinfo
  {pages} {3937} (\bibinfo {year} {1993})}\BibitemShut {NoStop}%
\bibitem [{\citenamefont {Hobza}, \citenamefont {Selzle},\ and\ \citenamefont
  {Schlag}(1994)}]{hobza_benzene_dimer_1994}%
  \BibitemOpen
  \bibfield  {author} {\bibinfo {author} {\bibfnamefont {P.}~\bibnamefont
  {Hobza}}, \bibinfo {author} {\bibfnamefont {H.~L.}\ \bibnamefont {Selzle}}, \
  and\ \bibinfo {author} {\bibfnamefont {E.~W.}\ \bibnamefont {Schlag}},\
  }\href {\doibase 10.1021/ja00087a041} {\bibfield  {journal} {\bibinfo
  {journal} {J. Am. Chem. Soc.}\ }\textbf {\bibinfo {volume} {116}},\ \bibinfo
  {pages} {3500} (\bibinfo {year} {1994})}\BibitemShut {NoStop}%
\bibitem [{\citenamefont {Jaffe}\ and\ \citenamefont
  {Smith}(1996)}]{jaffe_benzene_dimer_1996}%
  \BibitemOpen
  \bibfield  {author} {\bibinfo {author} {\bibfnamefont {R.~L.}\ \bibnamefont
  {Jaffe}}\ and\ \bibinfo {author} {\bibfnamefont {G.~D.}\ \bibnamefont
  {Smith}},\ }\href {\doibase 10.1063/1.472140} {\bibfield  {journal} {\bibinfo
   {journal} {J. Chem. Phys.}\ }\textbf {\bibinfo {volume} {105}},\ \bibinfo
  {pages} {2780} (\bibinfo {year} {1996})}\BibitemShut {NoStop}%
\bibitem [{\citenamefont {Hobza}, \citenamefont {Selzle},\ and\ \citenamefont
  {Schlag}(1996)}]{hobza_benzene_dimer_1996}%
  \BibitemOpen
  \bibfield  {author} {\bibinfo {author} {\bibfnamefont {P.}~\bibnamefont
  {Hobza}}, \bibinfo {author} {\bibfnamefont {H.~L.}\ \bibnamefont {Selzle}}, \
  and\ \bibinfo {author} {\bibfnamefont {E.~W.}\ \bibnamefont {Schlag}},\
  }\href {\doibase 10.1021/jp961239y} {\bibfield  {journal} {\bibinfo
  {journal} {J. Phys. Chem.}\ }\textbf {\bibinfo {volume} {100}},\ \bibinfo
  {pages} {18790} (\bibinfo {year} {1996})}\BibitemShut {NoStop}%
\bibitem [{\citenamefont {Gauss}\ and\ \citenamefont
  {Stanton}(2000)}]{gauss_stanton_2000}%
  \BibitemOpen
  \bibfield  {author} {\bibinfo {author} {\bibfnamefont {J.}~\bibnamefont
  {Gauss}}\ and\ \bibinfo {author} {\bibfnamefont {J.~F.}\ \bibnamefont
  {Stanton}},\ }\href {\doibase 10.1021/jp994408y} {\bibfield  {journal}
  {\bibinfo  {journal} {J. Phys. Chem. A}\ }\textbf {\bibinfo {volume} {104}},\
  \bibinfo {pages} {2865} (\bibinfo {year} {2000})}\BibitemShut {NoStop}%
\bibitem [{\citenamefont {Tsuzuki}\ \emph {et~al.}(2002)\citenamefont
  {Tsuzuki}, \citenamefont {Honda}, \citenamefont {Uchimaru}, \citenamefont
  {Mikami},\ and\ \citenamefont {Tanabe}}]{tsuzuki_benzene_dimer_2002}%
  \BibitemOpen
  \bibfield  {author} {\bibinfo {author} {\bibfnamefont {S.}~\bibnamefont
  {Tsuzuki}}, \bibinfo {author} {\bibfnamefont {K.}~\bibnamefont {Honda}},
  \bibinfo {author} {\bibfnamefont {T.}~\bibnamefont {Uchimaru}}, \bibinfo
  {author} {\bibfnamefont {M.}~\bibnamefont {Mikami}}, \ and\ \bibinfo {author}
  {\bibfnamefont {K.}~\bibnamefont {Tanabe}},\ }\href {\doibase
  10.1021/ja0105212} {\bibfield  {journal} {\bibinfo  {journal} {J. Am. Chem.
  Soc.}\ }\textbf {\bibinfo {volume} {124}},\ \bibinfo {pages} {104} (\bibinfo
  {year} {2002})}\BibitemShut {NoStop}%
\bibitem [{\citenamefont {Sinnokrot}, \citenamefont {Valeev},\ and\
  \citenamefont {Sherrill}(2002)}]{sinnokrot_benzene_dimer_2002}%
  \BibitemOpen
  \bibfield  {author} {\bibinfo {author} {\bibfnamefont {M.~O.}\ \bibnamefont
  {Sinnokrot}}, \bibinfo {author} {\bibfnamefont {E.~F.}\ \bibnamefont
  {Valeev}}, \ and\ \bibinfo {author} {\bibfnamefont {C.~D.}\ \bibnamefont
  {Sherrill}},\ }\href {\doibase 10.1021/ja025896h} {\bibfield  {journal}
  {\bibinfo  {journal} {J. Am. Chem. Soc.}\ }\textbf {\bibinfo {volume}
  {124}},\ \bibinfo {pages} {10887} (\bibinfo {year} {2002})}\BibitemShut
  {NoStop}%
\bibitem [{\citenamefont {Sinnokrot}\ and\ \citenamefont
  {Sherrill}(2004)}]{sinnokrot_benzene_dimer_2004}%
  \BibitemOpen
  \bibfield  {author} {\bibinfo {author} {\bibfnamefont {M.~O.}\ \bibnamefont
  {Sinnokrot}}\ and\ \bibinfo {author} {\bibfnamefont {C.~D.}\ \bibnamefont
  {Sherrill}},\ }\href {\doibase 10.1021/jp0469517} {\bibfield  {journal}
  {\bibinfo  {journal} {J. Phys. Chem. A}\ }\textbf {\bibinfo {volume} {108}},\
  \bibinfo {pages} {10200} (\bibinfo {year} {2004})}\BibitemShut {NoStop}%
\bibitem [{\citenamefont {Tauer}\ and\ \citenamefont
  {Sherrill}(2005)}]{tauer_additivity_2005}%
  \BibitemOpen
  \bibfield  {author} {\bibinfo {author} {\bibfnamefont {T.~P.}\ \bibnamefont
  {Tauer}}\ and\ \bibinfo {author} {\bibfnamefont {C.~D.}\ \bibnamefont
  {Sherrill}},\ }\href {\doibase 10.1021/jp0553479} {\bibfield  {journal}
  {\bibinfo  {journal} {J. Phys. Chem. A}\ }\textbf {\bibinfo {volume} {109}},\
  \bibinfo {pages} {10475} (\bibinfo {year} {2005})}\BibitemShut {NoStop}%
\bibitem [{\citenamefont {Podeszwa}, \citenamefont {Bukowski},\ and\
  \citenamefont {Szalewicz}(2006)}]{podeszwa_2006}%
  \BibitemOpen
  \bibfield  {author} {\bibinfo {author} {\bibfnamefont {R.}~\bibnamefont
  {Podeszwa}}, \bibinfo {author} {\bibfnamefont {R.}~\bibnamefont {Bukowski}},
  \ and\ \bibinfo {author} {\bibfnamefont {K.}~\bibnamefont {Szalewicz}},\
  }\href {\doibase 10.1021/jp064095o} {\bibfield  {journal} {\bibinfo
  {journal} {J. Phys. Chem. A}\ }\textbf {\bibinfo {volume} {110}},\ \bibinfo
  {pages} {10345} (\bibinfo {year} {2006})}\BibitemShut {NoStop}%
\bibitem [{\citenamefont {Grant~Hill}, \citenamefont {Platts},\ and\
  \citenamefont {Werner}(2006)}]{grant_benzene_dimer_2006}%
  \BibitemOpen
  \bibfield  {author} {\bibinfo {author} {\bibfnamefont {J.}~\bibnamefont
  {Grant~Hill}}, \bibinfo {author} {\bibfnamefont {J.~A.}\ \bibnamefont
  {Platts}}, \ and\ \bibinfo {author} {\bibfnamefont {H.-J.}\ \bibnamefont
  {Werner}},\ }\href {\doibase 10.1039/b608623c} {\bibfield  {journal}
  {\bibinfo  {journal} {Phys. Chem. Chem. Phys.}\ }\textbf {\bibinfo {volume}
  {8}},\ \bibinfo {pages} {4072} (\bibinfo {year} {2006})}\BibitemShut
  {NoStop}%
\bibitem [{\citenamefont {{R.A. DiStasio, Jr.}}\ \emph
  {et~al.}(2007)\citenamefont {{R.A. DiStasio, Jr.}}, \citenamefont {{von
  Helden}}, \citenamefont {Steele},\ and\ \citenamefont
  {Head-Gordon}}]{distasio_benzene_dimer_2007}%
  \BibitemOpen
  \bibfield  {author} {\bibinfo {author} {\bibnamefont {{R.A. DiStasio, Jr.}}},
  \bibinfo {author} {\bibfnamefont {G.}~\bibnamefont {{von Helden}}}, \bibinfo
  {author} {\bibfnamefont {R.~P.}\ \bibnamefont {Steele}}, \ and\ \bibinfo
  {author} {\bibfnamefont {M.}~\bibnamefont {Head-Gordon}},\ }\href {\doibase
  10.1016/j.cplett.2007.02.034} {\bibfield  {journal} {\bibinfo  {journal}
  {Chem. Phys. Lett.}\ }\textbf {\bibinfo {volume} {437}},\ \bibinfo {pages}
  {277} (\bibinfo {year} {2007})}\BibitemShut {NoStop}%
\bibitem [{\citenamefont {Janowski}\ and\ \citenamefont
  {Pulay}(2007)}]{janowski_benzene_dimer_2007}%
  \BibitemOpen
  \bibfield  {author} {\bibinfo {author} {\bibfnamefont {T.}~\bibnamefont
  {Janowski}}\ and\ \bibinfo {author} {\bibfnamefont {P.}~\bibnamefont
  {Pulay}},\ }\href {\doibase 10.1016/j.cplett.2007.09.003} {\bibfield
  {journal} {\bibinfo  {journal} {Chem. Phys. Lett.}\ }\textbf {\bibinfo
  {volume} {447}},\ \bibinfo {pages} {27} (\bibinfo {year} {2007})}\BibitemShut
  {NoStop}%
\bibitem [{\citenamefont {Lee}\ \emph {et~al.}(2007)\citenamefont {Lee},
  \citenamefont {Kim}, \citenamefont {{Jure\v{c}ka}}, \citenamefont
  {Tarakeshwar}, \citenamefont {Hobza},\ and\ \citenamefont
  {Kim}}]{lee_benzene_dimer_2007}%
  \BibitemOpen
  \bibfield  {author} {\bibinfo {author} {\bibfnamefont {E.~C.}\ \bibnamefont
  {Lee}}, \bibinfo {author} {\bibfnamefont {D.}~\bibnamefont {Kim}}, \bibinfo
  {author} {\bibfnamefont {P.}~\bibnamefont {{Jure\v{c}ka}}}, \bibinfo {author}
  {\bibfnamefont {P.}~\bibnamefont {Tarakeshwar}}, \bibinfo {author}
  {\bibfnamefont {P.}~\bibnamefont {Hobza}}, \ and\ \bibinfo {author}
  {\bibfnamefont {K.~S.}\ \bibnamefont {Kim}},\ }\href {\doibase
  10.1021/jp068635t} {\bibfield  {journal} {\bibinfo  {journal} {J. Phys. Chem.
  A}\ }\textbf {\bibinfo {volume} {111}},\ \bibinfo {pages} {3446} (\bibinfo
  {year} {2007})}\BibitemShut {NoStop}%
\bibitem [{\citenamefont {{Fern\'{a}ndez}}\ \emph {et~al.}(2007)\citenamefont
  {{Fern\'{a}ndez}}, \citenamefont {Pedersen}, \citenamefont {{S\'{a}nchez de
  Mer\'{a}s}},\ and\ \citenamefont {Koch}}]{fernandez_coupled_2007}%
  \BibitemOpen
  \bibfield  {author} {\bibinfo {author} {\bibfnamefont {B.}~\bibnamefont
  {{Fern\'{a}ndez}}}, \bibinfo {author} {\bibfnamefont {T.~B.}\ \bibnamefont
  {Pedersen}}, \bibinfo {author} {\bibfnamefont {A.}~\bibnamefont {{S\'{a}nchez
  de Mer\'{a}s}}}, \ and\ \bibinfo {author} {\bibfnamefont {H.}~\bibnamefont
  {Koch}},\ }\href {\doibase 10.1016/j.cplett.2007.05.017} {\bibfield
  {journal} {\bibinfo  {journal} {Chem. Phys. Lett.}\ }\textbf {\bibinfo
  {volume} {441}},\ \bibinfo {pages} {332} (\bibinfo {year}
  {2007})}\BibitemShut {NoStop}%
\bibitem [{\citenamefont {Bludsk\'{y}}\ \emph {et~al.}(2008)\citenamefont
  {Bludsk\'{y}}, \citenamefont {Rube\v{s}}, \citenamefont {Sold\'{a}n},\ and\
  \citenamefont {Nachtigall}}]{bludsky_2008}%
  \BibitemOpen
  \bibfield  {author} {\bibinfo {author} {\bibfnamefont {O.}~\bibnamefont
  {Bludsk\'{y}}}, \bibinfo {author} {\bibfnamefont {M.}~\bibnamefont
  {Rube\v{s}}}, \bibinfo {author} {\bibfnamefont {P.}~\bibnamefont
  {Sold\'{a}n}}, \ and\ \bibinfo {author} {\bibfnamefont {P.}~\bibnamefont
  {Nachtigall}},\ }\href {\doibase 10.1063/1.2890968} {\bibfield  {journal}
  {\bibinfo  {journal} {J. Chem. Phys.}\ }\textbf {\bibinfo {volume} {128}},\
  \bibinfo {pages} {114102} (\bibinfo {year} {2008})}\BibitemShut {NoStop}%
\bibitem [{\citenamefont {Pavone}, \citenamefont {Rega},\ and\ \citenamefont
  {Barone}(2008)}]{pavone_benzene_dimer_2008}%
  \BibitemOpen
  \bibfield  {author} {\bibinfo {author} {\bibfnamefont {M.}~\bibnamefont
  {Pavone}}, \bibinfo {author} {\bibfnamefont {N.}~\bibnamefont {Rega}}, \ and\
  \bibinfo {author} {\bibfnamefont {V.}~\bibnamefont {Barone}},\ }\href
  {\doibase 10.1016/j.cplett.2007.12.075} {\bibfield  {journal} {\bibinfo
  {journal} {Chem. Phys. Lett.}\ }\textbf {\bibinfo {volume} {452}},\ \bibinfo
  {pages} {333} (\bibinfo {year} {2008})}\BibitemShut {NoStop}%
\bibitem [{\citenamefont {{Pito\v{n}ak}}\ \emph {et~al.}(2008)\citenamefont
  {{Pito\v{n}ak}}, \citenamefont {{Neogr\'{a}dy}}, \citenamefont
  {{\v{R}ez\'{a}c}}, \citenamefont {{Jure\v{c}ka}}, \citenamefont {Urban},\
  and\ \citenamefont {Hobza}}]{pitonak_benzene_dimer_2008}%
  \BibitemOpen
  \bibfield  {author} {\bibinfo {author} {\bibfnamefont {M.}~\bibnamefont
  {{Pito\v{n}ak}}}, \bibinfo {author} {\bibfnamefont {P.}~\bibnamefont
  {{Neogr\'{a}dy}}}, \bibinfo {author} {\bibfnamefont {J.}~\bibnamefont
  {{\v{R}ez\'{a}c}}}, \bibinfo {author} {\bibfnamefont {P.}~\bibnamefont
  {{Jure\v{c}ka}}}, \bibinfo {author} {\bibfnamefont {M.}~\bibnamefont
  {Urban}}, \ and\ \bibinfo {author} {\bibfnamefont {P.}~\bibnamefont
  {Hobza}},\ }\href {\doibase 10.1021/ct800229h} {\bibfield  {journal}
  {\bibinfo  {journal} {J. Chem. Theory Comput.}\ }\textbf {\bibinfo {volume}
  {4}},\ \bibinfo {pages} {1829} (\bibinfo {year} {2008})}\BibitemShut
  {NoStop}%
\bibitem [{\citenamefont {Sherrill}, \citenamefont {Takatani},\ and\
  \citenamefont {Hohenstein}(2009)}]{sherrill_benzene_dimer_2009}%
  \BibitemOpen
  \bibfield  {author} {\bibinfo {author} {\bibfnamefont {C.~D.}\ \bibnamefont
  {Sherrill}}, \bibinfo {author} {\bibfnamefont {T.}~\bibnamefont {Takatani}},
  \ and\ \bibinfo {author} {\bibfnamefont {E.~G.}\ \bibnamefont {Hohenstein}},\
  }\href {\doibase 10.1021/jp9034375} {\bibfield  {journal} {\bibinfo
  {journal} {J. Phys. Chem. A}\ }\textbf {\bibinfo {volume} {113}},\ \bibinfo
  {pages} {10146} (\bibinfo {year} {2009})}\BibitemShut {NoStop}%
\bibitem [{\citenamefont {{Gr\"{a}fenstein}}\ and\ \citenamefont
  {Cremer}(2009)}]{grafenstein_benzene_dimer_2009}%
  \BibitemOpen
  \bibfield  {author} {\bibinfo {author} {\bibfnamefont {J.}~\bibnamefont
  {{Gr\"{a}fenstein}}}\ and\ \bibinfo {author} {\bibfnamefont {D.}~\bibnamefont
  {Cremer}},\ }\href {\doibase 10.1063/1.3079822} {\bibfield  {journal}
  {\bibinfo  {journal} {J. Chem. Phys.}\ }\textbf {\bibinfo {volume} {130}},\
  \bibinfo {pages} {124105} (\bibinfo {year} {2009})}\BibitemShut {NoStop}%
\bibitem [{\citenamefont {Anatole~von Lilienfeld}\ and\ \citenamefont
  {Tkatchenko}(2010)}]{anatole_2_and_3_2010}%
  \BibitemOpen
  \bibfield  {author} {\bibinfo {author} {\bibfnamefont {O.}~\bibnamefont
  {Anatole~von Lilienfeld}}\ and\ \bibinfo {author} {\bibfnamefont
  {A.}~\bibnamefont {Tkatchenko}},\ }\href {\doibase 10.1063/1.3432765}
  {\bibfield  {journal} {\bibinfo  {journal} {J. Chem. Phys.}\ }\textbf
  {\bibinfo {volume} {132}},\ \bibinfo {pages} {234109} (\bibinfo {year}
  {2010})}\BibitemShut {NoStop}%
\bibitem [{\citenamefont {Valdes}\ \emph {et~al.}(2008)\citenamefont {Valdes},
  \citenamefont {{Pluh\'{a}\v{c}kov\'{a}}}, \citenamefont {{Pitonak}},
  \citenamefont {{\v{R}ez\'{a}\v{c}}},\ and\ \citenamefont
  {Hobza}}]{valdes_benchmark_2008}%
  \BibitemOpen
  \bibfield  {author} {\bibinfo {author} {\bibfnamefont {H.}~\bibnamefont
  {Valdes}}, \bibinfo {author} {\bibfnamefont {K.}~\bibnamefont
  {{Pluh\'{a}\v{c}kov\'{a}}}}, \bibinfo {author} {\bibfnamefont
  {M.}~\bibnamefont {{Pitonak}}}, \bibinfo {author} {\bibfnamefont
  {J.}~\bibnamefont {{\v{R}ez\'{a}\v{c}}}}, \ and\ \bibinfo {author}
  {\bibfnamefont {P.}~\bibnamefont {Hobza}},\ }\href {\doibase
  10.1039/b719294k} {\bibfield  {journal} {\bibinfo  {journal} {Phys. Chem.
  Chem. Phys.}\ }\textbf {\bibinfo {volume} {10}},\ \bibinfo {pages} {2747}
  (\bibinfo {year} {2008})}\BibitemShut {NoStop}%
\bibitem [{\citenamefont {{M\o ller}}\ and\ \citenamefont
  {Plesset}(1934)}]{moller_plesset_1934}%
  \BibitemOpen
  \bibfield  {author} {\bibinfo {author} {\bibfnamefont {C.}~\bibnamefont {{M\o
  ller}}}\ and\ \bibinfo {author} {\bibfnamefont {M.~S.}\ \bibnamefont
  {Plesset}},\ }\href {\doibase 10.1103/PhysRev.46.618} {\bibfield  {journal}
  {\bibinfo  {journal} {Phys. Rev.}\ }\textbf {\bibinfo {volume} {46}},\
  \bibinfo {pages} {618} (\bibinfo {year} {1934})}\BibitemShut {NoStop}%
\bibitem [{\citenamefont {Baerends}, \citenamefont {Ellis},\ and\ \citenamefont
  {Ros}(1973)}]{baerends_1973}%
  \BibitemOpen
  \bibfield  {author} {\bibinfo {author} {\bibfnamefont {E.}~\bibnamefont
  {Baerends}}, \bibinfo {author} {\bibfnamefont {D.}~\bibnamefont {Ellis}}, \
  and\ \bibinfo {author} {\bibfnamefont {P.}~\bibnamefont {Ros}},\ }\href
  {\doibase 10.1016/0301-0104(73)80059-X} {\bibfield  {journal} {\bibinfo
  {journal} {Chem. Phys.}\ }\textbf {\bibinfo {volume} {2}},\ \bibinfo {pages}
  {41} (\bibinfo {year} {1973})}\BibitemShut {NoStop}%
\bibitem [{\citenamefont {Dunlap}, \citenamefont {Connolly},\ and\
  \citenamefont {Sabin}(1979)}]{dunlap_1979}%
  \BibitemOpen
  \bibfield  {author} {\bibinfo {author} {\bibfnamefont {B.~I.}\ \bibnamefont
  {Dunlap}}, \bibinfo {author} {\bibfnamefont {J.~W.~D.}\ \bibnamefont
  {Connolly}}, \ and\ \bibinfo {author} {\bibfnamefont {J.~R.}\ \bibnamefont
  {Sabin}},\ }\href {\doibase 10.1063/1.438728} {\bibfield  {journal} {\bibinfo
   {journal} {J. Chem. Phys.}\ }\textbf {\bibinfo {volume} {71}},\ \bibinfo
  {pages} {3396} (\bibinfo {year} {1979})}\BibitemShut {NoStop}%
\bibitem [{\citenamefont {Weigend}\ \emph {et~al.}(1998)\citenamefont
  {Weigend}, \citenamefont {H{\"a}ser}, \citenamefont {Patzelt},\ and\
  \citenamefont {Ahlrichs}}]{weigend_1998}%
  \BibitemOpen
  \bibfield  {author} {\bibinfo {author} {\bibfnamefont {F.}~\bibnamefont
  {Weigend}}, \bibinfo {author} {\bibfnamefont {M.}~\bibnamefont {H{\"a}ser}},
  \bibinfo {author} {\bibfnamefont {H.}~\bibnamefont {Patzelt}}, \ and\
  \bibinfo {author} {\bibfnamefont {R.}~\bibnamefont {Ahlrichs}},\ }\href
  {\doibase 10.1016/S0009-2614(98)00862-8} {\bibfield  {journal} {\bibinfo
  {journal} {Chem. Phys. Lett.}\ }\textbf {\bibinfo {volume} {294}},\ \bibinfo
  {pages} {143} (\bibinfo {year} {1998})}\BibitemShut {NoStop}%
\bibitem [{\citenamefont {Dunning}(1989)}]{dunning_cc-basis_1989}%
  \BibitemOpen
  \bibfield  {author} {\bibinfo {author} {\bibfnamefont {T.~H.}\ \bibnamefont
  {Dunning}},\ }\href {\doibase 10.1063/1.456153} {\bibfield  {journal}
  {\bibinfo  {journal} {J. Chem. Phys.}\ }\textbf {\bibinfo {volume} {90}},\
  \bibinfo {pages} {1007} (\bibinfo {year} {1989})}\BibitemShut {NoStop}%
\bibitem [{\citenamefont {Tukey}(1977)}]{tukey_exploratory_1977}%
  \BibitemOpen
  \bibfield  {author} {\bibinfo {author} {\bibfnamefont {J.~W.}\ \bibnamefont
  {Tukey}},\ }\href@noop {} {\emph {\bibinfo {title} {Exploratory data
  analysis}}},\ Addison-{Wesley} series in behavioral science\ (\bibinfo
  {publisher} {Addison-Wesley Pub. Co},\ \bibinfo {address} {Reading, MA},\
  \bibinfo {year} {1977})\BibitemShut {NoStop}%
\bibitem [{\citenamefont {Tkatchenko}\ \emph {et~al.}(2009)\citenamefont
  {Tkatchenko}, \citenamefont {{R. A. DiStasio Jr.}}, \citenamefont
  {Head-Gordon},\ and\ \citenamefont {Scheffler}}]{tkatchenko_jcp_mp2_2009}%
  \BibitemOpen
  \bibfield  {author} {\bibinfo {author} {\bibfnamefont {A.}~\bibnamefont
  {Tkatchenko}}, \bibinfo {author} {\bibnamefont {{R. A. DiStasio Jr.}}},
  \bibinfo {author} {\bibfnamefont {M.}~\bibnamefont {Head-Gordon}}, \ and\
  \bibinfo {author} {\bibfnamefont {M.}~\bibnamefont {Scheffler}},\ }\href
  {\doibase 10.1063/1.3213194} {\bibfield  {journal} {\bibinfo  {journal} {J.
  Chem. Phys.}\ }\textbf {\bibinfo {volume} {131}},\ \bibinfo {pages} {094106}
  (\bibinfo {year} {2009})}\BibitemShut {NoStop}%
\bibitem [{\citenamefont {Tkatchenko}, \citenamefont {{D. Alf\'{e}}},\ and\
  \citenamefont {Kim}(2012)}]{tkatchenko_catcher_2012}%
  \BibitemOpen
  \bibfield  {author} {\bibinfo {author} {\bibfnamefont {A.}~\bibnamefont
  {Tkatchenko}}, \bibinfo {author} {\bibnamefont {{D. Alf\'{e}}}}, \ and\
  \bibinfo {author} {\bibfnamefont {K.~S.}\ \bibnamefont {Kim}},\ }\href
  {\doibase 10.1021/ct300711r} {\bibfield  {journal} {\bibinfo  {journal} {J.
  Chem. Theory Comput.}\ }\textbf {\bibinfo {volume} {8}},\ \bibinfo {pages}
  {4317} (\bibinfo {year} {2012})}\BibitemShut {NoStop}%
\bibitem [{\citenamefont {Sygula}\ \emph {et~al.}(2007)\citenamefont {Sygula},
  \citenamefont {Fronczek}, \citenamefont {Sygula}, \citenamefont {Rabideau},\
  and\ \citenamefont {Olmstead}}]{sygula_catcher_2007}%
  \BibitemOpen
  \bibfield  {author} {\bibinfo {author} {\bibfnamefont {A.}~\bibnamefont
  {Sygula}}, \bibinfo {author} {\bibfnamefont {F.~R.}\ \bibnamefont
  {Fronczek}}, \bibinfo {author} {\bibfnamefont {R.}~\bibnamefont {Sygula}},
  \bibinfo {author} {\bibfnamefont {P.~W.}\ \bibnamefont {Rabideau}}, \ and\
  \bibinfo {author} {\bibfnamefont {M.~M.}\ \bibnamefont {Olmstead}},\ }\href
  {\doibase 10.1021/ja070616p} {\bibfield  {journal} {\bibinfo  {journal} {J.
  Am. Chem. Soc.}\ }\textbf {\bibinfo {volume} {129}},\ \bibinfo {pages} {3842}
  (\bibinfo {year} {2007})}\BibitemShut {NoStop}%
\bibitem [{\citenamefont {{M\"{u}ck}-Lichtenfeld}\ \emph
  {et~al.}(2010)\citenamefont {{M\"{u}ck}-Lichtenfeld}, \citenamefont {Grimme},
  \citenamefont {Kobryn},\ and\ \citenamefont
  {Sygula}}]{muck-lichtenfeld_catcher_2010}%
  \BibitemOpen
  \bibfield  {author} {\bibinfo {author} {\bibfnamefont {C.}~\bibnamefont
  {{M\"{u}ck}-Lichtenfeld}}, \bibinfo {author} {\bibfnamefont {S.}~\bibnamefont
  {Grimme}}, \bibinfo {author} {\bibfnamefont {L.}~\bibnamefont {Kobryn}}, \
  and\ \bibinfo {author} {\bibfnamefont {A.}~\bibnamefont {Sygula}},\ }\href
  {\doibase 10.1039/b925849c} {\bibfield  {journal} {\bibinfo  {journal} {Phys.
  Chem. Chem. Phys.}\ }\textbf {\bibinfo {volume} {12}},\ \bibinfo {pages}
  {7091} (\bibinfo {year} {2010})}\BibitemShut {NoStop}%
\bibitem [{\citenamefont {Le}\ \emph {et~al.}(2014)\citenamefont {Le},
  \citenamefont {Yanney}, \citenamefont {McGuire}, \citenamefont {Sygula},\
  and\ \citenamefont {Lewis}}]{le_thermodynamics_catcher_2014}%
  \BibitemOpen
  \bibfield  {author} {\bibinfo {author} {\bibfnamefont {V.~H.}\ \bibnamefont
  {Le}}, \bibinfo {author} {\bibfnamefont {M.}~\bibnamefont {Yanney}}, \bibinfo
  {author} {\bibfnamefont {M.}~\bibnamefont {McGuire}}, \bibinfo {author}
  {\bibfnamefont {A.}~\bibnamefont {Sygula}}, \ and\ \bibinfo {author}
  {\bibfnamefont {E.~A.}\ \bibnamefont {Lewis}},\ }\href {\doibase
  10.1021/jp5087152} {\bibfield  {journal} {\bibinfo  {journal} {J. Phys. Chem.
  B}\ }\textbf {\bibinfo {volume} {118}},\ \bibinfo {pages} {11956} (\bibinfo
  {year} {2014})}\BibitemShut {NoStop}%
\bibitem [{\citenamefont {Zabula}\ \emph {et~al.}(2014)\citenamefont {Zabula},
  \citenamefont {Sevryugina}, \citenamefont {Spisak}, \citenamefont {Kobryn},
  \citenamefont {Sygula}, \citenamefont {Sygula},\ and\ \citenamefont
  {Petrukhina}}]{zabula_catcher_2014}%
  \BibitemOpen
  \bibfield  {author} {\bibinfo {author} {\bibfnamefont {A.~V.}\ \bibnamefont
  {Zabula}}, \bibinfo {author} {\bibfnamefont {Y.~V.}\ \bibnamefont
  {Sevryugina}}, \bibinfo {author} {\bibfnamefont {S.~N.}\ \bibnamefont
  {Spisak}}, \bibinfo {author} {\bibfnamefont {L.}~\bibnamefont {Kobryn}},
  \bibinfo {author} {\bibfnamefont {R.}~\bibnamefont {Sygula}}, \bibinfo
  {author} {\bibfnamefont {A.}~\bibnamefont {Sygula}}, \ and\ \bibinfo {author}
  {\bibfnamefont {M.~A.}\ \bibnamefont {Petrukhina}},\ }\href {\doibase
  10.1039/c3cc49451a} {\bibfield  {journal} {\bibinfo  {journal} {Chem. Comm.}\
  }\textbf {\bibinfo {volume} {50}},\ \bibinfo {pages} {2657} (\bibinfo {year}
  {2014})}\BibitemShut {NoStop}%
\bibitem [{\citenamefont {Zhao}\ and\ \citenamefont
  {Truhlar}(2008)}]{zhao_catcher_2008}%
  \BibitemOpen
  \bibfield  {author} {\bibinfo {author} {\bibfnamefont {Y.}~\bibnamefont
  {Zhao}}\ and\ \bibinfo {author} {\bibfnamefont {D.~G.}\ \bibnamefont
  {Truhlar}},\ }\href {\doibase 10.1039/b717744e} {\bibfield  {journal}
  {\bibinfo  {journal} {Phys. Chem. Chem. Phys.}\ }\textbf {\bibinfo {volume}
  {10}},\ \bibinfo {pages} {2813} (\bibinfo {year} {2008})}\BibitemShut
  {NoStop}%
\bibitem [{\citenamefont {Waller}\ \emph {et~al.}(2012)\citenamefont {Waller},
  \citenamefont {Kruse}, \citenamefont {{M\"{u}ck}-Lichtenfeld},\ and\
  \citenamefont {Grimme}}]{waller_catcher_2012}%
  \BibitemOpen
  \bibfield  {author} {\bibinfo {author} {\bibfnamefont {M.~P.}\ \bibnamefont
  {Waller}}, \bibinfo {author} {\bibfnamefont {H.}~\bibnamefont {Kruse}},
  \bibinfo {author} {\bibfnamefont {C.}~\bibnamefont {{M\"{u}ck}-Lichtenfeld}},
  \ and\ \bibinfo {author} {\bibfnamefont {S.}~\bibnamefont {Grimme}},\ }\href
  {\doibase 10.1039/c2cs15244d} {\bibfield  {journal} {\bibinfo  {journal}
  {Chem. Soc. Rev.}\ }\textbf {\bibinfo {volume} {41}},\ \bibinfo {pages}
  {3119} (\bibinfo {year} {2012})}\BibitemShut {NoStop}%
\bibitem [{\citenamefont {Podeszwa}, \citenamefont {Cencek},\ and\
  \citenamefont {Szalewicz}(2012)}]{podeszwa_catcher_2012}%
  \BibitemOpen
  \bibfield  {author} {\bibinfo {author} {\bibfnamefont {R.}~\bibnamefont
  {Podeszwa}}, \bibinfo {author} {\bibfnamefont {W.}~\bibnamefont {Cencek}}, \
  and\ \bibinfo {author} {\bibfnamefont {K.}~\bibnamefont {Szalewicz}},\ }\href
  {\doibase 10.1021/ct300200m} {\bibfield  {journal} {\bibinfo  {journal} {J.
  Chem. Theory Comput.}\ }\textbf {\bibinfo {volume} {8}},\ \bibinfo {pages}
  {1963} (\bibinfo {year} {2012})}\BibitemShut {NoStop}%
\bibitem [{\citenamefont {Grimme}(2012)}]{grimme_supramolecular_2012}%
  \BibitemOpen
  \bibfield  {author} {\bibinfo {author} {\bibfnamefont {S.}~\bibnamefont
  {Grimme}},\ }\href {\doibase 10.1002/chem.201200497} {\bibfield  {journal}
  {\bibinfo  {journal} {Chem. Eur. J.}\ }\textbf {\bibinfo {volume} {18}},\
  \bibinfo {pages} {9955} (\bibinfo {year} {2012})}\BibitemShut {NoStop}%
\bibitem [{\citenamefont {Denis}(2013)}]{denis_catcher_2013}%
  \BibitemOpen
  \bibfield  {author} {\bibinfo {author} {\bibfnamefont {P.~A.}\ \bibnamefont
  {Denis}},\ }\href {\doibase 10.1039/c3ra45478a} {\bibfield  {journal}
  {\bibinfo  {journal} {RSC Adv.}\ }\textbf {\bibinfo {volume} {3}},\ \bibinfo
  {pages} {25296} (\bibinfo {year} {2013})}\BibitemShut {NoStop}%
\bibitem [{\citenamefont {Risthaus}\ and\ \citenamefont
  {Grimme}(2013)}]{risthaus_supramolecular_2013}%
  \BibitemOpen
  \bibfield  {author} {\bibinfo {author} {\bibfnamefont {T.}~\bibnamefont
  {Risthaus}}\ and\ \bibinfo {author} {\bibfnamefont {S.}~\bibnamefont
  {Grimme}},\ }\href {\doibase 10.1021/ct301081n} {\bibfield  {journal}
  {\bibinfo  {journal} {J. Chem. Theory Comput.}\ }\textbf {\bibinfo {volume}
  {9}},\ \bibinfo {pages} {1580} (\bibinfo {year} {2013})}\BibitemShut
  {NoStop}%
\bibitem [{\citenamefont {Hedberg}\ \emph {et~al.}(1991)\citenamefont
  {Hedberg}, \citenamefont {Hedberg}, \citenamefont {Bethune}, \citenamefont
  {Brown}, \citenamefont {Dorn}, \citenamefont {Johnson},\ and\ \citenamefont
  {De~Vries}}]{hedberg_c60_1991}%
  \BibitemOpen
  \bibfield  {author} {\bibinfo {author} {\bibfnamefont {K.}~\bibnamefont
  {Hedberg}}, \bibinfo {author} {\bibfnamefont {L.}~\bibnamefont {Hedberg}},
  \bibinfo {author} {\bibfnamefont {D.~S.}\ \bibnamefont {Bethune}}, \bibinfo
  {author} {\bibfnamefont {C.}~\bibnamefont {Brown}}, \bibinfo {author}
  {\bibfnamefont {H.}~\bibnamefont {Dorn}}, \bibinfo {author} {\bibfnamefont
  {R.~D.}\ \bibnamefont {Johnson}}, \ and\ \bibinfo {author} {\bibfnamefont
  {M.}~\bibnamefont {De~Vries}},\ }\href {\doibase
  10.1126/science.254.5030.410} {\bibfield  {journal} {\bibinfo  {journal}
  {Science}\ }\textbf {\bibinfo {volume} {254}},\ \bibinfo {pages} {410}
  (\bibinfo {year} {1991})}\BibitemShut {NoStop}%
\bibitem [{Note9()}]{Note9}%
  \BibitemOpen
  \bibinfo {note} {The PBE+MBD optimization was run in about 2.75 hours on 170
  Intel Xeon E5-2680 processors while the PBE+D3 optimization was run in about
  14 hours on 32 AMD Opteron 6376 Abu Dhabi processors.}\BibitemShut {Stop}%
\bibitem [{\citenamefont {Godec}, \citenamefont {Smith},\ and\ \citenamefont
  {Merzel}(2013)}]{godec_soft_2013}%
  \BibitemOpen
  \bibfield  {author} {\bibinfo {author} {\bibfnamefont {A.}~\bibnamefont
  {Godec}}, \bibinfo {author} {\bibfnamefont {J.~C.}\ \bibnamefont {Smith}}, \
  and\ \bibinfo {author} {\bibfnamefont {F.}~\bibnamefont {Merzel}},\ }\href
  {\doibase 10.1103/PhysRevLett.111.127801} {\bibfield  {journal} {\bibinfo
  {journal} {Phys. Rev. Lett.}\ }\textbf {\bibinfo {volume} {111}},\ \bibinfo
  {pages} {127801} (\bibinfo {year} {2013})}\BibitemShut {NoStop}%
\bibitem [{\citenamefont {Jinich}\ \emph {et~al.}(2014)\citenamefont {Jinich},
  \citenamefont {Rappoport}, \citenamefont {Dunn}, \citenamefont
  {Sanchez-Lengeling}, \citenamefont {Olivares-Amaya}, \citenamefont {Noor},
  \citenamefont {Even},\ and\ \citenamefont
  {Aspuru-Guzik}}]{jinich_quantum_2014}%
  \BibitemOpen
  \bibfield  {author} {\bibinfo {author} {\bibfnamefont {A.}~\bibnamefont
  {Jinich}}, \bibinfo {author} {\bibfnamefont {D.}~\bibnamefont {Rappoport}},
  \bibinfo {author} {\bibfnamefont {I.}~\bibnamefont {Dunn}}, \bibinfo {author}
  {\bibfnamefont {B.}~\bibnamefont {Sanchez-Lengeling}}, \bibinfo {author}
  {\bibfnamefont {R.}~\bibnamefont {Olivares-Amaya}}, \bibinfo {author}
  {\bibfnamefont {E.}~\bibnamefont {Noor}}, \bibinfo {author} {\bibfnamefont
  {A.~B.}\ \bibnamefont {Even}}, \ and\ \bibinfo {author} {\bibfnamefont
  {A.}~\bibnamefont {Aspuru-Guzik}},\ }\href {\doibase 10.1038/srep07022}
  {\bibfield  {journal} {\bibinfo  {journal} {Sci. Rep.}\ }\textbf {\bibinfo
  {volume} {4}},\ \bibinfo {pages} {7022} (\bibinfo {year} {2014})}\BibitemShut
  {NoStop}%
\bibitem [{\citenamefont {Er}\ \emph {et~al.}(2015)\citenamefont {Er},
  \citenamefont {Suh}, \citenamefont {Marshak},\ and\ \citenamefont
  {Aspuru-Guzik}}]{er_computational_2015}%
  \BibitemOpen
  \bibfield  {author} {\bibinfo {author} {\bibfnamefont {S.}~\bibnamefont
  {Er}}, \bibinfo {author} {\bibfnamefont {C.}~\bibnamefont {Suh}}, \bibinfo
  {author} {\bibfnamefont {M.~P.}\ \bibnamefont {Marshak}}, \ and\ \bibinfo
  {author} {\bibfnamefont {A.}~\bibnamefont {Aspuru-Guzik}},\ }\href {\doibase
  10.1039/C4SC03030C} {\bibfield  {journal} {\bibinfo  {journal} {Chem. Sci.}\
  }\textbf {\bibinfo {volume} {6}},\ \bibinfo {pages} {885} (\bibinfo {year}
  {2015})}\BibitemShut {NoStop}%
\bibitem [{\citenamefont {Towns}\ \emph {et~al.}(2014)\citenamefont {Towns},
  \citenamefont {Cockerill}, \citenamefont {Dahan}, \citenamefont {Foster},
  \citenamefont {Gaither}, \citenamefont {Grimshaw}, \citenamefont {Hazlewood},
  \citenamefont {Lathrop}, \citenamefont {Lifka}, \citenamefont {Peterson},
  \citenamefont {Roskies}, \citenamefont {Scott},\ and\ \citenamefont
  {Wilkens-Diehr}}]{xsede_2014}%
  \BibitemOpen
  \bibfield  {author} {\bibinfo {author} {\bibfnamefont {J.}~\bibnamefont
  {Towns}}, \bibinfo {author} {\bibfnamefont {T.}~\bibnamefont {Cockerill}},
  \bibinfo {author} {\bibfnamefont {M.}~\bibnamefont {Dahan}}, \bibinfo
  {author} {\bibfnamefont {I.}~\bibnamefont {Foster}}, \bibinfo {author}
  {\bibfnamefont {K.}~\bibnamefont {Gaither}}, \bibinfo {author} {\bibfnamefont
  {A.}~\bibnamefont {Grimshaw}}, \bibinfo {author} {\bibfnamefont
  {V.}~\bibnamefont {Hazlewood}}, \bibinfo {author} {\bibfnamefont
  {S.}~\bibnamefont {Lathrop}}, \bibinfo {author} {\bibfnamefont
  {D.}~\bibnamefont {Lifka}}, \bibinfo {author} {\bibfnamefont {G.~D.}\
  \bibnamefont {Peterson}}, \bibinfo {author} {\bibfnamefont {R.}~\bibnamefont
  {Roskies}}, \bibinfo {author} {\bibfnamefont {J.~R.}\ \bibnamefont {Scott}},
  \ and\ \bibinfo {author} {\bibfnamefont {N.}~\bibnamefont {Wilkens-Diehr}},\
  }\href {\doibase 10.1109/MCSE.2014.80} {\bibfield  {journal} {\bibinfo
  {journal} {Comput. Sci. Eng.}\ }\textbf {\bibinfo {volume} {16}},\ \bibinfo
  {pages} {62} (\bibinfo {year} {2014})}\BibitemShut {NoStop}%
\bibitem [{\citenamefont {Nalewajski}\ and\ \citenamefont
  {Parr}(2000)}]{nalewajski_2000}%
  \BibitemOpen
  \bibfield  {author} {\bibinfo {author} {\bibfnamefont {R.~F.}\ \bibnamefont
  {Nalewajski}}\ and\ \bibinfo {author} {\bibfnamefont {R.~G.}\ \bibnamefont
  {Parr}},\ }\href {\doibase 10.1073/pnas.97.16.8879} {\bibfield  {journal}
  {\bibinfo  {journal} {Proc. Natl. Acad. Sci. USA}\ }\textbf {\bibinfo
  {volume} {97}},\ \bibinfo {pages} {8879} (\bibinfo {year}
  {2000})}\BibitemShut {NoStop}%
\bibitem [{\citenamefont {{R. A. DiStasio Jr.}}\ \emph {et~al.}()\citenamefont
  {{R. A. DiStasio Jr.}}, \citenamefont {Ko}, \citenamefont {Santra},\ and\
  \citenamefont {Car}}]{distasio_unpublished}%
  \BibitemOpen
  \bibfield  {author} {\bibinfo {author} {\bibnamefont {{R. A. DiStasio Jr.}}},
  \bibinfo {author} {\bibfnamefont {H.-Y.}\ \bibnamefont {Ko}}, \bibinfo
  {author} {\bibfnamefont {B.}~\bibnamefont {Santra}}, \ and\ \bibinfo {author}
  {\bibfnamefont {R.}~\bibnamefont {Car}},\ }\href@noop {} {\enquote {\bibinfo
  {title} {Enabling \textit{ab initio} molcular dynamics with a self-consistent
  interatomic van der waals functional},}\ }\bibinfo {note}
  {Unpublished}\BibitemShut {NoStop}%
\bibitem [{\citenamefont {Bickley}(1941)}]{bickley_1941}%
  \BibitemOpen
  \bibfield  {author} {\bibinfo {author} {\bibfnamefont {W.~G.}\ \bibnamefont
  {Bickley}},\ }\href {http://www.jstor.org/stable/3606475} {\bibfield
  {journal} {\bibinfo  {journal} {Math. Gaz.}\ }\textbf {\bibinfo {volume}
  {25}},\ \bibinfo {pages} {19} (\bibinfo {year} {1941})}\BibitemShut {NoStop}%
\bibitem [{\citenamefont {Friswell}(1996)}]{friswell_1996}%
  \BibitemOpen
  \bibfield  {author} {\bibinfo {author} {\bibfnamefont {M.~I.}\ \bibnamefont
  {Friswell}},\ }\href {\doibase 10.1115/1.2888195} {\bibfield  {journal}
  {\bibinfo  {journal} {J. Vib. Acoust.}\ }\textbf {\bibinfo {volume} {118}},\
  \bibinfo {pages} {390} (\bibinfo {year} {1996})}\BibitemShut {NoStop}%
\bibitem [{\citenamefont {Andrew}\ and\ \citenamefont
  {Tan}(1998)}]{andrew_1998}%
  \BibitemOpen
  \bibfield  {author} {\bibinfo {author} {\bibfnamefont {A.~L.}\ \bibnamefont
  {Andrew}}\ and\ \bibinfo {author} {\bibfnamefont {R.~C.~E.}\ \bibnamefont
  {Tan}},\ }\href {\doibase 10.1137/S0895479896304332} {\bibfield  {journal}
  {\bibinfo  {journal} {{SIAM} J. Matrix Anal. Appl.}\ }\textbf {\bibinfo
  {volume} {20}},\ \bibinfo {pages} {78} (\bibinfo {year} {1998})}\BibitemShut
  {NoStop}%
\bibitem [{\citenamefont {Boyd}(1987)}]{boyd_1987}%
  \BibitemOpen
  \bibfield  {author} {\bibinfo {author} {\bibfnamefont {J.~P.}\ \bibnamefont
  {Boyd}},\ }\href {\doibase 10.1007/BF01061480} {\bibfield  {journal}
  {\bibinfo  {journal} {J. Sci. Comp.}\ }\textbf {\bibinfo {volume} {2}},\
  \bibinfo {pages} {99} (\bibinfo {year} {1987})}\BibitemShut {NoStop}%
\bibitem [{\citenamefont {Derevianko}, \citenamefont {Porsev},\ and\
  \citenamefont {Babb}(2010)}]{derevianko_2010}%
  \BibitemOpen
  \bibfield  {author} {\bibinfo {author} {\bibfnamefont {A.}~\bibnamefont
  {Derevianko}}, \bibinfo {author} {\bibfnamefont {S.~G.}\ \bibnamefont
  {Porsev}}, \ and\ \bibinfo {author} {\bibfnamefont {J.~F.}\ \bibnamefont
  {Babb}},\ }\href {\doibase 10.1016/j.adt.2009.12.002} {\bibfield  {journal}
  {\bibinfo  {journal} {At. Data Nucl. Data Tables}\ }\textbf {\bibinfo
  {volume} {96}},\ \bibinfo {pages} {323} (\bibinfo {year} {2010})}\BibitemShut
  {NoStop}%
\bibitem [{\citenamefont {Broyden}(1969)}]{broyden_bfgs_1969}%
  \BibitemOpen
  \bibfield  {author} {\bibinfo {author} {\bibfnamefont {C.~G.}\ \bibnamefont
  {Broyden}},\ }\href@noop {} {\bibfield  {journal} {\bibinfo  {journal}
  {Notices Amer. Math. Soc.}\ }\textbf {\bibinfo {volume} {16}},\ \bibinfo
  {pages} {670} (\bibinfo {year} {1969})}\BibitemShut {NoStop}%
\bibitem [{\citenamefont {Goldfarb}(1970)}]{goldfarb_bfgs_1970}%
  \BibitemOpen
  \bibfield  {author} {\bibinfo {author} {\bibfnamefont {D.}~\bibnamefont
  {Goldfarb}},\ }\href {\doibase 10.1090/S0025-5718-1970-0258249-6} {\bibfield
  {journal} {\bibinfo  {journal} {Math. Comput.}\ }\textbf {\bibinfo {volume}
  {24}},\ \bibinfo {pages} {23} (\bibinfo {year} {1970})}\BibitemShut {NoStop}%
\bibitem [{\citenamefont {Fletcher}(1970)}]{fletcher_bfgs_1970}%
  \BibitemOpen
  \bibfield  {author} {\bibinfo {author} {\bibfnamefont {R.}~\bibnamefont
  {Fletcher}},\ }\href {\doibase 10.1093/comjnl/13.3.317} {\bibfield  {journal}
  {\bibinfo  {journal} {Comp. J.}\ }\textbf {\bibinfo {volume} {13}},\ \bibinfo
  {pages} {317} (\bibinfo {year} {1970})}\BibitemShut {NoStop}%
\bibitem [{\citenamefont {Shanno}(1970)}]{shanno_bfgs_1970}%
  \BibitemOpen
  \bibfield  {author} {\bibinfo {author} {\bibfnamefont {D.~F.}\ \bibnamefont
  {Shanno}},\ }\href {\doibase 10.1090/S0025-5718-1970-0274029-X} {\bibfield
  {journal} {\bibinfo  {journal} {Math. Comput.}\ }\textbf {\bibinfo {volume}
  {24}},\ \bibinfo {pages} {647} (\bibinfo {year} {1970})}\BibitemShut
  {NoStop}%
\bibitem [{\citenamefont {Gygi}(2015{\natexlab{a}})}]{gygi_fmpd_repository}%
  \BibitemOpen
  \bibfield  {author} {\bibinfo {author} {\bibfnamefont {F.}~\bibnamefont
  {Gygi}},\ }\href {{http://fpmd.ucdavis.edu/potentials/index.htm}} {\enquote
  {\bibinfo {title} {{FPMD Pseudopotential Repository}},}\ } (\bibinfo {year}
  {Accessed Mar. 29, 2015}{\natexlab{a}})\BibitemShut {NoStop}%
\bibitem [{\citenamefont {Gygi}(2015{\natexlab{b}})}]{gygi_qso2upf}%
  \BibitemOpen
  \bibfield  {author} {\bibinfo {author} {\bibfnamefont {F.}~\bibnamefont
  {Gygi}},\ }\href {{http://eslab.ucdavis.edu/software/index.htm}} {\enquote
  {\bibinfo {title} {qso2upf},}\ } (\bibinfo {year} {Accessed Mar. 29,
  2015}{\natexlab{b}})\BibitemShut {NoStop}%
\bibitem [{\citenamefont {Kruse}\ and\ \citenamefont
  {Grimme}(2012)}]{kruse_geometrical_2012}%
  \BibitemOpen
  \bibfield  {author} {\bibinfo {author} {\bibfnamefont {H.}~\bibnamefont
  {Kruse}}\ and\ \bibinfo {author} {\bibfnamefont {S.}~\bibnamefont {Grimme}},\
  }\href {\doibase 10.1063/1.3700154} {\bibfield  {journal} {\bibinfo
  {journal} {J. Chem. Phys.}\ }\textbf {\bibinfo {volume} {136}},\ \bibinfo
  {pages} {154101} (\bibinfo {year} {2012})}\BibitemShut {NoStop}%
\bibitem [{\citenamefont {Weigend}\ and\ \citenamefont
  {Ahlrichs}(2005)}]{weigend_2005}%
  \BibitemOpen
  \bibfield  {author} {\bibinfo {author} {\bibfnamefont {F.}~\bibnamefont
  {Weigend}}\ and\ \bibinfo {author} {\bibfnamefont {R.}~\bibnamefont
  {Ahlrichs}},\ }\href {\doibase 10.1039/b508541a} {\bibfield  {journal}
  {\bibinfo  {journal} {Phys. Chem. Chem. Phys.}\ }\textbf {\bibinfo {volume}
  {7}},\ \bibinfo {pages} {3297} (\bibinfo {year} {2005})}\BibitemShut
  {NoStop}%
\bibitem [{\citenamefont {Eichkorn}\ \emph {et~al.}(1997)\citenamefont
  {Eichkorn}, \citenamefont {Weigend}, \citenamefont {Treutler},\ and\
  \citenamefont {Ahlrichs}}]{eichkorn_1997}%
  \BibitemOpen
  \bibfield  {author} {\bibinfo {author} {\bibfnamefont {K.}~\bibnamefont
  {Eichkorn}}, \bibinfo {author} {\bibfnamefont {F.}~\bibnamefont {Weigend}},
  \bibinfo {author} {\bibfnamefont {O.}~\bibnamefont {Treutler}}, \ and\
  \bibinfo {author} {\bibfnamefont {R.}~\bibnamefont {Ahlrichs}},\ }\href
  {\doibase 10.1007/s002140050244} {\bibfield  {journal} {\bibinfo  {journal}
  {Theor. Chem. Acc.}\ }\textbf {\bibinfo {volume} {97}},\ \bibinfo {pages}
  {119} (\bibinfo {year} {1997})}\BibitemShut {NoStop}%
\bibitem [{\citenamefont {Weigend}(2006)}]{weigend_2006}%
  \BibitemOpen
  \bibfield  {author} {\bibinfo {author} {\bibfnamefont {F.}~\bibnamefont
  {Weigend}},\ }\href {\doibase 10.1039/b515623h} {\bibfield  {journal}
  {\bibinfo  {journal} {Phys. Chem. Chem. Phys.}\ }\textbf {\bibinfo {volume}
  {8}},\ \bibinfo {pages} {1057} (\bibinfo {year} {2006})}\BibitemShut
  {NoStop}%
\bibitem [{\citenamefont {Kendall}\ and\ \citenamefont
  {Fr{\"u}chtl}(1997)}]{kendall_1997}%
  \BibitemOpen
  \bibfield  {author} {\bibinfo {author} {\bibfnamefont {R.~A.}\ \bibnamefont
  {Kendall}}\ and\ \bibinfo {author} {\bibfnamefont {H.~A.}\ \bibnamefont
  {Fr{\"u}chtl}},\ }\href {\doibase 10.1007/s002140050249} {\bibfield
  {journal} {\bibinfo  {journal} {Theor. Chem. Acc.}\ }\textbf {\bibinfo
  {volume} {97}},\ \bibinfo {pages} {158} (\bibinfo {year} {1997})}\BibitemShut
  {NoStop}%
\bibitem [{Note3()}]{Note3}%
  \BibitemOpen
  \bibinfo {note} {Podeszwa et al. used the PBE0 functional with Fermi-Amaldi
  asymptotic correction and with the Tozer-Handy splicing scheme, and an
  aug-cc-pVTZ Dunning basis with a monomer-centered ``plus'' basis set scheme.
  Their symmetry-adapted perturbation theory procedure used density fitting.
  They also used the vibrationally averaged monomer geometry
  ($r_{CC}=1.3965~\protect \text {\r A}$ and $r_{CH} = 1.085~\protect \text {\r
  A}$) from Tamagawa \protect \textit {et al. J. Mol. Struct.} 1976, \protect
  \textbf {30}, 243.}\BibitemShut {Stop}%
\end{thebibliography}%
\balance
\newpage

\clearpage
\onecolumngrid
\section{Electronic Supplementary Information}

\subsection{Symbol glossary}\label{appendix:glossary}
\begin{tabular}{ll }
\\\toprule
Symbol & Description \\\midrule
$\bm{\alpha}_a^{\rm free}(0)$ & static free-atom polarizability formed with $\alpha_a^{\rm free}(0)$ scalars on the diagonal\\
$\bm{\alpha}_a(0)$ &  Eq.~\eqref{TS_polarizability}: static bare polarizability, related to $\bm{\alpha}_a^{\rm free}(0)$ by the ratio $V_a/V_a^{\rm free}$ \\
$\bm{\alpha}_a(\mathbbm{i}\omega)$ &  Eq.~\eqref{TS_polarizability_lambda}: frequency-dependent bare polarizability tensor, calculated by Pad\'{e} approximant Eq.~\eqref{eq:alpha_pade}\\
$\alpha_a(\mathbbm{i}\omega)$ &  `isotropized' bare dipole polarizability scalar, calculated as $\alpha_a = \tfrac{1}{3}{\rm Tr} [\bm{\alpha}_a]$ \\
$\mathbf{A}(\mathbbm{i}\omega)$ & Eq.~\eqref{A_TS}: bare system polarizability tensor,  $3N\times 3N$ block diagonal matrix of $\bm{\alpha}_a(\mathbbm{i}\omega)$   \\
$\overline{\mathbf{A}}(\mathbbm{i}\omega)$ &  Eq.~\eqref{rsSCS}: screened system polarizability tensor, solved at complex frequency $\mathbbm{i}\omega$ using $ \left[\mathbf{A}^{-1} + \mathbf{T}_{SR}\right]^{-1}$ \\
$\overline{\bm{\alpha}}_a(0)$ & Eq.~\eqref{screened_static_polarizabilities}: screened static polarizability calculated by partial contraction of $\overline{\mathbf{A}}(0)$ \\
$\overline{\bm{\alpha}}_a(\mathbbm{i}\omega)$ & Eq.~\eqref{screened_polarizabilities}: screened frequency-dependent polarizability,   \\ & \qquad  calculated by partial contraction of $\overline{\mathbf{A}}(\mathbbm{i}\omega)$ \\
$\omega_a^{\rm free}$ & free-atom QHO excitation frequency,  computed as $\omega^{\rm free}_a = 4/3 \left( C_{6,aa}^{\rm free} / \left[ \alpha^{\rm free}_a(0) \right]^2 \right)$ \\
$C_{6,aa}^{\rm free}$ & free-atom $C_6$ coefficient (also called Hamaker constant)\\
$C_{6,aa}$ & bare effective atomic $C_6$ coefficient computed by weighting $C^{\rm free}_{6,aa}$ with $\left( V_a/V_a^{\rm free} \right)^2$ \\
$\overline{C}_{6,aa}$ & Eq.~\eqref{casimir-polder}: screened effective atomic $C_6$ coefficient \\
$\omega_a$ &  bare QHO excitation frequency, equals  $\omega^{\rm free}$ due to cancellation of $V_a/V_a^{\rm free}$ factors \\
$\overline{\omega}_a$ & Eq.~\eqref{screened_omega}:  screened QHO excitation frequency\\
$y_p$ & frequency grid for numerical integration \\
$g_p$ & weights for numerical integration \\
$ \sigma_a(\mathbbm{i}\omega)$ & Eq.~\eqref{qho_width}: QHO width, calculated from the bare  polarizability scalar \\
$ \Lambda_a(\mathbbm{i}\omega)$ & Eq.~\eqref{sigma_in_terms_of_Veff}: multiplicative prefactor to $V_a$ in defining  $\alpha_a$ \\
$V_a^{\rm free}$ & free-atom effective volume\\
$V_a$ & Eq.~\eqref{V_eff_defined}: Hirshfeld effective atomic volume \\
$ \Sigma_{ab}(\mathbbm{i}\omega)$ & Eq.~\eqref{Sigma}:  effective correlation length of the interaction potential,  \\ & \qquad  defined from QHO widths of  atoms $a$ and $b$ \\
$\bm{\mathscr{R}}_a$ & nuclear position of atom $a$ \\
$R_{ab}$ & internuclear distance between atoms $a$ and $b$ \\
$\mathbf{R}_{ab}$ & internuclear vector ($\bm{\mathscr{R}}_a -\bm{\mathscr{R}}_b$) between atoms $a$ and $b$  \\
${R}_{ab}^i$ & $i^{th}$ Cartesian component of the internuclear vector  \\
$\mathbf{r}$ & spatial position such as the argument of the electron density \\
$\zeta$ & ratio between interatomic separation $R$ and correlation length $\Sigma$ \\
$h(\zeta)$ & Eq.~\eqref{h_zeta}: function of $\zeta$ appearing in $\mathbf{T}$ \\
$ v(R,\mathbbm{i}\omega)$ & Eq.~\eqref{Coulomb_interaction_Gaussians}: Coulomb interaction between two QHO Gaussian  \\ & \qquad charge densities separated  by $R$ with frequency-dependent interaction  \\
$\mathbf{T}$ & Eq.~\eqref{T_GG}: dipole interaction tensor between  \quad  QHO Gaussian charge densities \\
$\mathbf{T}_{\rm dip}$ & Eq.~\eqref{T_dip}: dipole interaction tensor between  point dipoles\\
$\mathbf{T}_{SR}$ & Eq.~\eqref{T_SR}: short-range component of  $\mathbf{T}$,  evaluated using $f(Z^{\rm TS})$   \\
$\overline{\mathbf{T}}_{LR}$ & Eq.~\eqref{T_LR}: long-range component of  $\mathbf{T}$,     evaluated using  $f(\overline{Z}_{\rm vdW})$  \\
$f(Z)$ & Eq.~\eqref{f_damp}: damping function for range-separation of the dipole interaction tensor \\
$S_{ab}$ & sum of effective vdW radii scaled by $\beta$ \\
$Z_{ab}$ & ratio of interatomic separation $R_{ab}$ and $S_{ab}$  \\
$\mathcal{R}_{~a}^{\rm vdW,\,TS}$ & Eq.~\eqref{RvdW_TS}: effective vdW radii at the TS level \\
$ \overline{\mathcal{R}}_{\,a}^{\rm vdW} $&Eq.~\eqref{RvdW_SCS}: screened effective vdW radii \\ 
$\rho $ & total electronic charge density \\
$\rho_a^{\rm eff}$ & Eq.~\eqref{rho_eff}: Hirshfeld effective electron density assigned to atom $a$ \\ 
 $\rho_{\rm sad}$ & sum of spherical free-atom densities $\sum_b \rho_b^{\rm free}$  \\
$\bm{\mathcal{H}}^{\rm MBD}$ & Eq.~\eqref{H_MBD}:  MBD model Hamiltonian \\
$\mathbf{C}^{\rm MBD}$ & Eq.~\eqref{C_MBD}: MBD interaction matrix \\
$\bm{\mathcal{X}}$ & matrix of eigenvectors of  $\mathbf{C}^{\rm MBD}$ \\
$\bm{\lambda}$ & vector of eigenvalues of  $\mathbf{C}^{\rm MBD}$ \\
$\bm{\partial}_c$ & gradient with respect to nuclear position of  atom $c$,  equivalent to $\bm{\nabla}_{\bm{\mathscr{R}}_c}$  \\
$E_{\rm MBD}$ & MBD correlation energy \\
$\mathbf{F}^{\rm MBD}_c$ & MBD ionic forces: $-\bm{\nabla}_{\bm{\mathscr{R}}_c} E_{\rm MBD}$ 
\end{tabular}
\newpage

\subsection{Length Scale of Damping}\label{appendix:damping_lengthscale}
\begin{figure*}[!htbp]
\centering
\fbox{\includegraphics[height=5.12cm]{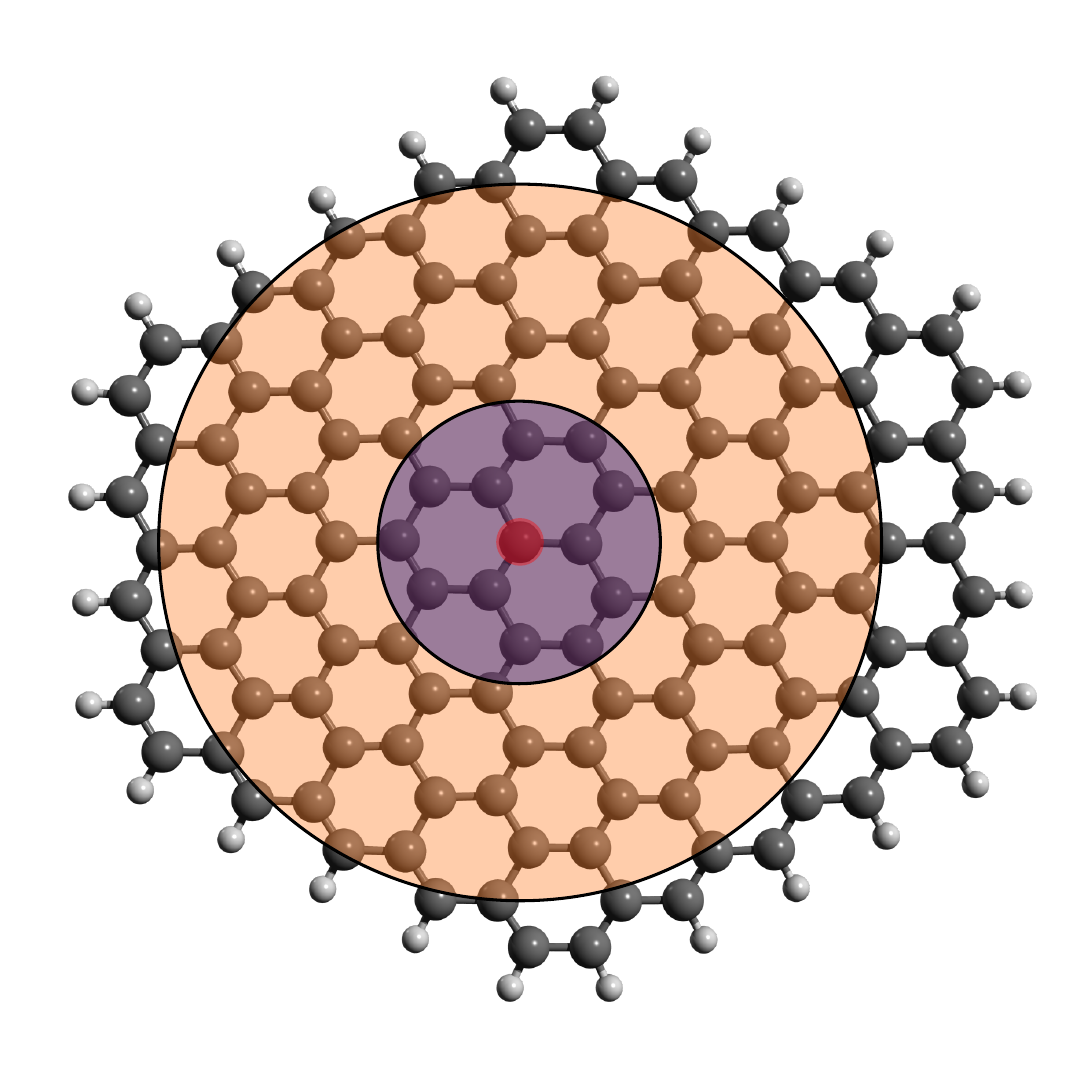}}\qquad\qquad
\fbox{\includegraphics[height=5.12cm]{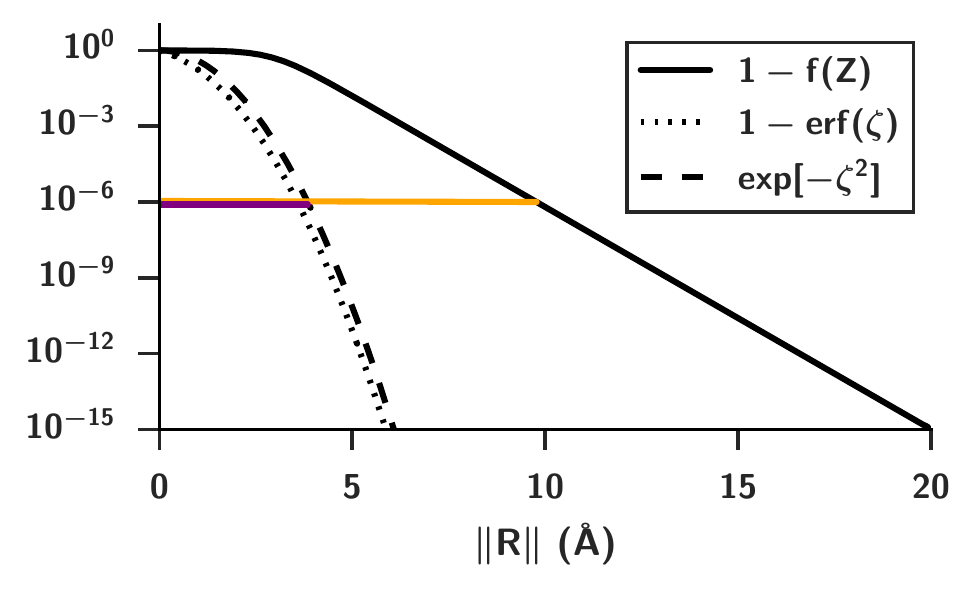}}
\caption{\textbf{Left:} Contours at $10^{-6}$ for damping functions $\exp\left[-\zeta^2\right]$ (purple) and $\left( 1-f\left[Z\right]\right)$ (orange), with $\| \mathbf{R} \|$ relative to the atom marked in red. Damping parameters $\Sigma \simeq 1.03$ and $S\simeq 2.96$ (cf. Eqs.~\eqref{Sigma} and~\eqref{S_ab}) were computed for a graphene nanoflake with the PBE functional.
\textbf{Right:} Comparison of the three damping functions with the same $10^{-6}$ contour indicated. 
The rapid decay of $\exp\left[-\zeta^2\right]$ relative to the Fermi damping function demonstrates that the short-range dipole-dipole interaction tensor $\mathbf{T}_{\rm SR}$ reduces to the frequency-independent $\mathbf{T}_{\rm dip}$ well before the long-range tensor $\overline{\mathbf{T}}_{\rm LR}$ has been fully ``turned on'' by the Fermi damping function (cf. Eqs.~(\ref{TGG_cartesian}, \ref{T_dip}, \ref{T_SR}, \ref{T_LR})). \label{fig:esi:damping_lengthscales}}
\end{figure*}

\subsection{Computation of \texorpdfstring{$\bm{\partial} V$}{dV}}\label{appendix:veff_derivative}
Nuclear coordinate forces within a fully self-consistent $\mathcal{O}(N)$ implementation of the Tkatchenko-Scheffler  (TS) scheme~\cite{tkatchenko_2009} were previously developed in \textsc{Quantum ESPRESSO} (\textsc{QE})~\cite{giannozzi_2009} by R. A. DiStasio Jr.~\cite{distasio_unpublished}  A subroutine of the \textsc{tsvdw} module computes the Hirshfeld partitioning into effective atomic volumes, $V_a$, and the derivatives of that volume, $\partial V_a$.
The Hirshfeld effective charge density of atom $a$ is:
\begin{eqnarray}
\rho^{\rm eff}_a(\mathbf{r}) = w_a(\mathbf{r})\, \rho(\mathbf{r})  = \frac{\rho_a^{\rm free }(\|\mathbf{r}-\bm{\mathscr{R}}_a\| )}{   \rho_{\rm sad}(\mathbf{r})   }\; \rho(\mathbf{r}) \label{rho_eff} ,
\end{eqnarray}
where $\rho(\mathbf{r})$ is the total molecular charge density and $\rho_{\rm sad}(\mathbf{r}) = \sum_b \rho_b^{\rm free}(\|\mathbf{r}-\bm{\mathscr{R}}_b\| )$ is the sum of free-atom densities.
The effective volume is then:
\begin{eqnarray}
V_a&=&   \int {\rm d}\mathbf{r} \;\|\mathbf{r}-\bm{\mathscr{R}}_a\|^3  \rho^{\rm eff}_a(\mathbf{r}).\label{V_eff_defined} 
\end{eqnarray}
Integrations on spherical atomic domains, such as in Eq.~\ref{V_eff_defined}, are computed on subsets of the real-space mesh. Using reference data for the free atom volumes, the radial grid cutoff value is determined for each species such that the free atom volume obtained by numerical integration up to this cutoff does not deviate from the reference value by more than 1.0\%.
The effective volume derivative is evaluated as
\begin{eqnarray}
 \bm{\partial}_c V_a &=& \int {\rm d}\mathbf{r} \;  	\|\mathbf{r}-\bm{\mathscr{R}}_a\|^3 \bm{\partial}_c \rho^{\rm eff}_a(\mathbf{r})  - 3\, \delta_{ca}  \int {\rm d}\mathbf{r} \; (\mathbf{r}-\bm{\mathscr{R}}_a)  \|\mathbf{r}-\bm{\mathscr{R}}_a\|  \rho^{\rm eff}_a(\mathbf{r})  \\
\bm{\partial}_c \rho^{\rm eff}_a(\mathbf{r}) &=&   \left[ \frac{ \rho_a^{\rm free}(\|\mathbf{r}-\bm{\mathscr{R}}_a\|)\,\rho(\mathbf{r})}{\left[ \rho_{\rm sad}(\mathbf{r}) \right]^2}  - \frac{\rho(\mathbf{r})}{\rho_{\rm sad}(\mathbf{r})}   \delta_{ca}  \right] \left[ \frac{ \mathbf{r}-\bm{\mathscr{R}}_c}{ \|\mathbf{r}-\bm{\mathscr{R}}_c\|} \right]  \frac{ \partial  \rho_c^{\rm free}(r)}{\partial r} 
\end{eqnarray}
Note that the free-atom density is spherically symmetric, which is why we reduce $\bm{\partial}_c \rho_c^{\rm free}(\|\mathbf{r}-\bm{\mathscr{R}}_c\|)$ to a spherical coordinate derivative $\partial \rho_c^{\rm free}/\partial r$.
Likewise, Eq.~\eqref{V_eff_defined} is evaluated by mapping the radial form of $\rho^{\rm eff}_a$ to an linear/equispaced grid, which is then interpolated using cubic splines. After interpolation, the derivative  $\bm{\partial}_c \rho^{\rm eff}_a$ at each grid point is evaluated by numerical differentiation using Bickley's 7-point formula.~\cite{bickley_1941} 

\subsection{Repeated Eigenvalues of \texorpdfstring{\textbf{C}\textsuperscript{MBD}}{C}}
In considering the derivative of $\lambda_p$, in Eq.~\eqref{derivative_of_lambda_distinct_eigenvalues} we assumed that $\mathbf{C}^{\rm MBD}$ had $3N$ distinct eigenvalues.  Due to numerical perturbations it is somewhat unlikely for $\mathbf{C}^{\rm MBD}$ to have repeated eigenvalues, but we cannot assume this \textit{a priori}. The procedure for taking derivatives of repeated eigenvalues of a real, symmetric matrix, like $\mathbf{C}^{\rm MBD}$, is essentially first order perturbation theory where the perturbation is the action of the derivative operator $\bm{\partial}_c$. Eigenvalue degeneracies are lifted by diagonalizing the perturbation in the degenerate subspace. For a more algorithmic discussion of repeated eigenvalue derivatives, see Friswell~\cite{friswell_1996} or Andrew \textit{et al.}~\cite{andrew_1998}. Since $\mathbf{C}^{\rm MBD}$ is real and symmetric, it is guaranteed to be diagonalizable with orthogonal eigenvectors.

\subsection{Importance of \texorpdfstring{$\bm{\partial} V$}{dV}}
\subsubsection{Benzene dimer}
\begin{figure}[!htbp]\centering
\fbox{\includegraphics[width=8.3cm]{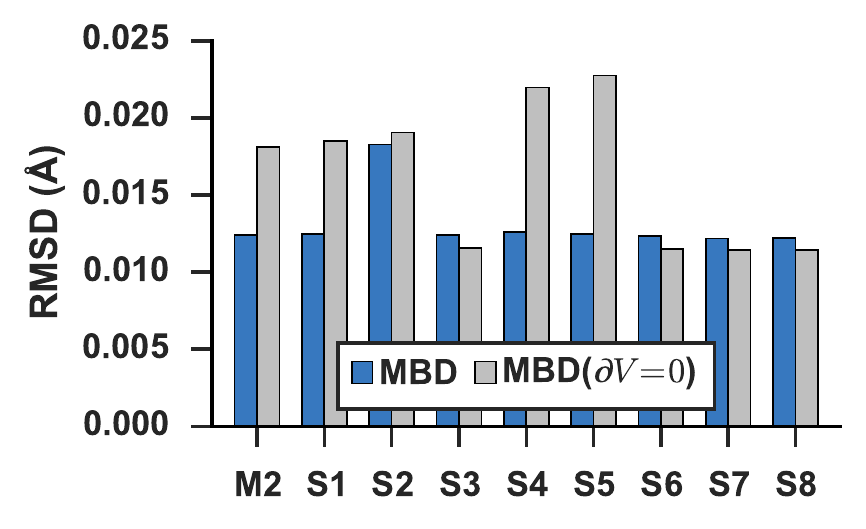}}
\caption{Root-mean-square-deviations (RMSD) in~\AA\ between the PBE+MBD and PBE/CCSD(T)~\cite{podeszwa_2006} optimized geometries of 9 benzene dimer configurations using the full MBD gradient (shown in blue), and the approximation where $\partial V$ contributions are set explicitly to zero (shown in grey).  \label{fig:esi:benzene_intermonomer_distance}}
\end{figure}

To analyze the importance of $\partial V$, we re-optimized the benzene dimer structures with $\bm{\partial} V$ terms set explicitly to zero.  As shown in Fig.~\ref{fig:esi:benzene_intermonomer_distance}, setting  $\bm{\partial} V=0$ slightly degrades the consistency of the PBE+MBD optimized geometries, but the final RMSDs are still quite good (all $<0.025$~\AA). The optimization of M1 with $\bm{\partial} V=0$ proved numerically unstable, and was unable to converge, so M1 is not included in the figure. The fact that the Hirshfeld gradients have a negligible impact on the benzene dimer optimizations is expected since the Hirshfeld effective atomic volumes only change when nearest neighbor atoms are moved.  In addition to being quite rigid, the benzene monomer is small enough that the range-separated MBD correction is largely turned off within the length scale of the monomer, which is where the Hirshfeld gradients could matter. 

\subsubsection{Polypeptides}
\begin{figure}[!htbp]\centering
\includegraphics[width=17.2cm]{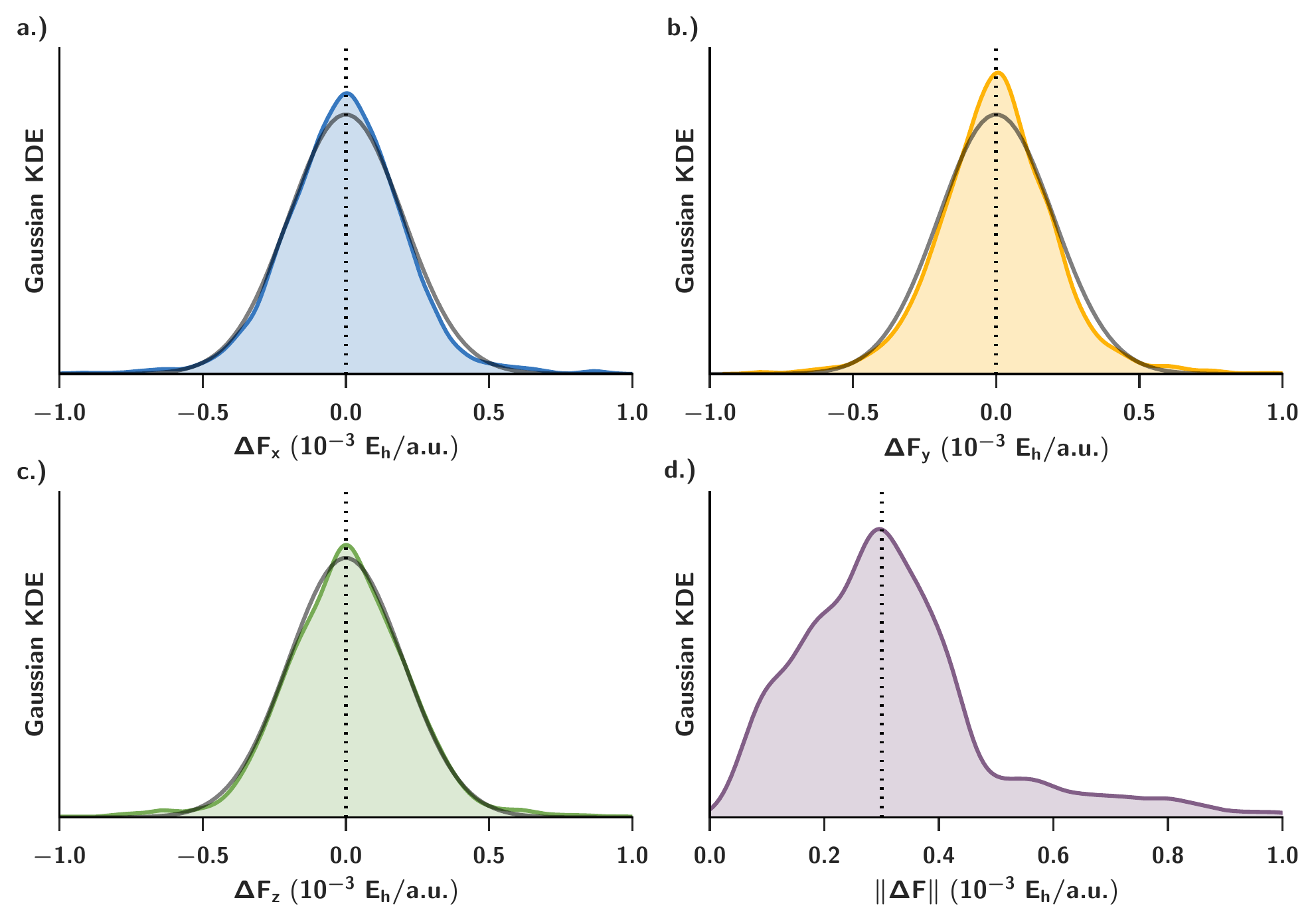} 
\caption{Gaussian kernel density estimates of the distributions of MBD forces acting on each atom at the optimized geometries of 76 tripeptide structures. {\bf a.)-c.)} Difference of the force components $\Delta \mathbf{F}_i = \mathbf{F}_i - \mathbf{F}_{i,\partial V=0}$, with a normal distribution $\mathcal{N}(0,0.2)$ and dotted line indicating zero mean superposed for reference. {\bf d.)} Norm of the difference of forces, $\| \Delta \mathbf{F} \| = \|  \mathbf{F} - \mathbf{F}_{\partial V=0} \|$, with the dotted line indicating that the peak occurs at $\sim0.3\times 10^{-3}~\rm E_h/a.u.$ 
\label{fig:esi:kde}
}
\end{figure}

We also performed single-point calculations on the optimized geometries of all 76 tripeptide structures to compare the MBD forces computed with and without the $\bm{\partial} V$ contributions.  The peptide structures are much more flexible than the benzene monomer and also have the opportunity for cooperative addition of the Hirshfeld volume gradients along the chain, \textit{i.e.} the local Hirshfeld volume gradients acting at the nearest neighbor level can propagate along the peptide chain and result in a larger change. In Fig.~\ref{fig:esi:kde} we visualize the deviation between the forces computed with full Hirshfeld volume gradients and those computed with $\bm{\partial} V=0$ in several ways: difference of individual force components $\Delta \mathbf{F}_i = \mathbf{F}_i - \mathbf{F}_{i,\partial V=0}$, norm of the difference of forces $\|  \mathbf{F} - \mathbf{F}_{\partial V=0} \|$, relative percentage error $\|\Delta \mathbf{F}\|/\|\mathbf{F}\|$, and distributions of the norms of forces $\|\mathbf{F}\|$ vs. $\|\mathbf{F}_\partial V=0\|$. 

\subsection{Structure of the \texorpdfstring{$\rm C_{60}@C_{60}H_{28}$}{C60@C60H28} Buckyball Catcher Host--Guest Complex}
\begin{figure}[!htbp]
\centering
\includegraphics[width=17.3cm]{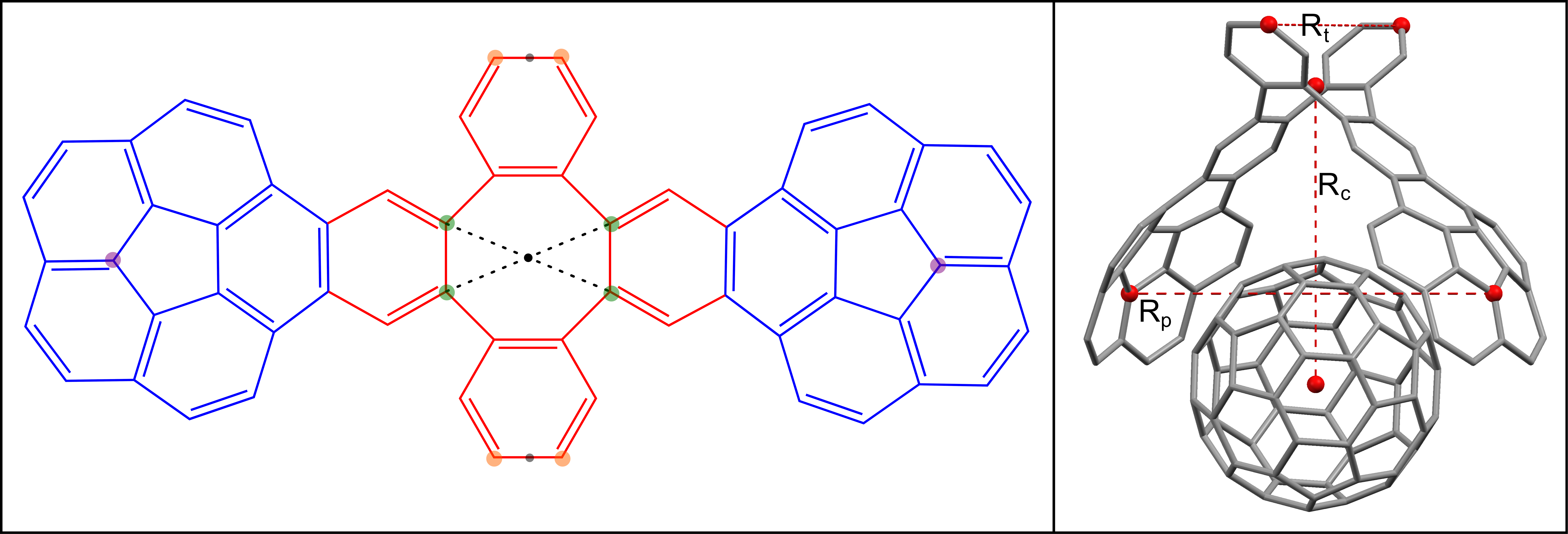} 
\caption{\textbf{Left:} 2D structure of the `buckyball catcher' \ce{C60H28}. The central tetrabenzocyclooctatetraene (TBCOT) tether (red) links the two corannulene bowls (blue). The orange circles mark the four atoms used to define the $R_t$ distance between the back ends of the TBCOT tether. The green circles mark the four atoms used to define the plane from which the distance to the \ce{C60} centroid, $R_c$, is measured. The purple circles mark the two atoms used to define the $R_p$ distance (C10e and C10e' in the notation of Ref.~\citenum{waller_catcher_2012}), which are the most separated atoms of the central five-membered rings of both corannulene subunits.   \textbf{Right:} 3D structure of the $\rm C_{60}@C_{60}H_{28}$ complex with the three distances $R_c$, $R_p$, and $R_t$ highlighted. \label{fig:esi:catcherstructure}} 
\end{figure}

In Fig.~\ref{fig:esi:catcherstructure} the 2D molecular structure of the buckyball catcher host and the 3D structure of the \ce{C60}@\ce{C60H28} host--guest complex with the three distances $R_c$, $R_p$, and $R_t$ are highlighted.  For each DFT-vdW optimized structure of the host, we report the $R_p$ and $R_t$ distances. All geometry optimizations of the \ce{C60}@\ce{C60H28} buckyball catcher host--guest complex started from the TPSS+D3/def2-TZVP structures in the S12L set.~\cite{grimme_supramolecular_2012} We optimized the complex, guest \ce{C60}, and conformers \textbf{a} and \textbf{b} of the host.  Structures of the complex, guest \ce{C60}, and host optimized with other functionals and vdW correction schemes can be found in the supplemental information of the following references:  Ref.~\citenum{grimme_supramolecular_2012}: TPSS+D3/def2-TZVP, Ref.~\citenum{muck-lichtenfeld_catcher_2010}: B97-D/TZVP, Ref.~\citenum{zhao_catcher_2008}: M06-2L/MIDI!. 

\newpage
\subsection{Self-Consistent Screening}\label{appendix:SCS_matrix_solution}
Self-consistent screening (SCS) is accomplished by solving the following non-homogeneous system of linear equations at a given complex frequency $\mathbbm{i}\omega$  (Eq. (17) in DiStasio et al.~\cite{distasio_2014}):
\begin{equation}
\overline{\mathbf{\alpha}}_a(\mathbbm{i}\omega) = \mathbf{\alpha}_a(\mathbbm{i}\omega) - \mathbf{\alpha}_a(\mathbbm{i}\omega) \sum_{b\neq a}^{N} \mathbf{T}_{ab}\;  \overline{\mathbf{\alpha}}_a(\mathbbm{i}\omega). 		\label{distasio_eq17}
\end{equation} 
To accomplish a range-separated self-consistent screening (rsSCS), we replace $\mathbf{T}$ with $\mathbf{T}_{\rm SR}$ (see Ref.~\citenum{ambrosetti_2014}).  Eq.~\eqref{distasio_eq17} can then be written as a matrix equation as:
\begin{equation}
\overline{\mathbf{A}} = \mathbf{A} - \mathbf{A}\, \mathbf{T}_{\rm SR} \, \overline{\mathbf{A}}  \label{rsSCS_matrix}
\end{equation} 
Note that $\zeta_{pp} = 0$ so $\mathbf{T}_{pp}=0$ naturally (see Eq.~\eqref{TGG_cartesian}).  Thus, the sum $\sum_{b\neq a} \mathbf{T}_{ab}\,  \overline{\mathbf{\alpha}}_a$ is accomplished by the product $\mathbf{T}_{\rm SR} \,\overline{\mathbf{A}}$.
Rearranging Eq.~\eqref{rsSCS_matrix} and then left multiplying by $\mathbf{A}^{-1}$ gives:
\begin{eqnarray}
\overline{\mathbf{A}} +\mathbf{A}\, \mathbf{T}_{\rm SR} \, \overline{\mathbf{A}}  &=& \mathbf{A} \\
\mathbf{A}^{-1}\left[\mathbbm{I}+\mathbf{A}\, \mathbf{T}_{\rm SR} \right] \overline{\mathbf{A}}  &=&\mathbf{A}^{-1} \mathbf{A} \\
 \left [\mathbf{A}^{-1} + \mathbf{T}_{\rm SR} \, \right]\overline{\mathbf{A}}    &=&\mathbbm{I}
\end{eqnarray}
Left multiplying by the inverse of the bracketed quantity yields:
\begin{equation}
 \overline{\mathbf{A}}  = \left [\mathbf{A}^{-1} + \mathbf{T}_{\rm SR} \, \right]^{-1}  
\end{equation}

\subsection{Derivation of \texorpdfstring{$\partial\mathbf{T}^{ij}$}{dT}}\label{appendix:derivation_of_TGG_derivative}
To break the derivative of $\mathbf{T}^{ij}$ into smaller pieces, we define some convenience functions:
\begin{eqnarray}
U &\equiv& {\rm erf}\left[\zeta\right] -  \frac{2}{\sqrt{\pi}}\zeta \exp\left[-\zeta^2\right]  \\
\mathbf{W}^{ij} &\equiv&   \left(\frac{R^i R^j}{R^5}\right) \, \frac{4}{\sqrt{\pi}}  \zeta^3  \exp\left[-\zeta^2\right] 	 \\
\mathbf{T}^{ij}_{\rm dip} &\equiv& - 3 \left(\frac{R^iR^j}{R^5} 	\right) + \frac{\delta_{ij}}{R^3}  
\end{eqnarray}
So in terms of these functions, $\mathbf{T}^{ij}$ is:
\begin{eqnarray}
 \mathbf{T}^{ij} &=& U \mathbf{T}^{ij}_{\rm dip} + \mathbf{W}^{ij}  \\
 \Rightarrow \bm{\partial} \mathbf{T}^{ij} &=& U \bm{\partial} \mathbf{T}^{ij}_{\rm dip} + \mathbf{T}_{\rm dip} \bm{\partial} U + \bm{\partial} \mathbf{W}^{ij}
\end{eqnarray}
The derivative of $\mathbf{T}_{\rm dip}^{ij}$ is given in Eq.~\eqref{Tdip_derivative}.
Note that we can write $ \bm{\partial}  \left(R^i R^j/R^5\right)$ in terms of  $ \bm{\partial} \mathbf{T}_{\rm dip}^{ij}$ as:
\begin{equation}
 \bm{\partial}  \left(\frac{R^i R^j}{R^5}\right)  =  -\frac{1}{3} \bm{\partial} \mathbf{T}_{\rm dip}^{ij} - \frac{\delta_{ij}}{R^4}\bm{\partial} R  	
\end{equation}
So the derivatives of $U$ and $\mathbf{W}^{ij}$ are:
\begin{eqnarray}
\bm{\partial} U &=&  \frac{4}{\sqrt{\pi}} \zeta^2 \exp\left[-\zeta^2\right]   \bm{\partial} \zeta \\
 \bm{\partial} \mathbf{W}^{ij}   &=&  \left(\frac{R^i R^j}{R^5}\right) \left[3   -2 \zeta^2 \right]   \frac{4}{\sqrt{\pi}}  \zeta^2\exp\left[-\zeta^2\right] \bm{\partial} \zeta  + \frac{4}{\sqrt{\pi}} \zeta^3 \exp\left[-\zeta^2\right]	  \left( -\frac{1}{3} \bm{\partial} \mathbf{T}_{\rm dip}^{ij} - \frac{\delta_{ij}}{R^4}\bm{\partial} R  \right) 
\end{eqnarray}
Now define $h(\zeta) \equiv \frac{4}{\sqrt{\pi}}\zeta^2  \exp\left[-\zeta^2\right] $. 
\begin{eqnarray}
\Rightarrow \bm{\partial} U &=&   h(\zeta) \bm{\partial} \zeta \\
\Rightarrow \bm{\partial} \mathbf{W}^{ij}   &=&  \left(\frac{R^i R^j}{R^5}\right) \left[3   -2 \zeta^2 \right]  h(\zeta) \bm{\partial} \zeta  + \zeta\,h(\zeta)	 \left(  -\frac{1}{3}  \bm{\partial} \mathbf{T}_{\rm dip}^{ij} - \frac{\delta_{ij}}{R^4}\bm{\partial} R  \right)	   
\end{eqnarray}
In terms of $h(\zeta)$ we can then write $\bm{\partial} \mathbf{T}^{ij}$ as:
\begin{eqnarray}
\bm{\partial} \mathbf{T}^{ij} &=&  \left[{\rm erf}\left[\zeta\right] - \frac{1}{2} \frac{h(\zeta)}{\zeta}\right] \bm{\partial} \mathbf{T}^{ij}_{\rm dip} 
 + \zeta\,h(\zeta)	 \left( -\frac{1}{3} \bm{\partial} \mathbf{T}_{\rm dip}^{ij} - \frac{\delta_{ij}}{R^4}\bm{\partial} R  \right)	   + \left[ \mathbf{T}_{\rm dip}   +  \left(\frac{R^i R^j}{R^5}\right) \left[3   -2 \zeta^2 \right] \right]  h(\zeta) \bm{\partial} \zeta  
\end{eqnarray}
Where the derivative of $\zeta_{ab}$ is in Eq.~\eqref{zeta_derivative}.

\subsection{Scaling of Gauss-Legendre Quadrature}\label{sec:GL-scaling}
To transform Gauss-Legendre quadrature from the interval $x_p \in [-1,1]$, to the semi-infinite interval $y_p \in [0,\infty)$, we map the abscissa $x_p$ and weights $w_p$ with an algebraic scaling:
 \begin{eqnarray}
y_p \in [0,\infty) \qquad y_p &=& L \frac{(1+x_p)}{(1-x_p)}\qquad x_p \in [-1,1] \\
g_p &=& - \frac{ 2 L }{(1-x_p)^2} w_p
 \end{eqnarray}
There are many different possible transformations to $[0,\infty)$, but the algebraic mapping is quite robust for quadrature of functions $f(x)$ that decay algebraically in $|x|$ as $x\to \infty$~\cite{boyd_1987}.  Since the isolated atom dynamic polarizability is expected to decay as $\alpha(\mathbbm{i}\omega) \propto 1/\omega^2$,~\cite{derevianko_2010} we found the algebraic scaling to be preferable, although other choices such as $y_p = L \tan\left(\tfrac{\pi}{2} \tfrac{(x_p+1)}{2}\right)$  also perform well. Since the number of Gauss-Legendre quadrature points, $n$, determines the number of self-consistent screening computations that must be performed to determine $\overline{\omega}_a$, the computational cost (and numerical error) of evaluating the MBD correlation energy can be varied by adjusting the number of quadrature points.   The quadrature error is also sensitive to the scale factor $L$.

Based on the available atomic dynamic polarizability reference data in Derevianko et al.~\cite{derevianko_2010} and the free atom reference quantities used in the TS method,~\cite{tkatchenko_2009} our quadrature method should be able to integrate a response function for excitation frequencies in the range of $\omega_i \sim 0.06$ (K) to $\omega_i \sim1.2$ (Ne).  In optimizing the number of Gauss-Legendre points, $n$, and the scale factor, $L$, we used the Casimir-Polder integral for a single excitation frequency dipole oscillator as a trial function with $\omega_a$ varying in the range $[10^{-2},10^2]$.
\begin{equation}
C_6(\omega_a) = \frac{3}{\pi} \int_0^\infty  \left[\frac{f}{[\omega_a^2 -(\mathbbm{i} \omega)^2 ]}\right]^2 { \rm d}(\mathbbm{i}\omega) = \frac{3}{4} \frac{f^2}{\omega_a^3}
\end{equation}
Using this trial function, we choose $n=20$ quadrature points and a scale factor $L=\tfrac{6}{10}$, which gives integration with a relative error less than $10^{-6}$ for all excitation frequencies in the range $[0.07,5]$ and also performed well in self-consistent screening computations across a  range of isolated atomic systems.

\subsection{Additional Computational Details}\label{sec:esi:additional_computational_details}

All geometry optimizations were performed using the quasi-Newton Broyden-Fletcher-Goldfarb-Shanno (BFGS) algorithm~\cite{broyden_bfgs_1969,goldfarb_bfgs_1970,fletcher_bfgs_1970,shanno_bfgs_1970},  with default parameters. 

\subsubsection{\textsc{Quantum ESPRESSO}}
\begin{figure*}[!htbp]
\centering
\fbox{\includegraphics[width=8.3cm]{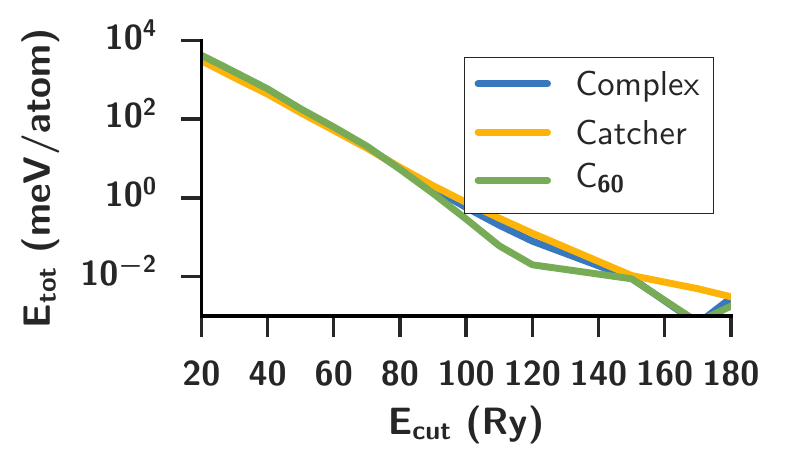}}\qquad 
\fbox{\includegraphics[width=8.3cm]{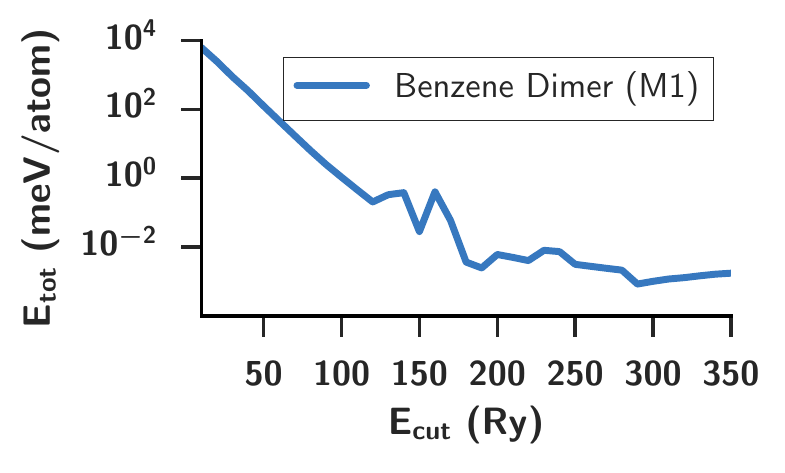}}
\caption{ Convergence of the total energy per atom with respect to kinetic energy wavefunction cutoff ($\rm E_{cut}$) of the planewave basis expansion for \textbf{Left:} simulations of the \ce{C60} catcher complex, \textbf{Right:} the benzene dimer (M1 configuration).\label{fig:esi:catcherconvergence}} 
\end{figure*}

Cartesian coordinate geometry optimizations in \textsc{Quantum ESPRESSO} (\textsc{QE})~\cite{giannozzi_2009} were performed in the \textsc{PWscf} module in large simple cubic unit cells. Table~\ref{esi:tolerances_table} gives details of the convergence tolerances, kinetic energy wavefunction cutoffs, and unit cell sizes used for each system.  Since \textsc{QE} uses Rydberg energy units ($\rm 1~Ry= \tfrac{1}{2}~E_h$), we report the tolerances in these units. The PBE functional~\cite{PBE_ref1,PBE_ref2} was used with Hamann-Schlueter-Chiang-Vanderbilt (HSCV) norm-conserving pseudopotentials~\cite{hamann_prl_1979,*bachelet_prb_1982,*vanderbilt_1985} obtained from the FPMD pseudopotential repository~\cite{gygi_fmpd_repository} (and converted to UPF format using a modified version of \textsc{qso2upf~v1.2}~\cite{gygi_qso2upf}).   All \textsc{QE} calculations were run at the $\Gamma$ point using a charge density cutoff of $\rho_{\rm cut} = \rm 4 E_{cut}$. PBE+MBD jobs used 20 quadrature points for the Casimir-Polder integration.
To ensure a fair comparison with our implementation of the MBD model, all TS calculations~\cite{tkatchenko_2009} were performed as \textit{a posteriori} corrections to the solution of the non-linear Kohn-Sham equations, \textit{i.e.} we turned off the self-consistent density updates from TS.   In Fig.~\ref{fig:esi:catcherconvergence} we present the results of convergence testing with respect to the kinetic energy cutoff in the planewave basis set expansion, showing that the total energy per atom was converged to better than 0.3 meV/atom for each system.

\begin{table}
\caption{Convergence tolerances and unit cell sizes used in \textsc{PWscf} geometry optimizations, reported in Rydberg atomic units ($\rm 1~Ry= \tfrac{1}{2}~E_h$)\label{esi:tolerances_table}}
\centering
\begin{tabular}{lccc}\toprule
			            & Benzene Dimer  & Peptides 		     & \ce{C60} Catcher  \\\midrule
$\rm E_{scf}~(Ry)$       &	$10^{-8}$	     &	$10^{-8}$		     &	$10^{-8}$	    	\\		      
$\rm E_{cut}~(Ry)$       &  400		     &	145			     &	110			\\
$\rm E_{tot}~(Ry)$        &  $10^{-8}$	     & $5\times 10^{-7}$  & $5\times10^{-7}$ \\
$\rm F_{tot}~(Ry/a_0)$ &  $10^{-4}$	     & $5\times 10^{-4}$  &  $5\times 10^{-4}$, \\
Cell Size $(\rm a_0)$    & 30                   & 30 			     & 50 \\
Grid Spacing (\AA)  & 0.04 & 0.07		     & 0.12 \\ 
\bottomrule
\end{tabular}
\end{table}

\subsubsection{Orca}
Redundant internal coordinate geometry optimizations in \textsc{Orca}~v3.03~\cite{neese_orca_2012} were performed with the PBE functional~\cite{PBE_ref1,PBE_ref2} with the atom-pairwise version of the D3 dispersion correction of Grimme \textit{et al}.~\cite{grimme_d3_2010}, using Becke-Johnson (BJ) damping.~\cite{grimme_bj_2011}  \textsc{Orca}~v3.03 implements D3 in the \textsc{DFTD3~v2.1r6} software, which does not contain analytical gradients of the three-body term. The geometric counterpoise correction (gCP) of Kruse~\textit{et al.} was employed in all \textsc{Orca} calculations.~\cite{kruse_geometrical_2012}  We employed the Ahlrichs def2-TZVP basis set~\cite{weigend_2005} coupled with an auxiliary Ahlrichs TZVP basis set~\cite{eichkorn_1997,weigend_2006} for the RI-J approximation.~\cite{baerends_1973,dunlap_1979,kendall_1997}  All calculations used the \textsc{g4} final integration grid.  All calculations used ``tight'' SCF tolerances; calculations on the benzene dimer and \ce{C60} catcher used ``tight'' optimization tolerances, while those of the peptides used default optimization tolerances. Table~\ref{esi:orca_tolerances} lists the tolerances corresponding to these two settings.  

\begin{table}
\caption{Convergance tolerances used in \textsc{Orca} geometry optimizations, reported in Hartree atomic units \label{esi:orca_tolerances}}
\centering
\begin{tabular}{lccc}\toprule
			            	   & Benzene Dimer  	& Peptides   		& \ce{C60} Catcher  \\\midrule
$\rm E_{scf}~(E_h)$       	  & $10^{-8}$	    	& $10^{-8}$  		& $10^{-8}$	    	\\	
$\rm E_{tot}~(E_h)$        	  & $10^{-6}$        	& $5\times10^{-6}$  	& $10^{-6}$ \\
$\rm F_{Max}~(E_h/a_0)$  & $10^{-4}$	     	& $3\times 10^{-4}$  &  $ 10^{-4}$ \\
$\rm F_{RMS}~(E_h/a_0)$ & $3\times10^{-5}$	& $10^{-4}$  		&  $3\times10^{-5}$ \\
Max Disp. $\rm(a_0)$ 	  & $10^{-3}$	 	& $4\times 10^{-3}$	&  $10^{-3}$ \\
RMS Disp. $\rm(a_0)$ 	  & $6\times10^{-4}$	& $2\times 10^{-3}$	& $6\times10^{-4}$ \\
\bottomrule
\end{tabular}
\end{table}

\newpage

\subsection{Cartesian Coordinates of Structures}
In the accompanying supplementary \textsf{.txt} files we provide the Cartesian coordinates (in~\AA) of all structures considered in the text. 

\subsubsection{Stationary points of the benzene dimer potential energy surface}
We consider ten configurations of the benzene dimer, which correspond to stationary points of the SAPT(DFT)\footnote{Podeszwa et al. used the PBE0 functional with Fermi-Amaldi asymptotic correction and with the Tozer-Handy splicing scheme, and an aug-cc-pVTZ Dunning basis with a monomer-centered ``plus'' basis set scheme. Their symmetry-adapted perturbation theory procedure used density fitting. They also used the vibrationally averaged monomer geometry ($r_{CC}=1.3965~\text{\AA}$ and $r_{CH} = 1.085~\text{\AA}$) from Tamagawa \textit{et al. J. Mol. Struct.} 1976, \textbf{30}, 243.} potential energy surface in Ref.~\citenum{podeszwa_2006}.  Using a fixed monomer geometry from the SDQ-MBPT(4)/cc-pVTZ results of Gauss and Stanton~\cite{gauss_stanton_2000}, Bludsk\'{y} \textit{et al.}~\cite{bludsky_2008} optimized these 10 configurations of the benzene dimer at the PBE/CCSD(T) level of theory, with an aug-cc-pVDZ basis set and counterpoise correction. 
The benzene dimer geometries may be found in the following files:
\begin{verbatim}
    benzene_monomer.txt      benzene_dimer_ccsd.txt    benzene_dimer_mbd.txt
    benzene_dimer_ts.txt     benzene_dimer_d3.txt
\end{verbatim}

\subsubsection{Secondary structure of isolated polypeptides}
We considered 76 conformers of the following 5 isolated small peptides, GFA, FGG, GGF, WG, and WGG, containing the residues glycine (G), alanine (A), phenylalanine (F), and tryptophan (W). 
Our starting geometries were taken from \href{http://www.begdb.com/index.php?action=oneDataset&id=9&state=show&order=ASC&by=name_m&method=}{www.begdb.com}, corresponding to the RI-MP2/cc-pVTZ optimized structures given in the supplemental information of Valdes \textit{et al.}~\cite{valdes_benchmark_2008}. Table~\ref{table:peptide_labels} gives the correspondence between the structure indexing scheme used in this work, the nomenclature of the \textsc{begdb} database and the nomenclature of  Valdes \textit{et al.}~\cite{valdes_benchmark_2008}
 Due to the ease of downloading structures from the \textsc{begdb} database, we only present our PBE+MBD and PBE+D3 optimized geometries in the accompanying text files.  
 The peptide geometries may be found in the following files:
\begin{verbatim}
    GFA_mbd.txt   FGG_mbd.txt   GGF_mbd.txt   WG_mbd.txt   WGG_mbd.txt
    GFA_ts.txt    FGG_ts.txt    GGF_ts.txt    WG_ts.txt    WGG_ts.txt
    GFA_d3.txt    FGG_d3.txt    GGF_d3.txt    WG_d3.txt    WGG_d3.txt
\end{verbatim}

\begin{table}[!htbp]
\caption{Peptide naming conventions in this work, the  \textsc{begdb} database, and Ref.~\citenum{valdes_benchmark_2008}}
\begin{tabular}{lll c lll}\label{table:peptide_labels}
\\\hline
This work & \textsc{begdb} & Ref.~\citenum{valdes_benchmark_2008} & \vline& This work & \textsc{begdb} & Ref.~\citenum{valdes_benchmark_2008}  \\\hline
0	&	252\_FGG55	&	FGG\_055	&	\vline&	38	&	228\_GGF08	&	GGF\_08\\
1	&	263\_FGG80	&	FGG\_080	&	\vline&	39	&	230\_GGF09	&	GGF\_09\\
2	&	253\_FGG99	&	FGG\_099	&	\vline&	40	&	225\_GGF10	&	GGF\_10\\
3	&	264\_FGG114	&	FGG\_114	&	\vline&	41	&	229\_GGF11	&	GGF\_11\\
4	&	257\_FGG215	&	FGG\_215	&	\vline&	42	&	224\_GGF12	&	GGF\_12\\
5	&	258\_FGG224	&	FGG\_224	&	\vline&	43	&	222\_GGF13	&	GGF\_13\\
6	&	255\_FGG252	&	FGG\_252	&	\vline&	44	&	221\_GGF14	&	GGF\_14\\
7	&	254\_FGG300	&	FGG\_300	&	\vline&	45	&	226\_GGF15	&	GGF\_15\\
8	&	265\_FGG357	&	FGG\_357	&	\vline&	46	&	214\_WGG01	&	WGG\_01\\
9	&	256\_FGG366	&	FGG\_366	&	\vline&	47	&	211\_WGG02	&	WGG\_02\\
10	&	259\_FGG380	&	FGG\_380	&	\vline&	48	&	209\_WGG03	&	WGG\_03\\
11	&	260\_FGG412	&	FGG\_412	&	\vline&	49	&	208\_WGG04	&	WGG\_04\\
12	&	261\_FGG444	&	FGG\_444	&	\vline&	50	&	210\_WGG05	&	WGG\_05\\
13	&	262\_FGG470	&	FGG\_470	&	\vline&	51	&	206\_WGG06	&	WGG\_06\\
14	&	266\_FGG691	&	FGG\_691	&	\vline&	52	&	215\_WGG07	&	WGG\_07\\
15	&	248\_GFA01	&	GFA\_01	&	\vline&	53	&	207\_WGG08	&	WGG\_08\\
16	&	239\_GFA02	&	GFA\_02	&	\vline&	54	&	217\_WGG09	&	WGG\_09\\
17	&	247\_GFA03	&	GFA\_03	&	\vline&	55	&	219\_WGG10	&	WGG\_10\\
18	&	251\_GFA04	&	GFA\_04	&	\vline&	56	&	216\_WGG11	&	WGG\_11\\
19	&	250\_GFA05	&	GFA\_05	&	\vline&	57	&	220\_WGG12	&	WGG\_12\\
20	&	245\_GFA06	&	GFA\_06	&	\vline&	58	&	218\_WGG13	&	WGG\_13\\
21	&	237\_GFA07	&	GFA\_07	&	\vline&	59	&	212\_WGG14	&	WGG\_14\\
22	&	242\_GFA08	&	GFA\_08	&	\vline&	60	&	213\_WGG15	&	WGG\_15\\
23	&	241\_GFA09	&	GFA\_09	&	\vline&	61	&	195\_WG01	&	WG\_01\\
24	&	238\_GFA10	&	GFA\_10	&	\vline&	62	&	194\_WG02	&	WG\_02\\
25	&	240\_GFA11	&	GFA\_11	&	\vline&	63	&	191\_WG03	&	WG\_03\\
26	&	244\_GFA12	&	GFA\_12	&	\vline&	64	&	204\_WG04	&	WG\_04\\
27	&	243\_GFA13	&	GFA\_13	&	\vline&	65	&	205\_WG05	&	WG\_05\\
28	&	249\_GFA14	&	GFA\_14	&	\vline&	66	&	193\_WG06	&	WG\_06\\
29	&	236\_GFA15	&	GFA\_15	&	\vline&	67	&	197\_WG07	&	WG\_07\\
30	&	246\_GFA16	&	GFA\_16	&	\vline&	68	&	202\_WG08	&	WG\_08\\
31	&	231\_GGF01	&	GGF\_01	&	\vline&	69	&	198\_WG09	&	WG\_09\\
32	&	234\_GGF02	&	GGF\_02	&	\vline&	70	&	192\_WG10	&	WG\_10\\
33	&	233\_GGF03	&	GGF\_03	&	\vline&	71	&	203\_WG11	&	WG\_11\\
34	&	227\_GGF04	&	GGF\_04	&	\vline&	72	&	201\_WG12	&	WG\_12\\
35	&	235\_GGF05	&	GGF\_05	&	\vline&	73	&	200\_WG13	&	WG\_13\\
36	&	232\_GGF06	&	GGF\_06	&	\vline&	74	&	196\_WG14	&	WG\_14\\
37	&	223\_GGF07	&	GGF\_07	&	\vline&	75	&	199\_WG15	&	WG\_15\\\hline
\end{tabular}
\end{table}

\subsubsection{ \texorpdfstring{$\rm C_{60}@C_{60}H_{28}$}{C60@C60H28} buckyball catcher host--guest complex}
 The buckyball catcher host--guest geometries may be found in the following files:
\begin{verbatim}
	catcher_host.txt
	catcher_monomer.txt
	catcher_complex.txt
\end{verbatim}



\end{document}